# Generation and Simulation of Yeast Microscopy Imagery with Deep Learning

Bachelor-Thesis of **Christoph Reich**
Darmstadt, October 8, 2020

**Supervisors:**
Prof. Dr. techn. Heinz Koeppl
Tim Prangemeier, M. Sc.
Christian Wildner, M. Sc.

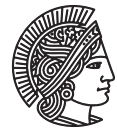

Department of Electrical Engineering and Information Technology
**Bioinspired Communication Systems Lab**

**Generation and Simulation of Yeast Microscopy Imagery with Deep Learning**

Bachelor-Thesis of **Christoph Reich**

**Supervisors:**
Prof. Dr. techn. Heinz Koeppl
Tim Prangemeier, M. Sc.
Christian Wildner, M. Sc.

Darmstadt, October 8, 2020

# Abstract

Time-lapse fluorescence microscopy (TLFM) is an important and powerful tool in synthetic biological research. Modeling TLFM experiments based on real data may enable researchers to repeat certain experiments with minor effort. This thesis is a study towards deep learning-based modeling of TLFM experiments on the image level. The modeling of TLFM experiments, by way of the example of trapped yeast cells, is split into two tasks. The first task is to generate synthetic image data based on real image data. To approach this problem, a novel generative adversarial network, for conditionalized and unconditionalized image generation, is proposed. The second task is the simulation of brightfield microscopy images over multiple discrete time-steps. To tackle this simulation task an advanced future frame prediction model is introduced. The proposed models are trained and tested on a novel dataset that is presented in this thesis. The obtained results showed that the modeling of TLFM experiments, with deep learning, is a proper approach, but requires future research to effectively model real-world experiments.

# Contents

# Notations

In this thesis, the notations listed below are used. The notations are similar to the notations used in the book Deep Learning by Goodfellow et al. [1].

**Numbers, Array, Tensors and Sets**

| | |
|---|---|
| $a$ | A scalar |
| $\boldsymbol{a}$ | A vector |
| $\boldsymbol{A}$ | A matrix |
| $\mathbf{A}$ | A tensor |
| $\boldsymbol{I}_n$ | Identity matrix with n rows and n columms |
| $\boldsymbol{I}$ | Identity matrix with dimensionality implied by context |
| $\boldsymbol{e}^{(i)}$ | Standard basis vector with a 1 at position $i$ |
| $\text{diag}(\boldsymbol{a})$ | A square, diagonal matrix with diagonal entries given by $\boldsymbol{a}$ |
| $\mathbb{A}$ | A set |
| $\mathbb{R}$ | Set of all real numbers |
| $\mathbb{R}^+$ | Set of all positive real numbers (without 0) |
| $\{0, 1, \ldots, n\}$ | The set of all integers between 0 and $n$ |
| $[a, b]$ | The real interval including $a$ and $b$ |
| $(a, b]$ | The real interval excluding $a$ but including $b$ |

**Indexing**

| | |
|---|---|
| $a_i$ | Element $i$ of vector $\boldsymbol{a}$, with indexing starting at 1 |
| $a_{-i}$ | All elements of vector $\boldsymbol{a}$ except for element $i$ |
| $A_{i,j}$ | Element $i, j$ of matrix $\boldsymbol{A}$ |
| $\boldsymbol{A}_{i,:} = \boldsymbol{A}(i,:)$ | Row $i$ of matrix $\boldsymbol{A}$ |
| $\boldsymbol{A}_{:,i} = \boldsymbol{A}(:,i)$ | Column $i$ of matrix $\boldsymbol{A}$ |
| $A_{i,j,k} = A(i,j,k)$ | Element $(i, j, k)$ of 3d tensor $\mathbf{A}$ |
| $\mathbf{A}_{:,:,i} = \mathbf{A}(:,:,i)$ | 2d slice of 3d tensor $\mathbf{A}$ |
| $\mathbf{A}_{1:4} = \mathbf{A}(1:4)$ | Sub-slice of tensor $\mathbf{A}$ |

**Linear Algebra Operations**

| | |
|---|---|
| $A^\mathsf{T}$ | Transpose of matrix $A$ |
| $A^+$ | Moore-Penrose pseudoinverse of $A$ |
| $A \odot B$ | Element-wise product of $A$ and $B$ |
| $\det(A)$ | Determinant of $A$ |
| $A * K$ | Convolution operation between the matrices $A$ and $K$ |
| $A *^\mathsf{T} K$ | Transposed convolution operation between the matrices $A$ and $K$ |
| $\operatorname{Tr}(X)$ | Trace of the matrix $X$ |

**Calculus**

| | |
|---|---|
| $\frac{dy}{dx}$ | Derivative of $y$ with respect to $x$ |
| $\frac{\partial y}{\partial x}$ | Partial derivative of $y$ with respect to $x$ |
| $\nabla_x y$ | Gradient of $y$ with respect to $\boldsymbol{x}$ |
| $\nabla_X y$ | Matrix derivatives of $y$ with respect to $X$ |
| $\nabla_\mathbf{X} y$ | Tensor containing derivatives $y$ with respect to $\mathbf{X}$ |

**Probability**

| | |
|---|---|
| $\mathbb{E}[\boldsymbol{x}]$ | Mean of the vector $\boldsymbol{x}$ |
| $\mathbb{E}[f(x)]$ | Expectation of the function $f(x)$ |
| $\operatorname{Var}[\boldsymbol{x}]$ | Variance of the vector $\boldsymbol{x}$ |
| $\operatorname{Cov}[X]$ | Covariance of the matrix $X$ |
| $\operatorname{Std}(\mathbf{X})$ | Standard deviation of the tensor $\mathbf{X}$ |
| $\mathcal{N}(\boldsymbol{x}; \boldsymbol{\mu}, \Sigma)$ | Gaussian distribution over $\boldsymbol{x}$ with mean $\boldsymbol{\mu}$ and covariance $\Sigma$ |
| $\mathcal{N}(0, 1)$ | Gaussian normal distribution with a mean of 0 and a variance of 1 |
| $U(\boldsymbol{x}; 0, 1)$ | Uniform distribution over $\boldsymbol{x}$ in the interval $[0, 1)$ |
| $U(0, 1)$ | Uniform distribution with an interval of $[0, 1)$ |
| $\operatorname{KL}(P_1 \| P_2)$ | Kullback-Leibler divergence of $P_1$ and $P_2$ |
| $P(a)$ | A probability distribution over a discrete variable |
| $p(a)$ | A probability distribution over a continuous variable |
| $\mathbb{E}_{x \sim p}[f(x)]$ | Expectation of $f(x)$ with respect to the distribution $p$ |

**Functions**

| | |
|---|---|
| $f : \mathbb{A} \to \mathbb{B}$ | The function $f$ with domain $\mathbb{A}$ and range $\mathbb{B}$ |
| $f(\mathbf{X}; \mathbf{\Theta})$ | A function of $\mathbf{X}$ parametrized by $\mathbf{\Theta}$ |
| $\log(x)$ | Natural logarithm of $x \in \mathbb{R}$ |
| $\exp(x)$ | Exponential function of $x \in \mathbb{R}$ |
| $\max(x, y)$ | Maximum function of $x, y \in \mathbb{R}$ |

**Datasets**

| | |
|---|---|
| $\mathbb{X}$ | A set of training examples |
| $\boldsymbol{x}^{(i)}$ | The $i$-th example (input) from a dataset |
| $\boldsymbol{y}^{(i)}$ | The target associated with $\boldsymbol{x}^{(i)}$ for supervised learning |
| $\boldsymbol{X}$ | The $m \times n$ matrix with input example $\boldsymbol{x}^{(i)}$ in row $\boldsymbol{X}_{i,:}$ |
| $\mathbb{V}$ | A set of validation examples |

# 1 Introduction

This chapter introduces the motivation and objectives of this thesis. Further, a brief introduction to each chapter and the proposed algorithms is given.

## 1.1 Motivation

Modeling biological experiments, especially time-lapse fluorescence microscopy (TLFM) [2, 3, 4] experiments, on the image-level based on real data, would speed up biological research. Experiments could be repeated based on data from a past experiment, with minor effort, compared to repeating a TLFM experiment in the laboratory, which is very laborious and time-consuming. In the future, it may become possible to combine data from multiple experiments to model a new synthetic experiment.
Deep learning is a widely used tool throughout many different domains. Examples of deep models in biology include the successful application to the inverse protein folding problem where an attention-based model outperformed all recent non-deep learning approaches in terms of speed and accuracy [5]. Other examples include the fast and precise segmentation of trapped yeast cells in time-lapse fluorescence microscopy images [2, 3] and thus resolves a bottleneck in online monitoring of TLFM experiments [6, 7]. Deep learning, despite its widespread success in tasks like image generation [8, 9, 10, 11, 12, 13], optical flow estimation [14, 15, 16, 17, 18] or future frame prediction [19, 20, 21, 22, 23], has yet to be applied to modeling TLFM experiments, on the image-level. Employing deep learning to this problem could be the key towards a possible solution and this thesis should be the first step towards such a deep learning-based solution.
In this thesis, the task of modeling a TLFM experiment is split into two problems, namely generation, and simulation. The task of generation includes both the conditionalized and unconditionalized generation of synthetic images, similar to real images. The task of simulation involves the prediction of multiple future microscopy frames over discrete time-steps, based on given past frames. The resulting models can also be further used to reinforce the robustness existing models for cell segmentation by enlarging existing datasets. Generated synthetic images, for example, could be used for semi-supervised learning [24, 25, 26, 27, 28, 29]. Further, a simulation model, which predicts future frames, could be used for label propagation [30] to generate new synthetic labels of a given dataset.

## 1.2 Objective and Structure of this Thesis

The task of modeling TLFM experiments on the image-level with deep learning is separated into two tasks, namely generation, and simulation. The decision to split the task of modeling TLFM into generation and simulation subtasks is based on the recent advance of deep learning in the tasks of image generation [31, 13] and future frame prediction [22, 23], in terms of the achieved image quality. This thesis examines whether generating and simulating trapped yeast cells on the image-level, based on deep learning, is a suitable approach to modeling TLFM experiments. The examination includes the development and testing of two deep learning algorithms, for generation and simulation, respectively. Further, it is evaluated what future work needs to be done to combine the task of generation and simulation to model TLFM experiments in a real-world setting.

This thesis is structured as follows: Chapter 2 introduces the necessary fundamentals for the later proposed models. An introduction to deep learning, deep feed-forward network networks, and convolutional neural networks, is given. More recent generative adversarial networks are introduced. Finally, the topics of semi-supervised learning, semantic segmentation, and optical flow are presented.

Chapter 3 summarizes recent state-of-the-art methods in the fields of generative adversarial networks and future frame prediction. Especially, the StyleGAN 2 [13] model, the SDC-Net [22], and the general and adaptive robust loss function by Barron [32], is introduced.

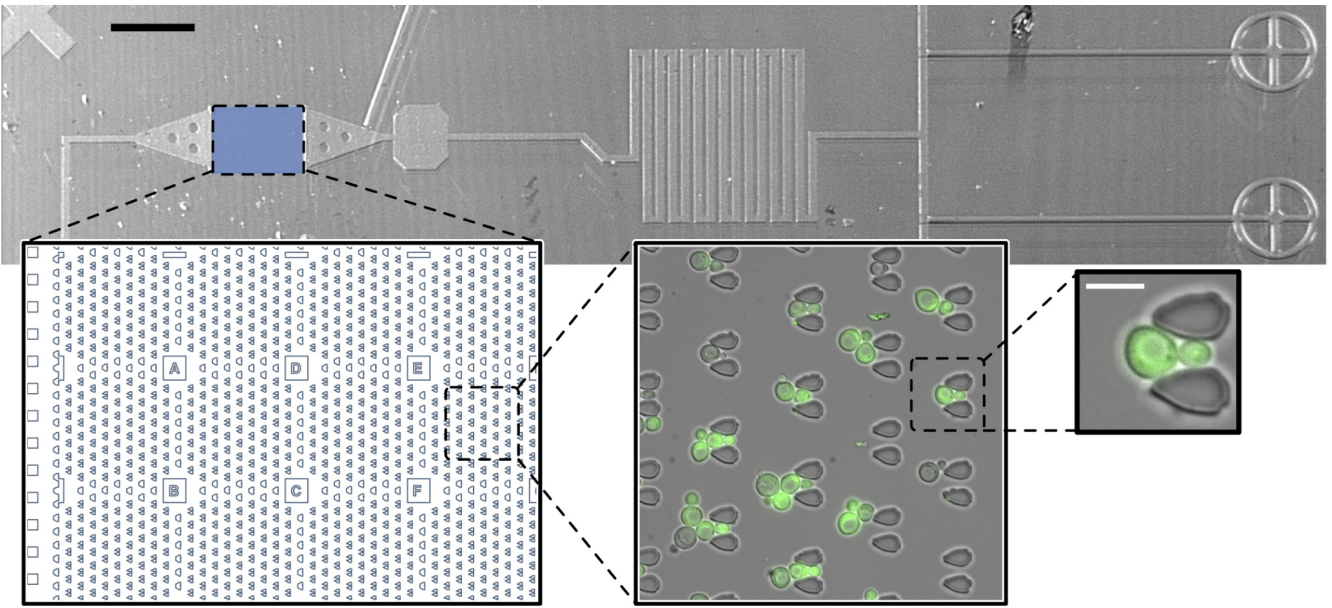

**Figure 1.1:** Microscope images and trap chamber (blue). The trap chamber includes approximately 1000 traps. Multiple specimen images are recorded by a microscopy for different positions. One exemplary specimen image is shown on the right and includes multiple trapped yeast cells and the corresponding fluorescent of the cells in green. One example of a small image including one trap is shown right beside the big specimen image; white scale bar 10$\mu$m. Image taken from [6, 7] by T. Prangemeier, C. Reich and colleagues.

In chapter 4 a novel trapped yeast cell time-series dataset is proposed. This dataset was produced from real-world experimental data of a TLFM experiment, including trapped yeast cells. A semantic of these trapped yeast cells can be seen in figure 1.1. The proposed dataset is later used to perform adversarial image generation and future frame prediction.

Chapter 5 proposes a novel conditionalized and unconditionalized approach for adversarial microscopy image generation. The proposed method is first introduced on a theoretical basis and then qualitative and qualitative evaluated with the trapped yeast cell dataset.

In chapter 6 an advanced method for simulating sequences of microscopy images, over time, is proposed. This method is a further development of the SDC-Net and is also evaluated on the trapped yeast cell dataset.

Finally, in chapter 7 and 8, the proposed methods are discussed, if they are suitable for modeling TLFM experiments, and a final conclusion is drawn.

# 2 Fundamentals

This chapter introduces the fundamentals of deep learning, neural networks, generative adversarial networks, semi supervised learning as well as semantic segmentation. All these topics are anchored or related to the field of deep learning. Deep learning is a subset of machine learning. These relations are visualized in the figure 2.1 below.

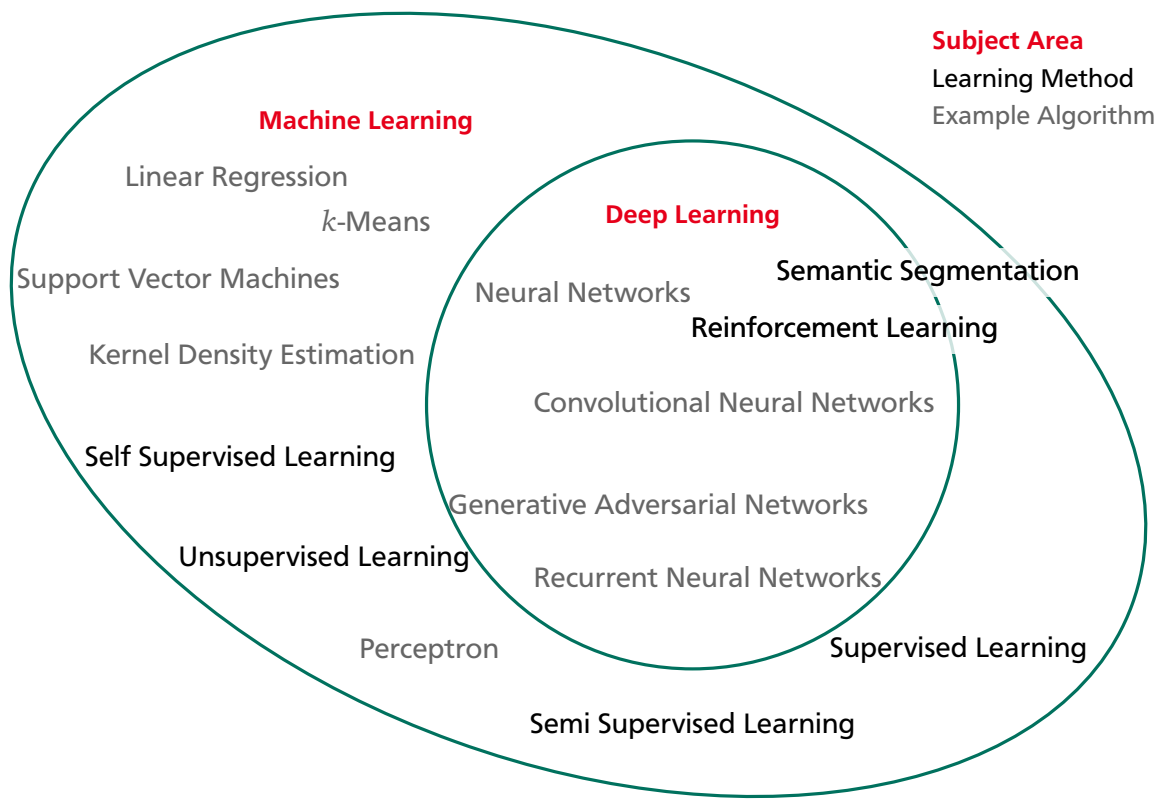

**Figure 2.1:** Venn diagram showing the fields of machine learning and deep learning.

For a more detailed introduction to the topic of deep learning see Goodfellow et al. [1] or Reich [33].

## 2.1 Deep Learning

Deep learning is the active research field that deals with deep learnable architectures to solve the central problem in representation learning by learning intermediate representations. In contrast to classical machine learning, deep learning can learn feature representations of data. These feature representations can are used, in the lower layers of the deep learning architectures, to learn a specific mapping.[1]

Classical machine learning typically relies on handcrafted feature extraction, as illustrated in figure 2.2. This feature extraction step needs a good human knowledge of the overall prob-

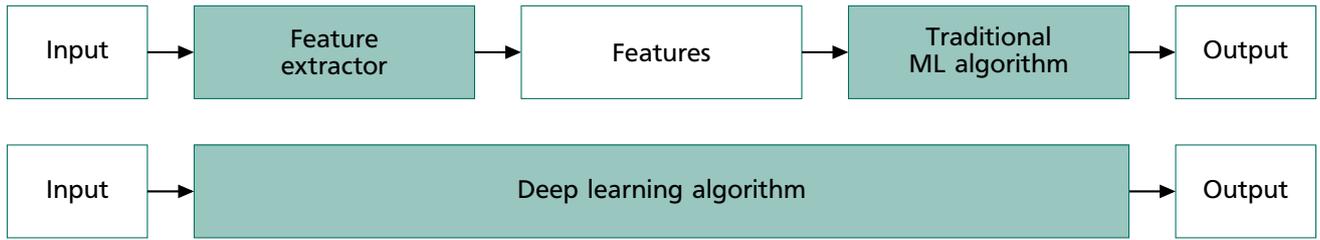

**Figure 2.2:** Deep learning algorithm vs. a class machine learning algorithm. [1]

lem to extract the important features. However, human knowledge is often the performance bottleneck of classical machine learning algorithms because a human can only extract low dimensional features. Furthermore, relying on a handcrafted feature extraction step means that the whole algorithm can not be considered as a real general-purpose algorithm, which is desired to build artificial intelligence algorithms. A deep learning algorithm, like a deep neural network, however, is able to learn this feature representation by itself. Furthermore, a deep learning algorithm can learn these feature representations in a high-dimensional space, which can typically not be explained by a human anymore. Since one deep learning algorithm can be applied to many different problems with minor or no adaptations, the algorithm can be considered as a general-purpose algorithm. [1, 34, 35]

Deep learning algorithms especially convolutional neural networks showed, in the past few years, groundbreaking results in different kinds of use cases throughout computer science. Inter alia deep learning algorithms are the state of the art solution for image classification [36], image segmentation [37], or object detection [38]. In 2016 a deep learning algorithm was able to beat the professional player Lee Sedol in the game of go [39]. This achievement was groundbreaking since the game of go is considered highly more complex than for example the game of chess. However, these use cases can be typically learned supervised or by reinforcement learning. Some use cases are very hard, or even not possible, to be solved by classical learning strategies, including supervised learning, unsupervised learning, or reinforcement learning. For instance, learning a generative model to sample photo-realistic images of faces form a random noise input is very hard to tackle by traditional learning techniques. To solve problems of these kinds a different learning technique so-called generative adversarial learning was proposed. [1, 40, 8]

## 2.2 Deep Feedforward Neural Networks

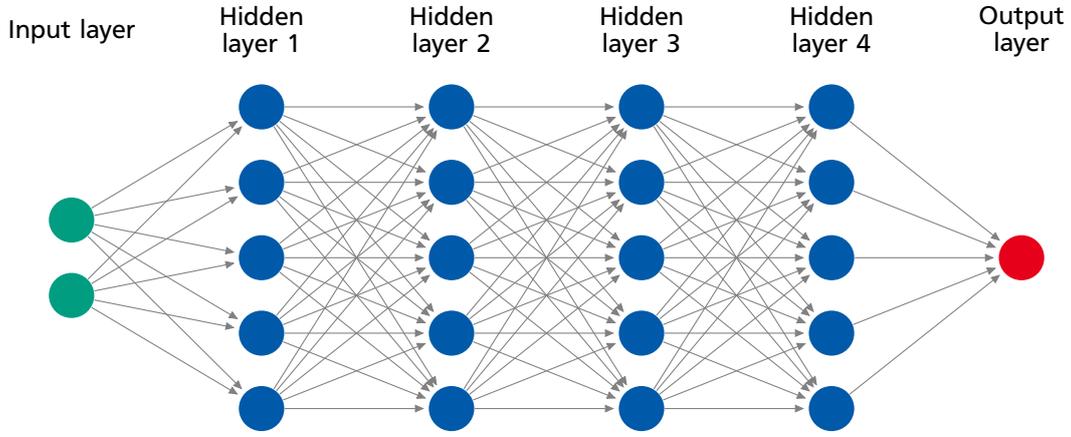

**Figure 2.3:** Graphical representation of a feed forward neural network with four hidden layers. [1]

Deep feedforward neural networks (DFFNN), also called multilayer perceptrons, are considered as the most basic deep learning model. A feedforward neural network is typically utilized to learn a specific function approximation $f^*$. This function can be for example a regression mapping or a classification mapping. Typically a DFFNN is trained in a supervised setting. A DFFNN can be expressed as the mapping $y = f(x;\theta)$, with the learnable parameters $\theta$. These parameters are learned during training, to approximate $f^*$. [1]

### 2.2.1 Mathematical Formulation of a Deep Feedforward Neural Network

The basic operation in a deep forward neural is a vector-matrix multiplication of the input or feature vector $x$ with a matrix of learnable parameters $\theta$. This vector-matrix multiplication is followed by a learnable bias addition with the vector $b$ and a non-linear activation function $g$, to be able to learn non-linear mappings. This combination of operation is typically described as a layer of deep feedforward neural network and can be described mathematically as: [1]

$$f : \mathbb{R}^n \to \mathbb{R}^m, \quad f(x;\theta,b) = g(x^\intercal \theta + b), \quad x \in \mathbb{R}^n, \quad \theta \in \mathbb{R}^{m \times n}, \quad b \in \mathbb{R}^m. \tag{2.1}$$

A deep feedforward neural network consists of at least two layers, which are chained together. This architecture is illustrated in figure 2.3. Thus, a two-layer DFFNN can be expressed as $f(x) = f^{(2)}(f^{(1)}(x))$. The whole mapping $y = f(x;\Theta)$, with the set of parameters $\Theta = \{\theta_1 \in \mathbb{R}^{m \times n}, b_1 \in \mathbb{R}^m, \theta_2 \in \mathbb{R}^{k \times m}, b_2 \in \mathbb{R}^k\}$, can thus be described as: [1]

$$f : \mathbb{R}^n \to \mathbb{R}^k, \quad f(x;\Theta) = g\left(g(x^\intercal \theta_1 + b_1)^\intercal \theta_2 + b_2\right). \tag{2.2}$$

Theoretically, it can be proven that a DFFNN, with at least two layers, is, under additional constraints, a universal function approximator. [1]

### 2.2.2 Training of a Deep Feedforward Neural Network

The phrase of training refers to the process of optimizing the parameters Θ of the deep neural network. Commonly the goal when optimizing deep neural networks is to minimize the empirical risk for a given training dataset. However, other minimization or maximization tasks are possible. To learn the parameters which minimize the empirical risk typically a supervised setting is utilized. In this supervised setting first, the output of the deep neural network is computed, for a mini-batch of input training examples. This process is called forward-pass. Then the loss value for the output predictions and the corresponding labels is computed. Since the goal is to minimize this loss mini-batch stochastic gradient descent is used. To apply a gradient descent method the gradients of each parameter to be optimized have to be computed. This is done by applying backpropagation to the deep feedforward neural network since each of the operations included in the network is differentiable. [41, 34, 35, 1]

## 2.3 Convolutional Neural Networks

Convolutional neural networks (CNN) are the state-of-the-art deep learning architecture for computer vision tasks. In contrast to classical deep feedforward neural networks, deep convolutional neural networks are able to handle efficiently grid-like data e. g. images, videos, or time-series data. However, convolutional neural networks are no recent invention, the Turing Award winner Yann LeCun et al. invented the convolutional neural network back in 1989 [41]. Since problems like the vanishing gradient problem [42, 36] in deep neural networks are solved convolutional neural networks became the state-of-the-art in many areas such as image classification [36, 43], image segmentation [44, 45, 37], adversarial image generation [9, 11, 12, 13], or object detection [46, 47, 48, 38].

Convolutional neural networks are utilizing a discreet convolution operation over given grid-like input data by applying a learnable kernel, to learn feature representations of the input data. Utilizing convolutions instead of linear transformations, as a normal feedforward neural network does, solves multiple issues regarding the linear operation. In a linear operation, as introduced in 2.1, the input must be represented in vector form. When applying a linear operation to grid-like data, like for example an image, this image has to be reshaped to match the needed vector structure. When reshaping, however, the spatial information of the image gets lost. Furthermore, a linear layer is computational very expensive since for each input neuron (element in the input vector) a conection to each output neuron has to be utilized. In CNN's this issue is solved by the idea of parameter sharing because the same learnable kernel is applied over the whole input data. [41, 1, 49]

### 2.3.1 Mathematical Formulation of a Convolutional Neural Network

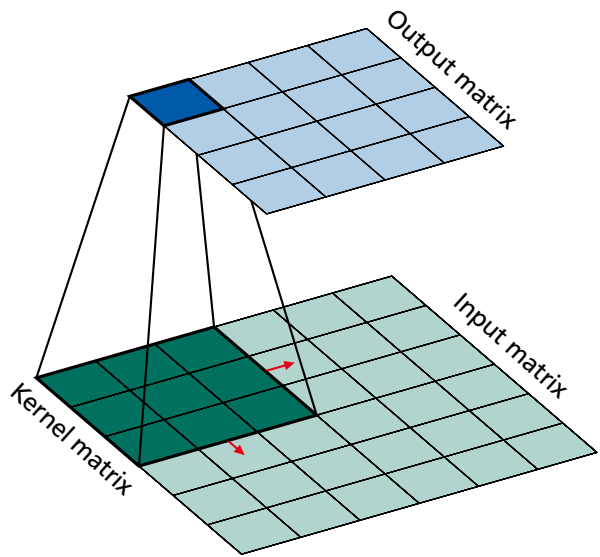

**Figure 2.4:** Visualization of a 2d convolution with a kernel size of 3 × 3, a stride of one, and one input and one output feature channel.

Convolutional operations can be applied to different grid-like data. In this mathematical formulation, only 2d convolutions are introduced since 2d convolutions are the most common form convolutions utilized in a CNN. The idea of a 2d convolutions can be simply adapted to 1d or 3d. [1, 50, 36]

Similar to a deep feedforward neural network a convolutional neural network also consists of layers. Each convolutional layer includes the discreet 2d convolution, between the input tensor **X** and the learnable kernel $\boldsymbol{\theta}$ followed by a bias addition with the vector $\boldsymbol{b} \in \mathbb{R}^{c_{\text{out}}}$ and a non-linear activation function $g$. This whole 2d convolution layer can be mathematically described as: [1, 41, 49]

$$f : \mathbb{R}^{c_{\text{in}} \times h_1 \times w_1} \to \mathbb{R}^{c_{\text{out}} \times h_2 \times w_2}, \quad f(\mathbf{X}; \boldsymbol{\theta}, \boldsymbol{b}) = g(\mathbf{X} * \boldsymbol{\theta} + \boldsymbol{b}). \tag{2.3}$$

The 2d convolution operation $\mathbf{X} * \boldsymbol{\theta}$, in a CNN, can be specified by the equation

$$Y\left(c_{\text{out}_i}, h, w\right) = \sum_{l=1}^{c_{\text{in}}} \sum_{i=-1}^{1} \sum_{j=-1}^{1} \theta(l, i, j) X(l, h+i, w+j). \tag{2.4}$$

Where the tensor **Y** represents the output feature map of the convolution, and second and third sum a kernel size of $3 \times 3$ with the corresponding kernel tensor $\boldsymbol{\theta} \in \mathbb{R}^{c_{\text{in}} \times 3 \times 3}$. Furthermore, the dimensions $c_{\text{in}}$ and $c_{\text{out}}$ represents the number of input and output feature maps. These dimensions are utilized to control the number of features in each layer. If a network utilizes a high number of feature channels the networks is commonly consider as a wide network. [1, 41, 51, 49]

The activation function $g$ in each convolution layer is crucial to enable the whole network to learn non-linear mappings. Commonly a Leaky rectified linear unit (Leaky ReLU) $\max(\alpha x, x)$, $\alpha \in \mathbb{R}$ [52] is used as a standard baseline activation function but more advance activation functions has been published in recent years. However, especially in classification tasks, different activations function are utilized in the final layer of the model. If for example, a multi-class classification problem is present a Softmax activation $\exp(x_j)/\sum_{i \in I} \exp(x_i)$ is typically utilized as the final activation to achieve a probability vector as the output. [1, 49, 53]

### 2.3.2 Additional Layers of a Convolutional Neural Network

A typical CNN utilizes besides the convolution layer also other kinds of layers. These additional layers are briefly described and visualized in the following sections. [1]

#### 2.3.2.1 Linear Layer

Some CNN architectures like the famous LeNet [49] utilizing linear layers in the final stage of the network architecture. This is often done in classification networks since a probability vector regarding the one-hot classification is desired. For more information regarding the linear layer see section 2.2.1. [1, 49]

#### 2.3.2.2 Normalization Layer

Normalization layers are used to reduce the covariance shift in the feature maps inside the deep neural network. Furthermore, normalization layers, like batch normalization, can accelerate the training of deep models. And can also improve the generalization ability of the network. As mentioned one, and foremost the most common normalization layer in deep neural networks is batch normalization. In batch normalization, the whole batch gets normalized to a mean of zero and a standard deviation of one by applying the formula

$$\mathbf{X}_{i,\text{new}} = \gamma_i \frac{\mathbf{X}_i - \mathbb{E}[\mathbf{X}_i]}{\text{Std}(\mathbf{X}_i)} + \beta_i. \quad (2.5)$$

Where the vectors $\boldsymbol{\gamma}$ and $\boldsymbol{\beta}$ are channel-wise learnable parameters which are optimized during training. Furthermore, the batch statistics are accumulated at training time by applying the running average because computing the mean and standard derivation is computationally expensive. At inference time the tracked statistics are then applied in the batch normalization layer for faster inference. Besides batch normalization, other normalization techniques, like instance normalization [54] or layer normalization [55] has been proposed. [56, 1]
The original batch normalization paper claims that normalizing the intermediate feature activations reduces the internal covariance shift, as mentioned earlier. But recent work has claimed differently. Therefore, batch normalization leads to a smoothing of the loss surface which results in a faster training. [57]

### 2.3.2.3 Pooling Layer

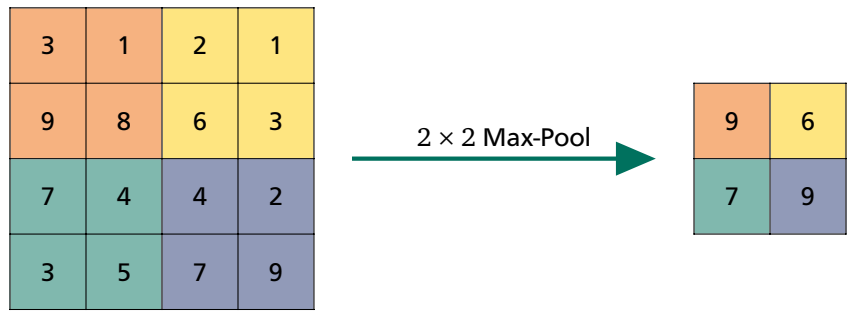

**Figure 2.5:** Visualization of a max-pooling operation with a kernel size of two and a stride of two. The input feature map is show on the left and the output feature map on the right.

Pooling layers are a way to make a CNN invariant to the position of certain features by downscaling the feature maps. A pooling layer is typically used after the activation step of a convolutional layer. In a pooling layer, the input feature maps are downscaled, typically by a factor of two. These downscaling is achieved with different methods. The most common pooling layer is max-pooling, where the input feature map by a pre-defined kernel and stride factor gets downsampled. This kernel returns only the current biggest overlapping value in the input feature map. A visualization of this operation can be seen in figure 2.5. However, other pooling operations like average pooling are available. [1, 49]

### 2.3.2.4 Upsampling Layer

In many computer vision tasks like semantic segmentation, encoder-decoder architectures are used. In these encoder-decoder architectures, like the famous U-Net architecture [45], first, the input image gets transformed by the encoder to a latent vector by reducing the spatial dimensions and increasing the feature channels. The decoder path, however, reverses this trend, since the goal is to produce for example the corresponding semantic segmentation, from the latent vector. To achieve the desired spatial dimensions typically multiple upsampling layers are utilized in the encoder path. This upsampling layer consists of a non-learnable upsampling operation. The most basic upsampling operation is nearest neighbor upsampling, which is visualized in figure 2.6. Nearest neighbor upsampling is utilized in many deep neural networks and achieved good performance, however, in some use cases like super-resolution, more advanced upsampling operations like bilinear upsampling are utilized in each upsampling layer. [37, 45, 58]

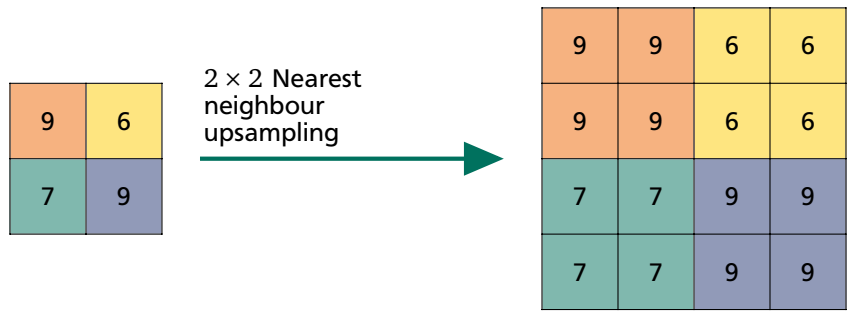

**Figure 2.6:** Visualization of a nearest neighbor upsampling operation with a upscaling factor of two. The input feature map is show in the left and the output feature map on the right.

#### 2.3.2.5 Transposed Convolution Layer

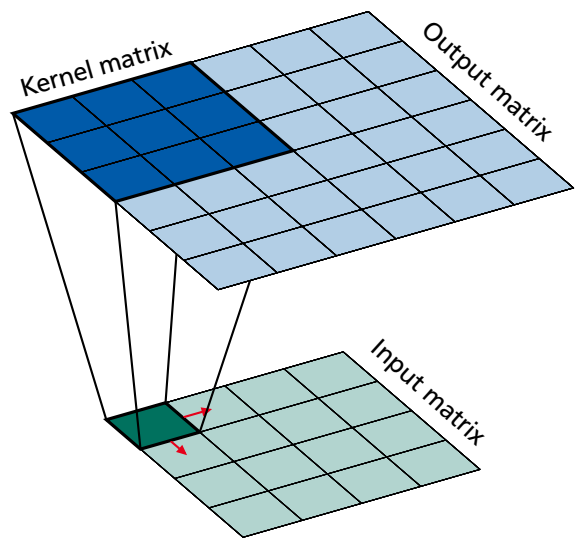

**Figure 2.7:** Visualization of the transposed convolution with a kernel size of 3 × 3, a stride of one, and one input and one output feature channel.

In contrast to the non-learnable upsampling layer, the transposed convolution introduces the idea of a learnable upsampling operation. The transposed convolution applies a kernel to one element in the input matrix to produce the values in the output matrix, corresponding to the filter and the filter position, which can be observed in figure 1. This operation can also be transferred to a normal convolution where dilation is applied to the input matrix. [37]
Using transposed convolution especially in lower layers of a deep convolutional neural network can lead to a problem called checkerboard artifacts. This problem describes the issue that an upsampled image by a transposed convolution often induces checkerboard patterns which is typically not desired in natural images. To solve this issue it is common practice to replace transposed convolutions in the higher layers with non-learnable upsampling layers. [59, 22]

### 2.3.3 Training of a Convolutional Neural Network

A deep convolutional neural network is trained in the same gradient-based fashion as a classical deep feedforward neural network. In a CNN, which aims to perform empirical risk minimization, also backpropagation is performed to compute the gradients of each parameter of the network. This can be done since each layer introduced earlier, is differentiable with respect to its inputs and parameters. To optimize the parameters of the CNN, a first-order optimization method is used. For more information regarding the training process see section 2.2.2. [1]

#### 2.3.3.1 Optimization of a Convolutional Neural Network

As briefly mentioned earlier CNNs, and DNN in general, are commonly optimized during training with a first-order optimization (gradient-based) algorithm. To be precise the foremost common choice of optimizers is mini-batch stochastic gradient descent. Mini-batches are utilized since, computing the gradient of the whole network over the whole dataset is commonly impractical. The state-of-the-art choice for a mini-batch stochastic gradient-based optimization method is the Adam optimizer [60], which combines the idea of momentum and an adaptive learning rate, see algorithm 1.

However, also other state-of-the-art optimizers are available, like for example Adagrad [61] or RMSprop [62]. And also advanced versions of the Adam optimizers, like AdamW [63] or AdamP [64], have been proposed in the past. Especially the AdamP showed strong performance improvements, through a variety of different deep learning use-cases like ImageNet [65] or audio classification, over the classical Adam optimizer. AdamP extends the classical Adam optimizer by a novel projection-based approach that regularizes the momentum-induced norm growth. However, the proposed projection-based approach can also be applied to other optimizers like standard stochastic gradient decent. [64]

The used function $\Pi_{\theta_n}$ in the algorithm 2 is the, by the AdamP [64], proposed projection onto the tangent space $\Pi_{\theta_n}(\Delta\theta_n) = \Delta\theta_n - (\theta_n \Delta\theta_n)\theta_n$, for the input gradient $\Delta\theta_n$ and the parameter $\theta_n$, to be optimized. For a more detailed description see the AdamP paper by Heo et al. [64].

**Algorithm 1** Adam optimizer algorithm. [1, 60]
___
**Require:** Learning rate $\epsilon_{lr}$ (Suggested default: 0.001)
**Require:** Exponential decay rates for moment estimates, $\beta_1$ and $\beta_2$ in [0, 1) (Suggested defaults: 0.9 and 0.999 respectively)
**Require:** Small constant $\delta$ used for numerical stabilization (Suggested default: $10^{-8}$)
**Require:** Initial parameters $\boldsymbol{\theta}$
**Require:** Initialize 1st and 2nd moment variables $\mathbf{S} = \mathbf{0}$ and $\mathbf{R} = \mathbf{0}$
**Require:** Initialize time step $t = 0$
 1: **while** stopping criterion not met **do**
 2:     Sample a mini-batch of size $m$ from the training set $\mathbb{X}$ with corresponding targets
 3:     Compute gradient for each parameter: $G_n \leftarrow \frac{1}{m} \nabla_{\boldsymbol{\theta}} \sum_i L\left(f\left(\boldsymbol{x}^{(i)}; \boldsymbol{\theta}\right), \boldsymbol{y}^{(i)}\right)$
 4:     Update time step $t \leftarrow t + 1$
 5:     Update biased first moment estimate: $S_n \leftarrow \beta_1 S_n + (1-\beta_1) G_n$
 6:     Update biased second raw moment estimate: $R_n \leftarrow \beta_2 R_n + (1-\beta_2) G_n \odot G_n$
 7:     Correct bias in first moment: $\hat{S}_n \leftarrow \frac{S_n}{1-\beta_1^t}$
 8:     Correct bias in second raw moment: $\hat{R}_n \leftarrow \frac{R_n}{1+\beta_2^t}$
 9:     Compute update: $\Delta\theta_n \leftarrow -\frac{\hat{S}_n}{\sqrt{\hat{R}_n}+\delta}$
10:     Apply update: $\theta_n \leftarrow \theta_n + \epsilon_{lr}\Delta\theta_n$
11: **end while**
___

## 2.4 Generative Adversarial Networks

In 2014 Ian J. Goodfellow et al. proposed in the paper "Generative Adversarial Nets" a new gradient-based learning technique called generative adversarial learning. This technique enables typically a neural network to learn to generate a desired output samples form a random noise input. [8]

As shown in figure 2.8 above a generative adversarial network (GAN) consists of a generator network $G$ and a discriminator network $D$. Those networks are "playing" a min max game, which is inspired by game theory. In the first step, the generator $G$ network produces a fake output for a random noise input. This fake output gets feed into the discriminator network $D$. Now the discriminator $D$ is trained to detect the input as fake. The generator $G$ however, is trained with the gradients of the discriminator $D$ with the goal to fool discriminator $D$ in his decision. In the second step a real data sample form the dataset $\mathbb{X}$, which includes real data that wants to be learned from the generator, is fed into the discriminator $D$ network. Then the discriminator is trained to detect the real sample as real. These two steps are repeated until the whole system eventuality converges. However, convergence is not guaranteed. This training process is typically called generative adversarial learning. [8]

### 2.4.1 Mathematical Formulation and Analysis of Generative Adversarial Learning

The two player min max game, also called generative adversarial learning, described previously can be formalized mathematically as: [8]

**Algorithm 2** AdamP optimizer algorithm [64]. Changes to the classical Adam [60] optimizer algorithm shown in green ■. [64]

**Require:** Learning rate $\epsilon_{lr}$ (Suggested default: 0.001)
**Require:** Exponential decay rates for moment estimates, $\beta_1$ and $\beta_2$ in [0, 1] (Suggested defaults: 0.9 and 0.999 respectively)
**Require:** Small constant $\delta$ used for numerical stabilization (Suggested default: $10^{-8}$)
**Require:** Scale invariance threshold $\lambda$ (Suggested default: 0.1)
**Require:** Initial parameters $\boldsymbol{\theta}$
**Require:** Initialize 1st and 2nd moment variables $\mathbf{S} = \mathbf{0}$ and $\mathbf{R} = \mathbf{0}$
**Require:** Initialize time step $t = 0$
1: **while** stopping criterion not met **do**
2:     Sample a mini-batch of size $m$ from the training set $\mathbb{X}$ with corresponding targets
3:     Compute gradient for each parameter: $G_n \leftarrow \frac{1}{m} \nabla_{\boldsymbol{\theta}} \sum_i L\left(f\left(\boldsymbol{x}^{(i)}; \boldsymbol{\theta}\right), \boldsymbol{y}^{(i)}\right)$
4:     Update time step $t \leftarrow t + 1$
5:     Update biased first moment estimate: $S_n \leftarrow \beta_1 S_n + (1 - \beta_1) G_n$
6:     Update biased second raw moment estimate: $R_n \leftarrow \beta_2 R_n + (1 - \beta_2) G_n \odot G_n$
7:     Correct bias in first moment: $\hat{S}_n \leftarrow \frac{S_n}{1 - \beta_1^t}$
8:     Correct bias in second raw moment: $\hat{R}_n \leftarrow \frac{R_n}{1 + \beta_2^t}$
9:     Compute update: $\Delta \theta_n \leftarrow -\frac{\hat{S}_n}{\sqrt{\hat{R}_n} + \delta}$
10:     **if** $-\theta_t G_t < \lambda$ **then**
11:         Apply update: $\theta_n \leftarrow \theta_n + \epsilon_{lr} \Pi_{\theta_n}(\Delta \theta_n)$
12:     **else**
13:         Apply update: $\theta_n \leftarrow \theta_n + \epsilon_{lr} \Delta \theta_n$
14:     **end if**
15: **end while**

$$\min_G \max_D V(D, G) = \min_G \max_D \mathbb{E}_{\boldsymbol{x} \sim p_{\text{data}}}[\log D(\boldsymbol{x})] + \mathbb{E}_{\boldsymbol{z} \sim p_{\boldsymbol{z}}(\boldsymbol{z})}[\log(1 - D(G(\boldsymbol{z})))] \quad (2.6)$$

Where $G$ is the generator and $D$ the discriminator. Additionally, $p_{\text{data}}$ is the distribution of the data, to learn, V the value function and $\boldsymbol{z}$ a random vector sampled form a distribution $p_{\boldsymbol{z}}(\boldsymbol{z})$. Here typically a normal distribution $\mathcal{N}(0, 1)$ or a uniform distribution $U(0, 1)$ is used in practice. [8]

To optimize parameters of the generator and discriminator network a first-order optimization algorithm is used. The most basic first-order optimization algorithm used to train neural networks is stochastic gradient descent. However, more advanced methods, based on stochastic gradient descent, were published in the past few years. These include, for example, the widely used Adam algorithm which was proposed in 2014. [1, 60]

Optimizing a deep neural network is a non-convex optimization problem. Therefor is, the process of optimizing the formula 2.6 also a non-convex problem. Furthermore, the goal of optimizing the formula 2.6 is to converge to a Nash equilibrium between the generator and the discriminator network. This is different from other learning techniques like supervised learning, where it is desired to converge to the global minimum of a given loss function. Additionally, this equilibrium between the generator and the discriminator network can be seen mathematically

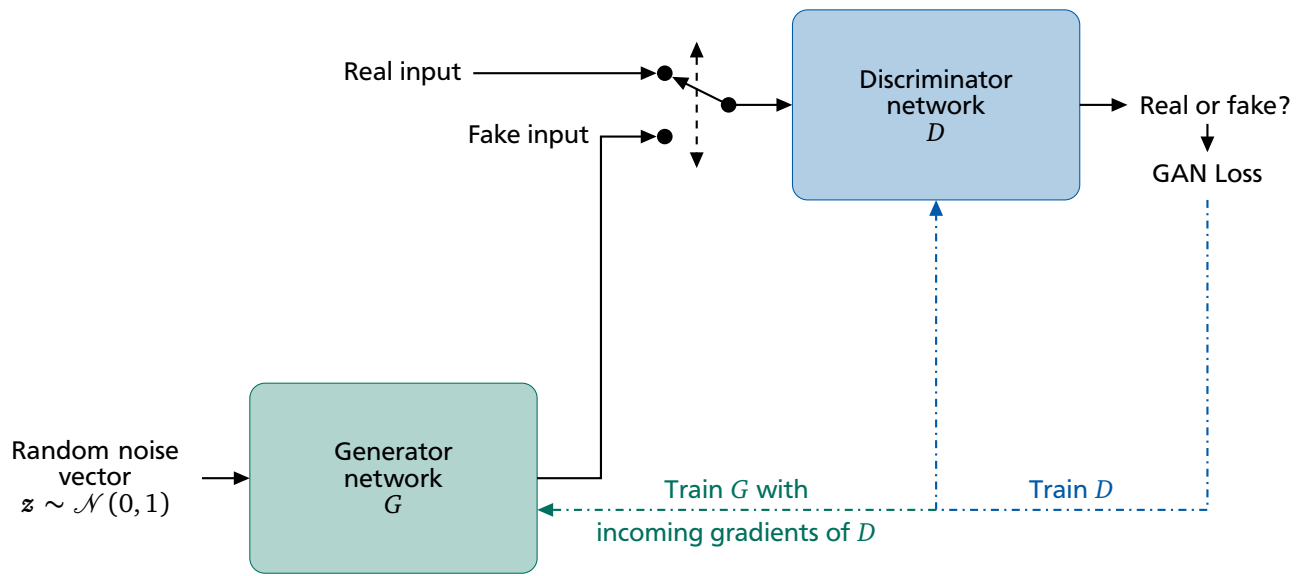

**Figure 2.8:** Illustration of a generative adversarial network architecture.

as a saddle point of the value function V. Reaching this saddle point is usually very challenging, which results often in an unstable training process. [8, 1]

If the generator and discriminator networks are successfully trained implies that the optimal saddle point of the value function *V* has been reached. Reaching the optimal saddle point of the value function results in a generator network which is able to model the mapping from a random space to the distribution of the given dataset $\mathbb{X}$. This means that the generator network is able to generate synthetic data strongly related to the given real data. The discriminator network however converges to a state where the probability output of the network is always $\frac{1}{2}$, regardless of whether the input is sampled form the dataset $\mathbb{X}$ or generated by the generator network. This means that the discriminator is completely unsure if the input is a real data sample or a fake on. [8]

#### 2.4.1.1 Formal definition of the Generative Adversarial Learning process

The training process described earlier can be formalized formally by using (mini-batch) stochastic gradient descent. However, the gradient-based updates of the learnable parameters, included in the generator and discriminator network, can be done by any standard gradient-based learning algorithm. Furthermore, the number of epochs performed while training and the mini-batch size used are hyperparameters, which has to be set beforehand. Some GAN implementations are using multiple update steps of the discriminator network before performing one update step of the generator network. This procedure, however, is not formalized in the following algorithm 3 because the use of multiple update steps is typically rare. [8]

#### 2.4.1.2 Proof of the Global Optimality of $p_G = p_{\text{data}}$

As briefly described previously the optimum of the formula 2.6 is reached if the generator is able to mimic the distribution $p_{\text{data}}$ of the dataset $\mathbb{X}$. This can be proven as follows: [8]

**Algorithm 3** Generative adversarial learning process with (mini-batch) stochastic gradient decent. [8]

**Require:** Mini-batch size $m$ and number of training iterations $k$
1: **for** $k$ training iterations **do**
2:     Sample mini-batch of $m$ noise samples $\{z^{(1)}, \ldots, z^{(m)}\}$ from $p_z$
3:     Sample mini-batch of $m$ samples for the given dataset $\mathbb{X}$
4:     Update the parameters of the discriminator network $D$ by ascending its stochastic gradient:

$$\nabla_{\Theta_D} \frac{1}{m} \sum_{i=1}^{m} \left[ \log D\left(x^{(i)}; \Theta_D\right) + \log\left(1 - D\left(G\left(z^{(i)}; \Theta_G\right); \Theta_D\right)\right)\right].$$

5:     Sample mini-batch of $m$ noise samples $\{z^{(1)}, \ldots, z^{(m)}\}$ from $p_z$
6:     Update the parameters of the generator network $G$ by descending its stochastic gradient:

$$\nabla_{\Theta_G} \frac{1}{m} \sum_{i=1}^{m} \log\left(1 - D\left(G\left(z^{(i)}; \Theta_G\right); \Theta_D\right)\right).$$

7: **end for**

**Proposition 1.** *In case of a fixed generator G, the optimal discriminator D is:* [8]

$$D_{opti, G_{fixed}}(x) = \frac{p_{data}(x)}{p_{data}(x) + p_g(x)}. \tag{2.7}$$

*Proof.* For any given generator $G$, the training criterion is to maximize the quantity $V(G, D)$ by the discriminator. [8]

$$V(G, D) = \int_x p_{\text{data}}(x) \log(D(x)) \, dx + \int_z p_z(z) \log(1 - D(g(z))) \, dz \tag{2.8}$$

$$= \int_x p_{\text{data}}(x) \log(D(x)) + p_g(x) \log(1 - D(x)) \, dx \tag{2.9}$$

To derive formula 2.9 form formula 2.8 the change of variables is applied. [8, 66]
The function $y \to a \log(y) + b \log(1 - y)$, under the integral, reaches its maximum in $[0, 1]$ at $\frac{a}{a+b}$, for $a, b \in \mathbb{R} \setminus \{0\}$. [8] ∎

The training objective for the discriminator $D$ can be seen as finding the maximum of the log likelihood for deriving the conditional probability $p(Y = y|x)$. Where $Y$ represents the condition whether $x$ is form $p_{\text{data}}$ (with $y = 1$) ore from $p_g$ (with $y = 0$). So formula 2.6 can be rewritten as:

$$C(G) = \max_D V(G, D) \tag{2.10}$$
$$= \mathbb{E}_{x \sim p_{\text{data}}} \left[ \log D_{\text{opti},G_{\text{fixed}}}(x) \right] + \mathbb{E}_{z \sim p_z(z)} \left[ \log \left( 1 - D_{\text{opti},G_{\text{fixed}}}(G(z)) \right) \right] \tag{2.11}$$
$$= \mathbb{E}_{x \sim p_{\text{data}}} \left[ \log D_{\text{opti},G_{\text{fixed}}}(x) \right] + \mathbb{E}_{x \sim p_g(z)} \left[ \log \left( 1 - D_{\text{opti},G_{\text{fixed}}}(x) \right) \right] \tag{2.12}$$
$$= \mathbb{E}_{x \sim p_{\text{data}}} \left[ \log \frac{p_{\text{data}}(x)}{p_{\text{data}}(x) + p_g(x)} \right] + \mathbb{E}_{x \sim p_g(z)} \left[ \log \frac{p_g(x)}{p_g(x) + p_{\text{data}}(x)} \right] \tag{2.13}$$

**Theorem 1.** *If and only if $p_G = p_{data}$ the global minimum of the virtual training criterion $C(G) = \max_D V(G,D)$ in reached. If this minimum is achieved, the training criterion reaches the value of $-\log(4)$. [8]*

*Proof.* If $p_G = p_{\text{data}}$ than followed from formula 2.7 $D_{\text{opti},G_{\text{fixed}}} = \frac{1}{2}$. Furthermore, $C(G) = \log\left(\frac{1}{2}\right) + \log\left(\frac{1}{2}\right) = -\log(4)$. ∎

### 2.4.2 Common problems of Generative Adversarial Networks

The training process of a generative adversarial network is often prone to some issues. This is mainly because instead of a minimum or maximum, the goal is to reach a saddle point. As mentioned in section 2.4.1 this saddle point represents a Nash equilibrium between the generator and discriminator network. This property can cause issues while training. [8, 67, 68, 69]

#### 2.4.2.1 Non-convergence of the Generative Adversarial Network

Generative adversarial networks often have the issue of non-convergence while training. This is due to the property of the non-convex two-player min max game and the optimization by gradient descent. If each player (network) is able to successfully minimize respectively maximize its performance, can this update step have a bad impact on the other player (network) simultaneously. Which can result again on the other hand in a possible decrease to the overall performance. [8, 67]
It has been shown that the min max game can converge while using simultaneous gradient descent if the updates are made in function space. In practice, updating the generator and discriminator network is done in the parameter space. This means that the proof of convergence can not be applied, in practice. [8, 67]
These facts can lead to a non-convergence or an oscillation of the overall performance. [8, 67]
  Tuning the hyperparameters of the generative adversarial network can often be the key to prevent non-convergence. However, finding the right hyperparameters can be quite hard or very computationally expensive, in case of an automatic hyperparameter search.

#### 2.4.2.2 Mode Collapse of the Generator Network

Another major problem of generative adversarial networks is mode collapse also known as the Helvetica scenario. Mode collapse describes the problem that the generator network only maps the input vectors $z$ to the same output point. This results in an optimized value function $V$, and results in a generator that is not able to capture the full variance of the data distribution $p_{\text{data}}$. This problem is visualized bellow in figure 2.10. [67]

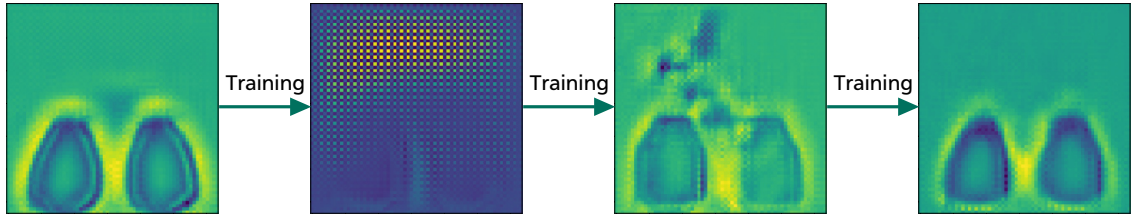

**Figure 2.9:** Example of an oscillating GAN while training.

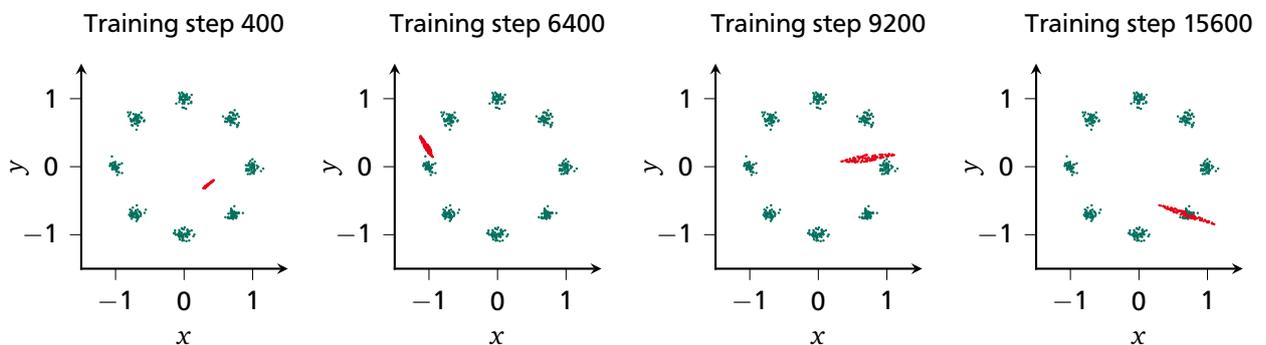

**Figure 2.10:** Example of a generative adversarial network lacking by mode collapse in $\mathbb{R}^2$. Samples of the data distribution $p_{\text{data}}$ in green ■ and the learned generator distribution $p_G$ in red ■. Full experiment specification and additional plots can be seen in the appendix 9.3.

To overcome the problem of mode collapse a variety of methods were published. Like for example the mini-batch discriminator, which looks at multiple generated samples at the same time to force the network to learn a variance in the generated samples. Additionally, different normalization methods were introduced to prevent the generator network to lack from mode collapse. More on methods for preventing mode collapse are introduced in chapter Improved Generative Adversarial Network Methods. [8, 67, 70]

#### 2.4.2.3 Diminished Gradient Problem and Unbalance between the Generator and Discriminator Network

As visualized in figure 2.8, the generator gets trained with the gradients of the discriminator. If the discriminator gets too good, the gradients of the generator can possibly diminish. If this problem occurs a unbalance between the discriminator and generator is present. On the other hand, it can happen that the generator is overfitting if an unbalanced training process is present. So a balanced training process, where the discriminator and generator are both learning equally fast, is desired. [8, 70, 67]

### 2.4.3 Evaluation of Generative Adversarial Networks

Evaluating the sample quality of a GAN generator network is a highly non-trivial task. If for example, a GAN achieves a good likelihood it can not be concluded that it also generates good samples. This means a GAN which generates good samples is also not guaranteed to have a good likelihood. [67]
In recent years multiple evaluation approaches of GANs have been published. The common state-of-the-art metrics, to evaluate GANs, for image generation are the Inception score (IS) [71] and the Fréchet Inception Distance (FID) [72]. Both of these evaluation methods rely on the famous InceptionNet [73, 74], which is pre-trained for image classification, on ImageNet [65]. Studies showed a high correlation between human judgment and a good IS or FID score, in terms of image quality. [71, 72, 73, 12]

#### 2.4.3.1 Inception Score

The Inception Score [71] is mathematically defined as:

$$\text{IS}(G) = \exp(\mathbb{E}_{\bm{x}}\left[\text{KL}(p(y|\bm{x}=G(\bm{z}))||p(y))\right]). \tag{2.14}$$

Where $G$ represents the generator network, to be evaluated. Furthermore, $p(y|x)$ represents the probability distribution for samples, produced by the generator model, predicted by a pre-trained InceptionNet, and $p(y)$ represents the marginalized probability of $p(y|x)$. A higher score indicates better quality samples. [71, 73]

#### 2.4.3.2 Fréchet Inception Distance

In contrast to the Inception score, where the probability output of the InceptionNet is utilized, the Fréchet Inception Distance [72] extracts intermediate feature activations from one layer to compute an evaluation score. To be precise, the feature activations of layer 7c of the Inception-Net are extracted for all a fixed number of real samples. Then the same features are extracted for the same number of generated fake samples. These feature activations are then used to model a multivariate Gaussian distribution with the mean $\bm{\mu}$ and the covariance $\bm{\Sigma}$. These statistics are then compared by the following formula to derive the FID score. [72]

$$\text{FID}(G, \mathbb{X}) = \left\|\bm{\mu}_{\mathbb{X}} - \bm{\mu}_{G}\right\|_{2}^{2} + \text{Tr}\left(\bm{\Sigma}_{\mathbb{X}} + \bm{\Sigma}_{G} - 2(\bm{\Sigma}_{\mathbb{X}}\bm{\Sigma}_{G})^{\frac{1}{2}}\right) \tag{2.15}$$

Here $G$ stands for the generator model and $\mathbb{X}$ for the dataset of real samples. The vectors $\bm{\mu}_{\mathbb{X}}$ and $\bm{\mu}_{G}$ are the mean of the multivariant Gaussian modeling the feature activations from the Inception net. Furthermore, $\bm{\Sigma}_{\mathbb{X}}$ and $\bm{\Sigma}_{G}$ are the corresponding covariance matrices of the real and generated samples. In general, a lower mean FID score corresponds to better image quality and diversity. [72]

### 2.4.4 Comparison to other Deep Generative Models

Besides generative adversarial networks, a verity of different deep generative models has been proposed. The foremost commonly used deep generative adversarial model, besides GANs, are variational autoencoders [75, 76]. In the following table, the most important advantages and disadvantages of VAEs and GANs are mentioned. [1, 8]

In recent years also hybrid architectures of GANs and VAEs have been proposed. But nonetheless, generative adversarial networks based architectures are state-of-the-art in terms of sample quality. An overview of the advantages and disadvantages of GANs vs. VAEs can be seen in figure 2.1. [67, 13, 77]

**Table 2.1:** Overview of the advantages and disadvantages of GANs [8, 67] vs. VAEs [75].

| Method | Advantages | Disadvantages |
| --- | --- | --- |
| GAN [8] | • State-of-the-art sample quality (e. g. generated image sample) | • Possible unstable training resulting in a non-convergence |
| VAE [76] | • General principled approach<br>• Stable training<br>• Latent features can sometimes be human interpretable | • Lower quality samples compared to GANs<br>• Maximizes lower bound of likelihood |

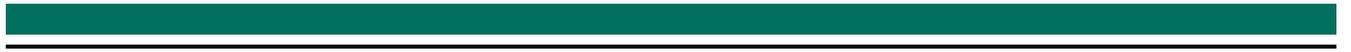

## 2.5 Semi Supervised Learning

In supervised learning a label for every training sample is needed. Labeling every sample of a training dataset can be often very time consuming and costly. Semi-supervised learning, however, deals with the case where a partly labeled dataset is present. [27, 24]

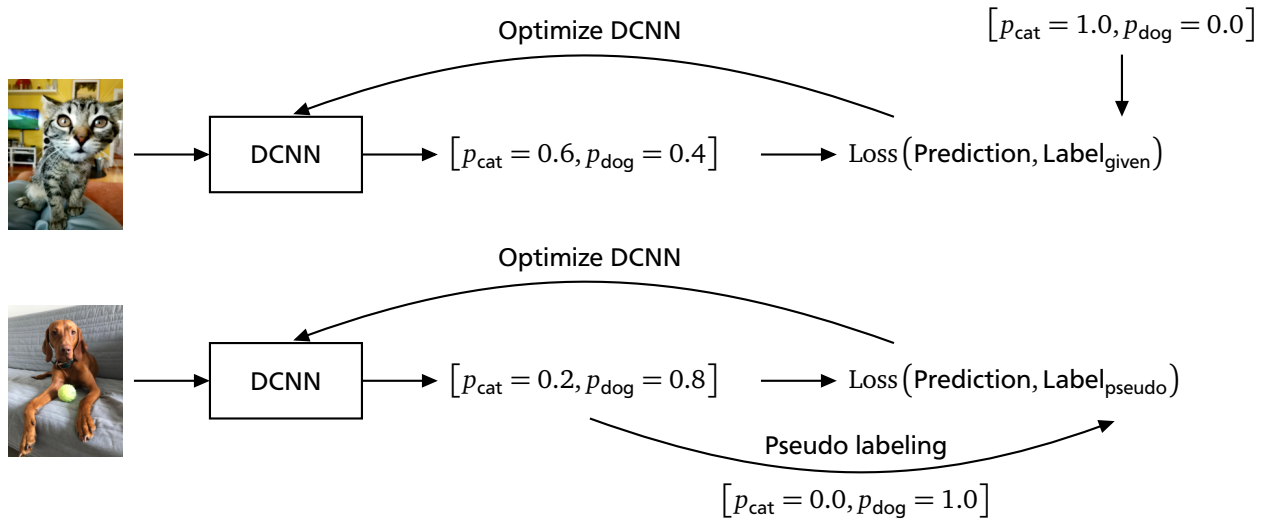

**Figure 2.11:** Illustration of the semi-supervised pseudo labeling training process. The supervised training stage is shown at the top. The pseudo labeling and training stage is shown below.

To explain the process of semi-supervised learning (pseudo labeling), consider the case of classifying images of cats and dogs. For a few images, a label (cat or dog) is given. However, the dataset also includes a lot of images where no label is present. Now a convolutional neural network should be trained to classifying the given images. In semi-supervised learning, first, the model gets trained supervised on the labeled part of the given dataset. After training for a given number of iterations, on the labeled dataset, the unlabeled images are feed into the model. While feeding every unlabeled image to the model a pseudo label is generated. This means, if the model predicts, with a probability of $p_{\text{cat}} > p_{\text{dog}}$, that a given unlabeled image includes a cat, these images are labeled as a cat image. If this process is done, for every unlabeled image, the model is trained on the whole labeled, respectively partly pseudo labeled, the dataset for a given number of iterations. [27, 24]
In the process of producing the pseudo label often a condition is used, in real-world applications. This condition is normally a certainty, which the prediction of the model has to achieve. Because if the model is completely unsure, the training on the produced pseudo label can lead to a decline in the model performance. This condition can often be a crucial hyperparameter that highly affects the overall model performance. However, with the previously describe semi-supervised training process it is often possible to build a more robust and accurate model, with a partly labeled dataset, instead of only using the labeled part of the dataset. [26, 27, 24] However, also advanced semi-supervised methods [78, 29, 28, 25] have been proposed. The MixMatch [28] method for semi-supervised learning for example relays on data augmentation, prediction averaging/sharpening, and MixUp. Semi-supervised learning, however, stays an active area of research.

## 2.6 Semantic Segmentation

Semantic segmentation is the task of classifying a given image (gray-scale or color image) pixel-wise to different categories. Since every pixel gets classified, semantic segmentation can be seen as a special type of classification problem. Typically the problem of semantic segmentation is tackled by a deep convolutional neural network, which is trained in a supervised setup. [45, 37]

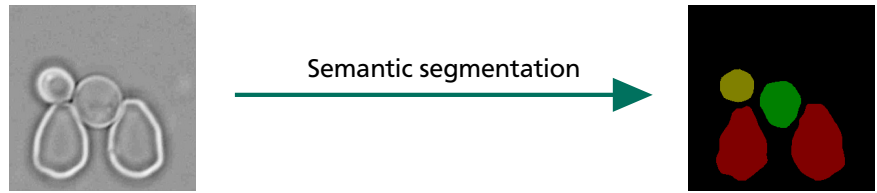

**Figure 2.12:** Brightfield microscopy image, including cells and traps, and the corresponding semantic segmentation label.

In the past years, a verity of deep learning models was published, to perform end-to-end semantic segmentation. These deep learning models outperformed more classical models, in more or less all use-cases. The first fully convolutional neural network for semantic segmentation [37] was published by J. Long et al. in 2015. Later in the same year, O. Ronneberger et al. published a paper of a convolutional neural network specially developed for biological and medical imaging. This network is called U-Net [45] and is able to perform semantic segmentation on very small datasets, with often not more than a few dozen training elements. [37, 45]
In the past years, multiple advanced U-Net architectures have been published. One of the most influential advanced U-Net architectures is the 3d U-Net [79, 50] which enables 3d semantic segmentation of 3-dimensional data. However, also other developments in the field of deep learning have influenced the architecture of the U-Net like for example the residual U-Net [80] which utilizes ResNet-like [43] building blocks.

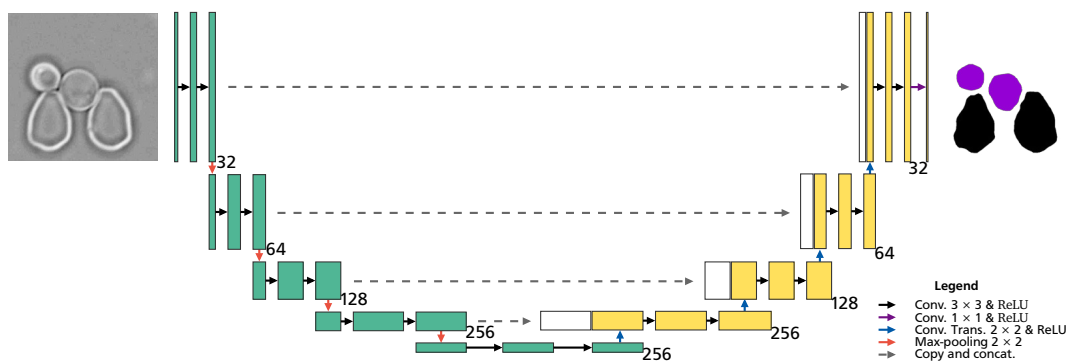

**Figure 2.13:** Typical U-Net architecture for semantic segmentation. Brightfield microscopy image on the left and the resulting multi class semantic segmentation on the right. The encoder path of the U-Net indicated in green ■ and the decoder path in yellow ■. [6]

In the past, even challenging semantic segmentation tasks have been solved with encoder-decoder-based CNNs. Christoph Reich, Tim Prangemeier et al. proposed a U-Net based solution for accurate multi-class semantic segmentation of cells in microstructures [6]. However, also more advanced semantic segmentation models beyond U-net like the DeepLab3+ [81], have been recently proposed.

## 2.7 Optical Flow

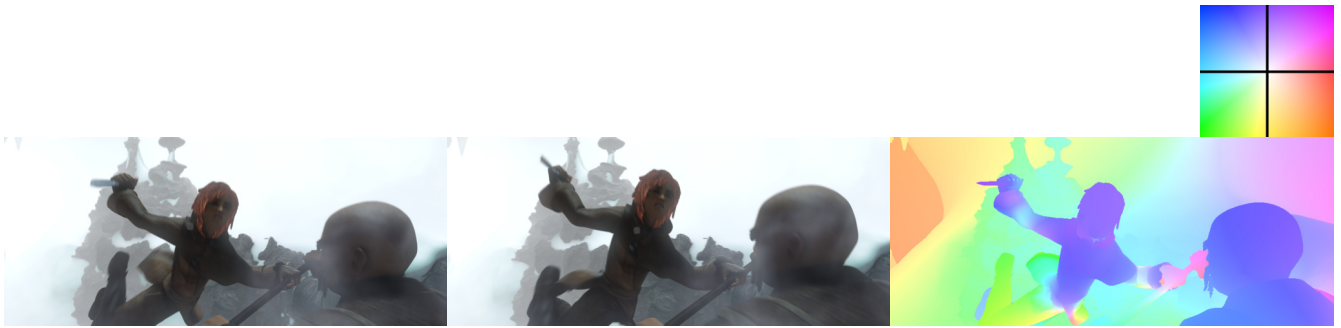

**Figure 2.14:** Two frames of the MPI sintel flow dataset [82] and the corresponding ground truth optical flow encoded in the classical color wheel (top) introduced in [83].

The optical flow describes the motion of each pixel from one frame to another. Estimation of this optical flow is considered as a general and challenging version of motion estimation in computer vision (CV). The concept of a flow field describing the motion of objects in images, the optical flow, was first introduced by James J. Gibson in the 1940s. Mathematically the optical flow $\boldsymbol{F} \in \mathbb{R}^{h \times w \times 2}$, for the images $\boldsymbol{I}_1 \in \mathbb{R}^{h \times w}$ and $\boldsymbol{I}_2 \in \mathbb{R}^{h \times w}$ describes the change of each pixel $\boldsymbol{I}_1(i + \boldsymbol{F}(i,j,1), j + \boldsymbol{F}(i,j,2))$ to the subsequent image. [84, 85]

### 2.7.1 Optical Flow Estimation

Estimation of the optical flow is a non-trivial task. Traditional computer vision methods like the Lucas-Kanade method or the Gunner Farneback algorithm relies on multiple assumptions. The foremost important assumptions are, first that the pixel intensities of an object don't change in consecutive frames. And secondly, that neighboring pixels, in an image, have similar motion. However, estimating an optical flow approximation, with the previously mentioned methods, is computationally expensive and often leads to inadequate approximation. [84, 85]
To overcome the limitations of traditional computer vision algorithms, deep convolutional neural networks have been applied to the problem of optical flow estimation. The first CNN approach, FlowNet [14], showed great improvement in the speed and accuracy of optical flow estimation compared to traditional methods. Since the publication of the FlowNet in 2015 many more supervised CNN approaches, like the PWC-Net [16], FlowNet 2 [15] or MaskFlownet [17], have been proposed. Also, unsupervised approaches [86, 87, 88, 89, 90] have been published in the past years, which can even perform on par with supervised methods like FlowNet 2 [18]. Today convolutional neural networks are the state-of-the-art method for optical flow estimation [16, 17, 18].
Recent work has also extended the idea of optical flow estimation with convolutional neural networks to estimations a whole scene flow, based on monocular imagery [91].

2.7 Optical Flow                                                                                                          25

# 3 Related Work

## 3.1 Improved Generative Adversarial Network Methods

Since the invention of generative adversarial networks by Ian Goodfellow back in 2015, many more advance GAN architectures and loss functions for adversarial learning has been published. Mainly for two reasons. First, to adopt the general GAN architecture to special use-cases or second, to overcome the previously described issues of GANs in section 2.4.2. This chapter introduces the Wasserstein GAN, the progressive growing GAN architecture and the StyleGAN architecture. [8, 10, 11]

### 3.1.1 Wasserstein Generative Adversarial Networks

The Wasserstein GAN an improved training algorithm for generative adversarial networks, based on the Wasserstein distance, rather than on the log-likelihood. This new training algorithm can prevent the generative adversarial network from mode collapse and is able to stabilize the whole adversarial training process. [10]

#### 3.1.1.1 Wasserstein Distance

The Wasserstein distance, also called Earth Mover's distance, is a measurement of the distance between two probability distributions. This distance can be seen informal as the minimum energy cost required to move one probability distribution to a shape of another probability distribution. To visualize this informal idea of the Wasserstein distance consider the following case: Given two discrete unnormalized probability distributions $P$ and $Q$, which are further defined as: [10, 92]

$$P = \begin{cases} P_1 &= 4 \\ P_2 &= 1 \\ P_3 &= 2 \\ P_4 &= 4 \end{cases}, \quad Q = \begin{cases} Q_1 &= 1 \\ Q_2 &= 3 \\ Q_3 &= 4 \\ Q_4 &= 3 \end{cases}. \tag{3.1}$$

Now the distributions $P$ and $Q$ should be transformed to match each other. This process can be visualized as followed: [10, 92]

As can be seen from figure 3.1 the costs for all movements are: $\delta_1 = 3$, $\delta_2 = 1$, $\delta_3 = 1$. So the resulting Wasserstein distance is $W = \sum |\delta_i| = 5$.
Mathematically the Wasserstein distance is defined, for two continuous probability distributions $p_1$ and $p_2$, as: [10, 92]

$$W(p_1, p_2) = \inf_{\gamma \sim \Pi(p_1, p_2)} \mathbb{E}_{(x,y) \sim \gamma} \|x - y\|_2. \tag{3.2}$$

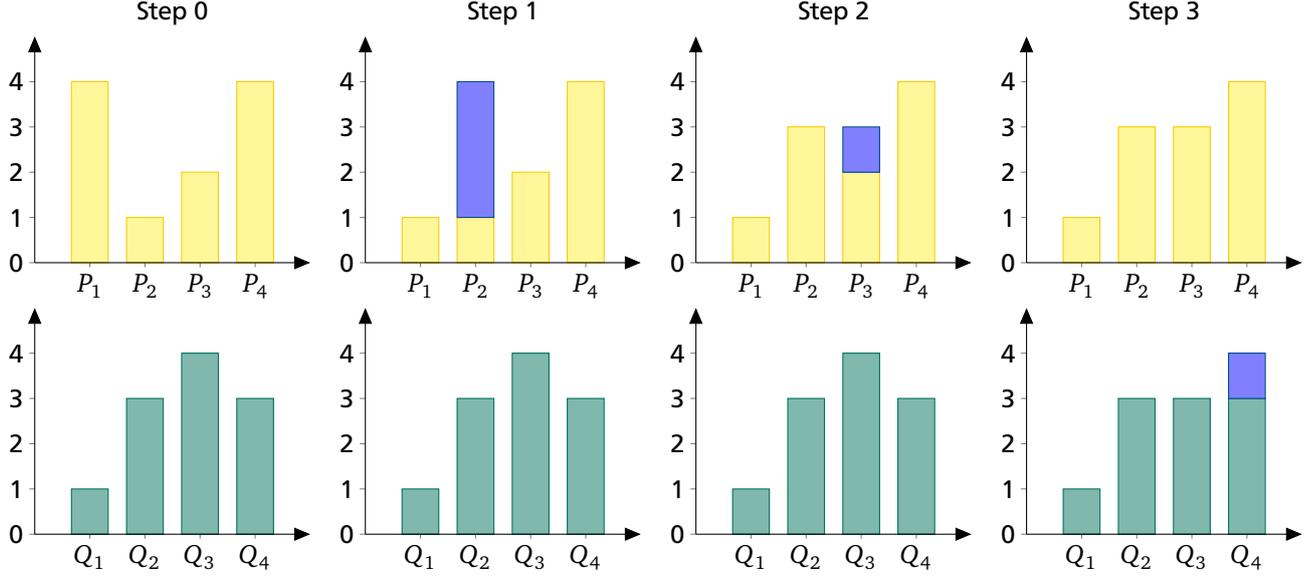

**Figure 3.1:** Visualization of the informal Wasserstein distance estimation. Currently moved blocks visualized in blue. [92]

Where $\Pi(p_1, p_2)$ represents all possible joint distributions between $p_1$ and $p_2$. Each joint distributions represents a possible transformation between $p_1$ and $p_2$. [92]

#### 3.1.1.2 Wasserstein value function

The Wasserstein value function for the discriminator and generator network, which can be derived from the Wasserstein distance, is defined as: [10]

$$\min_D V(D, G) = -\mathbb{E}_{\boldsymbol{x} \sim p_{\text{data}}}[D(\boldsymbol{x})] + \mathbb{E}_{\boldsymbol{z} \sim p_{\boldsymbol{z}(\boldsymbol{z})}}[D(G(\boldsymbol{z}))] \tag{3.3}$$

$$\min_G V(D, G) = -\mathbb{E}_{\boldsymbol{z} \sim p_{\boldsymbol{z}(\boldsymbol{z})}}[D(G(\boldsymbol{z}))]. \tag{3.4}$$

However, this value function violates the Lipschitz continuity. Due to the proof, which derives the Wasserstein value function, the discriminator mapping has to be 1-Lipschitz continues. For information regarding this proof see Arjovsky et al. [10].
A simple way to ensure Lipschitz continuously is weight clipping, which was presented in the Wasserstein value function. Weight clipping can lead to multiple issues while training the Wasserstein GAN. In case if the clipping parameter is too large, training to reach an equilibrium, between the generator and discriminator, can take a long time. If the clipping parameter is too small gradients tend to vanish, which also hurts the performance. However, in practice, when setting the clipping parameter to 0.01, the Wasserstein GAN, with gradient clipping, showed great performance improvements, when utilizing a deep convolutional GAN architecture (DCGAN). [10, 9]
To solve the issues of gradient clipping and setting the required hyperparameter Gulrajani et al. [93] proposed a gradient penalty to the Wasserstein value function to ensure Lipschitz continuity. With this extension, the value function for the discriminator changes to: [93, 10]

$$\min_D V(D, G) = -\mathbb{E}_{\boldsymbol{x} \sim p_{\text{data}}}[D(\boldsymbol{x})] + \mathbb{E}_{\boldsymbol{z} \sim p_{\boldsymbol{z}(\boldsymbol{z})}}[D(G(\boldsymbol{z}))] + \lambda \mathbb{E}_{\boldsymbol{z} \sim p_{\boldsymbol{z}(\boldsymbol{z})}}\left[(\|\nabla \boldsymbol{z} D(\boldsymbol{z})\|_2 - 1)^2\right]. \quad (3.5)$$

Empirical results showed, better performance and a more stable training when utilizing the Wasserstein value function with gradient penalty (WGAN-GP). This holds also when using different generator and discriminator architecture, then the DCGAN architecture. However, when utilizing the gradient penalty term a considerably large computational cost is added to the whole architecture. [93]

### 3.1.2 Progressive Growing Generative Adversarial Networks

The paper "Progressive Growing Of GANs For Improved Quality, Stability, and Variation" published in 2018 by Karras et al. [11] (Nvidia research), proposed a new training methodology for generative adversarial networks. The basic idea of the proposed architecture (ProGAN) is to grow progressively both the generator and the discriminator network. This process speeds up the training and is able to stabilize the adversarial training process. Furthermore, the generator network is able to produce images with very high variation. [11]

#### 3.1.2.1 Growing architecture

Common GAN architectures like the DCGAN struggle generating high-resolution images. To tackle the issues occurring when generating high-resolution images the growing ProGAN architecture has been developed. In contrast to other GAN architectures, the ProGAN starts the training process by generating small images. Typically a starting resolution of 4 × 4 is chosen. After learning to generate these low-resolution images a convolutional block, for the next resolution stage, is added to the generator and to the discriminator, which is visualized in figure 3.2. This process is repeated until the typical resolution of 1024 × 1024 is achieved. The insertion process of new convolutional blocks to higher the current resolution leads to a stable training and amazing sample quality. [11]

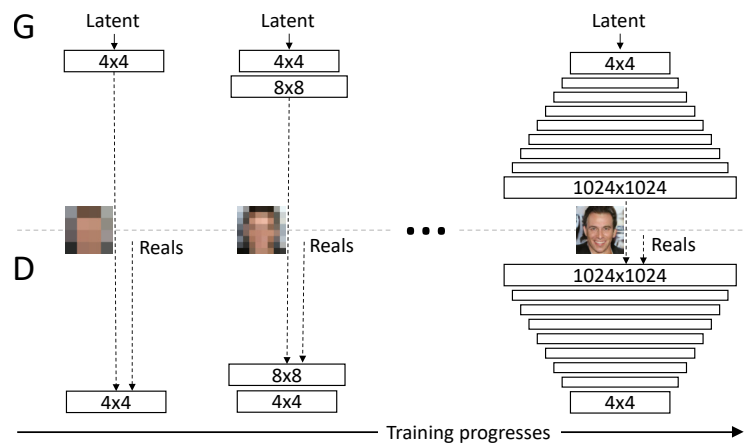

**Figure 3.2:** Architecture of the ProGAN generator and discriminator network over the training time. [11]

However, only adding the new blocks to the architecture and retraining the grown architecture again can cause issues in terms of performance. For this reason, the ProGAN architecture utilizes an adaptive residual insertion process of new blocks. This residual insertion process is shown in figure 3.3.

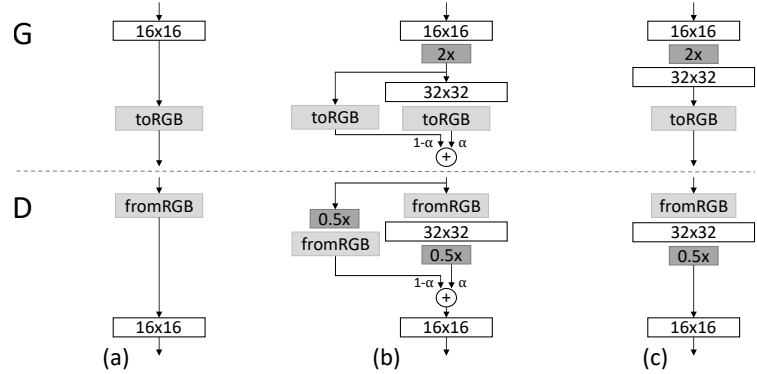

**Figure 3.3:** Residual insertion process proposed in the ProGAN paper. Architecture in the resolution stage 16 × 16 in (a). Architecture at the start of the next 32 × 32 resolution stage with residual insertion process in (b). Architecture in the 32 × 32 stage after finishing the residual insertion process in (c). [11]

When reaching a new resolution stage a new generator and discriminator convolution block including each two 3 × 3 convolutions followed each by a Leaky ReLU activation function, is initialized. Furthermore, in the generator, a nearest neighbor upsampling operation is utilized. In the discriminator, however, an average pooling is utilized. The output of the generator is then constructed with the upsampled output of the previous resolution stage with the output of the new stage. To combine the predicted samples simple weighted addition with the factor $\alpha$ is used. This weighting between the previous stage and the new stage decreases in a linear fashion. When the factor $\alpha$ reaches the value zero the residual insertion process is finished. This residual insertion process is also utilized in the discriminator network, however, here the input to the previous stage is constructed. One issue of this residual insertion process is however that a new hyperparameter, which describes the number of iteration it takes to complete the insertion process. [11]

To stabilize the adversarial training process, even more, an equalized extension to convolutions and linear layers has been introduced. Typically the weights in a deep neural network are initialized by a specific method like He's initialization. Instead of initializing the weights by He's method each learnable weights in the ProGAN architecture are scaled at runtime by the factor $1/c_{\text{He}}$. This factor is the per-layer normalization form He's method which is defined as $c_{\text{He}} = \sqrt{2/c_{\text{in}}}$. Where $c_{\text{in}}$ is the number of input units to the corresponding learnable operation. Furthermore, all weights are initialized by sampling from $\mathcal{N}(0, 1)$. [11]

A common problem in GANs are overshooting signal magnitudes which leads to an unhealthy competition between the generator and the discriminator network. To prevent this issue the ProGAN utilizes a local response normalization layer in the after each convolutional layer in the generator network. [11] The ProGAN model is trained with the Wasserstein GAN loss with gradient penalty. Additionally, the regularization term $\epsilon_{\text{drift}}\mathbb{E}_{\boldsymbol{x}\sim p_{\text{data}}}\left[D(\boldsymbol{x})^2\right]$, where $\epsilon_{\text{drift}} = 0.001$, is added to the discriminator loss, to prevent the discriminator output from drifting too far away from zero. [11]

To achieve a high variation in the generated samples the ProGAN architecture utilizes a mini-batch standard deviation layer in the lowest block of the discriminator network. This layer computes the standard deviation for each batch instance and concatenates the reshaped standard deviation tensor channel-wise with the original feature tensor. [11, 70] The generator weights for predicting samples are achieved by utilizing an exponential running average over all learnable parameters during the training. This method enables the generator to produce even better quality samples. [11]

For more information regarding the architecture, hyperparameters or the training process see [11].

#### 3.1.2.2 Results

The ProGAN paper proposed results for a verity of datasets, however, the most stunning results have been achieved on the CelebA-HQ dataset (1024 × 1024), which includes images of celebrity faces. The generating samples achieved an Inception score of 8.80, which outperformed all previous state-of-the-art methods at this time. [11]

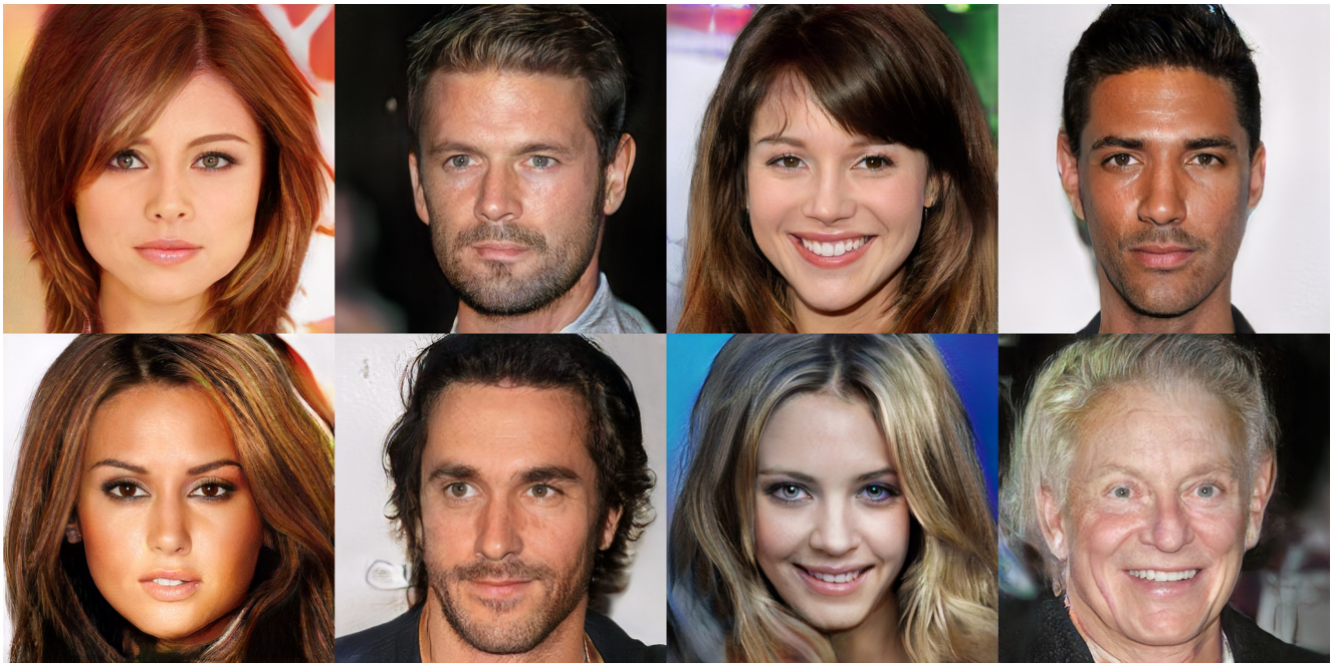

**Figure 3.4:** Original proposed results of the ProGAN architecture on the CelebA-HQ (1024 × 1024) dataset. [11]

### 3.1.3 Style-Based Generator Architecture for Generative Adversarial Networks (StyleGAN)

The paper "A Style-Based Generator Architecture for Generative Adversarial Networks" published in 2019 by Karras et al. [12] (Nvidia research), proposed a new generator architecture for GANs. This new generator architecture is able to learn the separation of low and high-level features of an image. This key property facilitates the generator also to perform a task described as style transfer. The resulting GAN architecture achieved state-of-the-art performance in terms of the sample quality. Furthermore, it also achieves smooth samples when interpolation in the latent space. [12]

### 3.1.3.1 Style Based Architecture

In a traditional GAN generator network the latent vector $z \sim p_z(z)$ gets processed though all convolutional layers. This process can be observed in figure 3.5a. [12, 11]

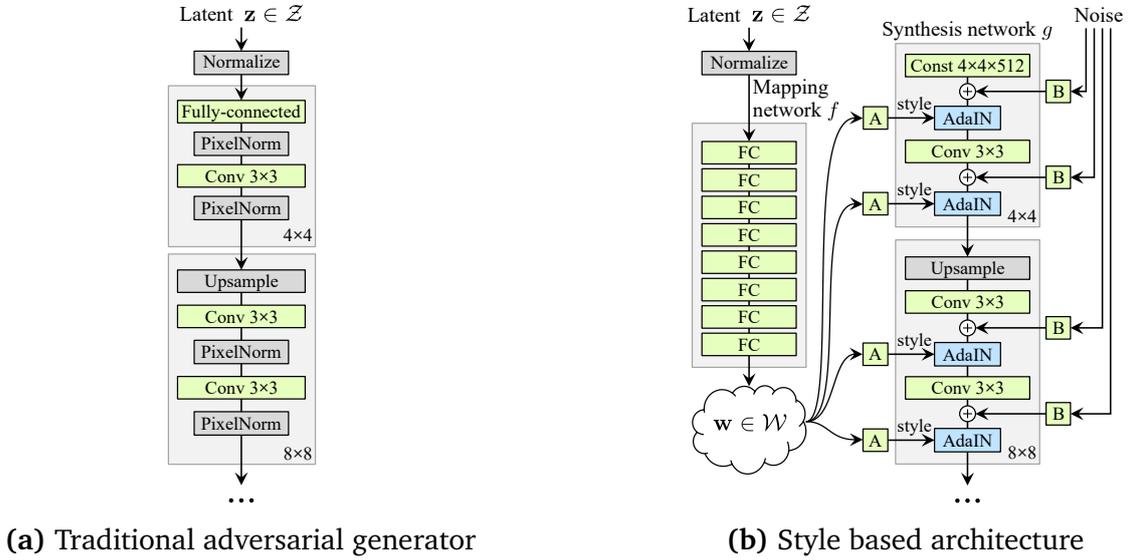

**(a)** Traditional adversarial generator    **(b)** Style based architecture

**Figure 3.5:** Traditional adversarial generator network which feeds the latent vector though all convolutional layers in **(a)**. Illustration of the style based architecture in **(b)**. The mapping network $f$ can be seen on the left and the convolutional synthesis generator $g$, which gets controlled by the latent space $w$ through adaptive instance normalization, on the right. The block indicated with an A represents a leaned affine transformation. Each block indicated with an B stands for a leaned channel-wise scaling of the noise input. [11, 12]

The style based generator network, however, introduced another idea. In the StyleGAN generator, a constant, however, a learnable tensor is fed into the convolutional path. The latent vector $z$ is, however, processed by a separate mapping network $f$. The idea of this mapping network is to transform the latent vector $z$ into an intermediate latent space $W$. The intermediated feature space is based on the deliberation that the distribution of the features represented in the training set differs from the distribution of $z$. This mapped latent vector is further used in each convolutional block. The StyleGAN architecture maps first the latent tensor with an affine transformation to a so-called style tensor. This style tensor parameterizes than an adaptive instance normalization layer (AdaIN), which changes the distribution of the features in the convolutional path. This change in the distribution of the feature activation influences which image features are included in the final image sample. [12]
The AdaIN operation which incorporates the latent vector into the convolutional features of the synthesis network $g$ is defined as:

$$\text{AdaIN}(\mathbf{X}_i, \mathbf{y}) = \mathbf{y}_{s,i} \frac{\mathbf{X}_i - \mathbb{E}[\mathbf{X}_i]}{\text{Std}(\mathbf{X}_i)} + \mathbf{y}_{b,i} \tag{3.6}$$

where $\mathbf{X}_i$ represents each feature map, which are normalized separately. Than the normalized features are scales and biased by utilizing the parts of the style vector $\mathbf{y}$. This the style vector $\mathbf{y}$ has twice the number of feature maps then the number of features in $\mathbf{X}_i$. [12]

Besides the changed generator architecture and smaller adoptions like noise inputs, which can be seen in figure 3.5b, mixing regularization and bilinear up and downsampling layers, the main training setup, progressive growing architecture, and the discriminator of the ProGAN remains. [12, 11]

#### 3.1.3.2 Results

Besides the empirical results of the StyleGAN architecture, the StyleGAN paper proposes a new Flicker-Faces-HQ FFHQ dataset, which consists of 70,000 high-resolution (1024 × 1024) images including human faces. This dataset includes much more variation than the CelebA-HQ dataset. [12]

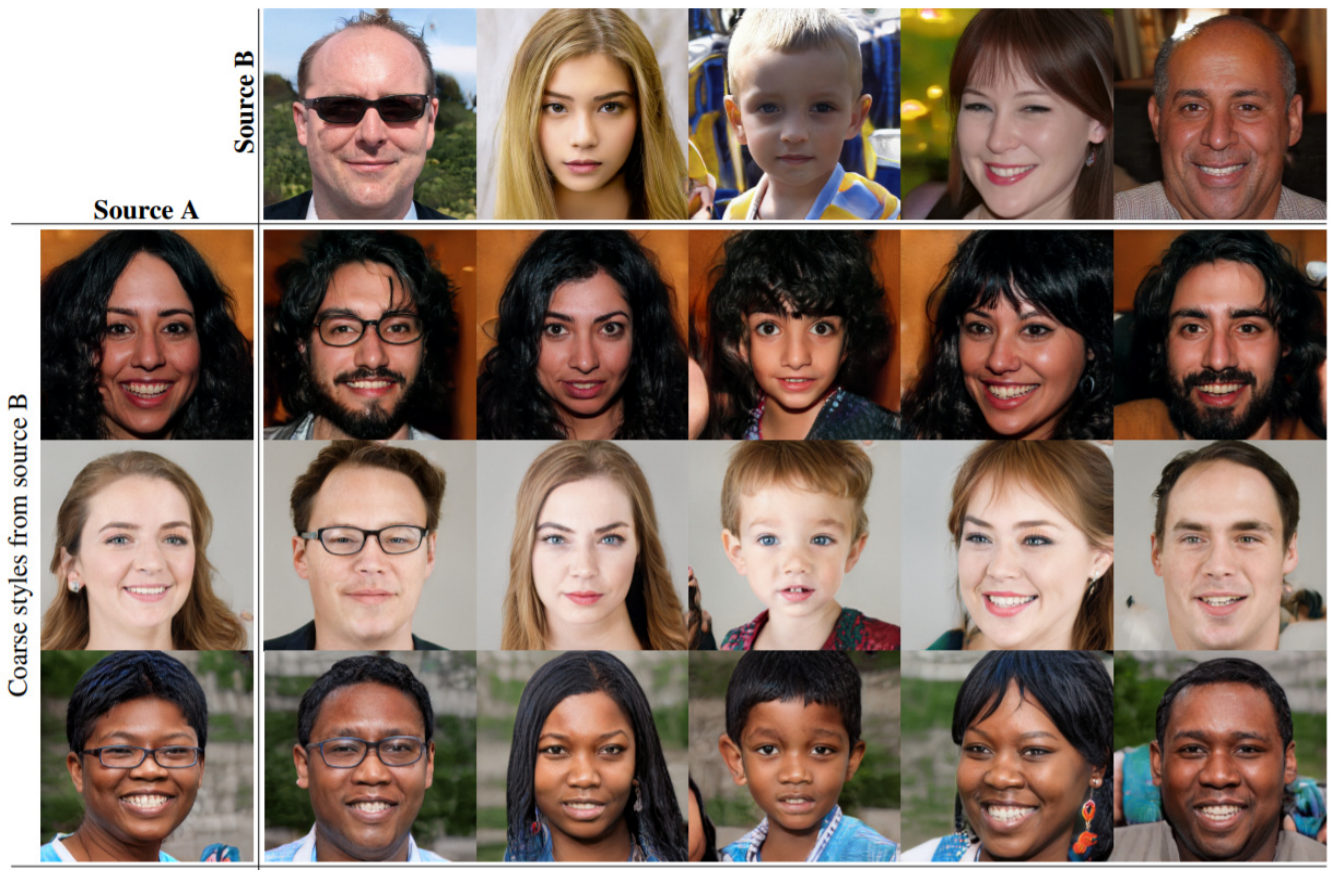

**Figure 3.6:** Results of the StyleGAN model trained on the FFHQ dataset with a resolution of 1024 × 1024. Generated source images can be seen in the left column and in the top row. The remaining images are generated by performing style transfer with the corresponding source images. [12]

The StyleGAN architecture has been tested in multiple settings. Starting by the ProGAN architecture and adding gradually the new proposed StyleGAN features. The full StyleGAN architecture outperformed the ProGAN architecture both on the CelebA-HQ and the FFHQ dataset, achieving state-of-the-art-performance. The StyleGAN reached an FID score of 5.17 on the CelebA-HQ dataset compared to the score of 7.79 of the ProGAN baseline architecture. On the FFHQ dataset, the StyleGAN architecture achieved an FID score of 4.40 and the ProGAN architecture a score of 8.04. [12, 11]

The StyleGAN paper proposes also qualitative results on style transfer. In style transfer, the style vectors of two previously generated images are combined to generate a new image. This image has a high similarity to the source images. This experiment also shows that styles in the lower resolution stages of the generator network correspond to lower lever image features and styles in the higher resolution stages corresponds on the other hand with higher level features. Some of these style transfer results are visualized in figure 3.6. [12]

### 3.1.4 Improved Style-Based Generator Architecture (StyleGAN 2)

The paper "Analyzing and Improving the Image Quality of StyleGAN" published in 2019 by Karras et al. [13] (Nvidia research), analyzes the original StyleGAN architecture. Additionally, a new improved style bases generative adversarial network architecture (StyleGAN 2) is proposed. This new architecture does not rely on a progressive growing generator and discriminator model anymore. Furthermore, a new regularization method is proposed alongside a new replacement for the adaptive instance normalization operation. This improved architecture leads to state-of-the-art results for image generation. [13]

#### 3.1.4.1 Improved Style Based Architecture

The use of adaptive instance normalization can cause water like droplet artifacts, in images generated by the StyleGAN architecture. The StyleGAN 2 architecture replaces the adaptive instance normalization layers with a so-called weights demodulation, of the convolutional weights. This weights demodulation is described as: [13]

$$W'_{ijk} = W_{ijk} s_i \tag{3.7}$$

$$W''_{ijk} = \frac{W'_{ijk}}{\sqrt{\sum_{i,k} {W'_{ijk}}^2 + \epsilon}} \tag{3.8}$$

where **W** is the weights tensor of the convolution, $s_i$ the corresponding style vector, and **W**″ the resulting modulated and normalized weight tensor.

The development of this weights demodulation, which can be seen in figure 3.7, is based on the idea to base normalization on the statistics expected of the input feature maps, however without explicit forcing. This redesign presents the generator to produce images without the characteristic artifacts of the AdaIN while preserving full controllability by the style vector. [13] When utilizing a progressive growing architecture often strong location prefereces of details can occur. This issue happens for example when generating images of human faces, in the case of teeth or eyes, when interpolating in the latent space. Previously proposed architectures like the MSG-GAN [94] adapted the original StyleGAN architecture to remove progressive growing. Empirical results have been shown that utilizing skip connections or residual connections lead to better image quality then progressive growing. Furthermore, the best performance on the FFHQ and the LSUN Car dataset has been achieved when using skip connections in the generator network and residual connections in the discriminator network. [13]
The StyleGAN 2 architecture also applies a lazy regularization technique while training. This means regularizations like the $R_1$ regularization method or the path length regularization are not applied in each training step. In all training runs of the StyleGAN 2, when utilizing lazy regularization, the regularizations has been used every 16 mini-batches. [12]

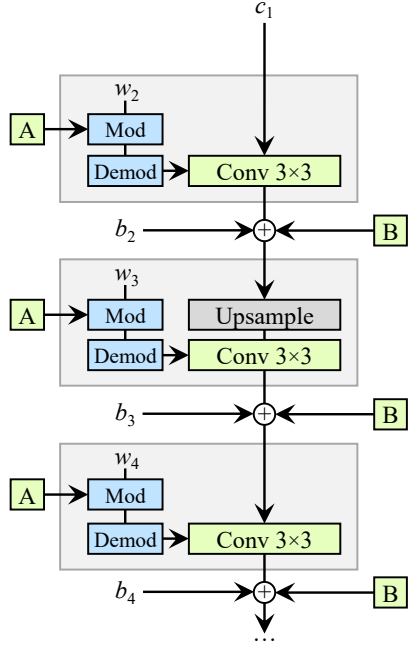

**Figure 3.7:** StyleGAN 2 architecture with weight demodulation instead of the adaptive instance normalization as in the original StyleGAN. [13, 12]

A surplus of path distortion in the generator network in an evident for bad local conditioning. It is desired that, small displacements in the latent space yield to changes of equal magnitude in the generated image. To regularize the generator model to follow this desired behavior the path length regularization has been proposed which is defined as: [13]

$$\mathbb{E}_{\boldsymbol{w},\boldsymbol{y}\sim\mathcal{N}(0,\boldsymbol{I})}\left[\left\|\boldsymbol{J}_{\boldsymbol{w}}^{\intercal}\boldsymbol{y}\right\|_{2}-a\right]^{2}. \quad (3.9)$$

Where $\boldsymbol{w}$ is the latent vector of the mapping network $f$, $\boldsymbol{J} = \frac{\partial G(\boldsymbol{w})}{\partial \boldsymbol{w}}$ the Jacobian matrix and $\boldsymbol{y}$ a random vector sampled from $\mathcal{N}(0,\boldsymbol{I})$. The constant $a$ is estimated during training dynamically as the long-runnning exponential moving average of $\left\|\boldsymbol{J}_{\boldsymbol{w}}^{\intercal}\boldsymbol{y}\right\|$. To use standard backpropagation $\boldsymbol{J}_{\boldsymbol{w}}^{\intercal}\boldsymbol{y}$ is computed as $\nabla \boldsymbol{w}\left(G(\boldsymbol{w})\,\boldsymbol{y}\right)$, which is more efficient to estimate. [13]

#### 3.1.4.2 Results

Empirical results on different datasets showed that the StyleGAN 2 outperforms the original StyleGAN on all datasets. The StyleGAN 2 achieved, for example, an FID score of 2.84 on the FFHQ dataset, compared to the FID score of 4.40 for the original StyleGAN.

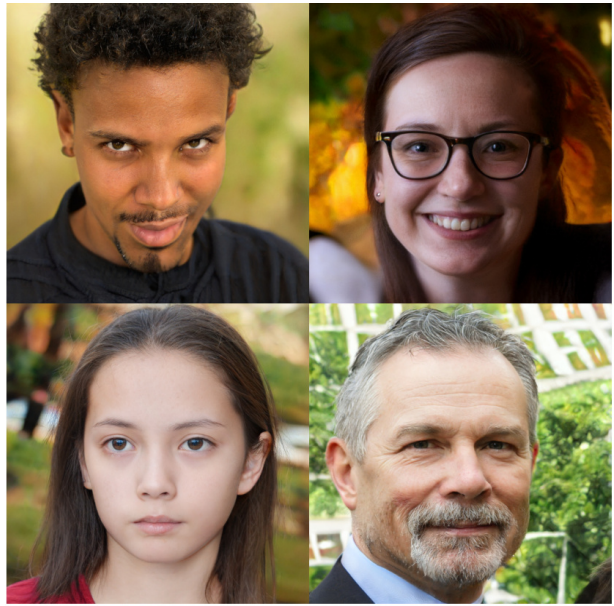

**Figure 3.8:** Hand-picked results of the StyleGAN 2 model trained on the FFHQ dataset with a resolution of 1024 × 1024. [13]

## 3.2 Future frame prediction

Future frame prediction, also called video prediction, is the problem setting of inferring future frames based on a given video sequence from the past. This task has many use-cases like future state estimation of autonomous driving cars by video analysis or label propagation to boost semantic segmentation. Also, the problem of predicting a microscopy image, based on microscopy images from previous time steps, can be seen as a future frame prediction problem. [19, 22]
The task of future frame prediction is, in particular, challenging since a video prediction model has to capture how objects are moving and as well as their displacement affects. All common approaches to future frame prediction rely on deep neural networks, however, video prediction is still an active area of research. [19, 22]
Current and past approaches to video prediction can be clustered into three classes. First, direct approaches, kernel-based methods [95], and flow-based methods [20]. Direct approaches directly predict the pixels of the next video frame. Video frames predicted by direct approaches are, however, often blurry and long-range motions are captured poorly. Flow-based methods also called vector-based methods often produces speckle noise in the predicted video frame. The class of kernel-based methods are typically able to capture small motions. The big problem with kernel-based methods is that they are typically unable to model large motions. [22]

### 3.2.1 Deep Multi-Scale Video Prediction Beyond Mean Square Error

The multi-scale video prediction approach by Mathieu et al. [19], also called BeyondMSE, introduces an unsupervised method for direct future frame prediction. Instead of learning the next frame by simply applying a mean squared error, which often leads to blurry predictions image, an adversarial multi-scale architecture in combination with an image gradient difference loss function is proposed. The BeyondMSE approach showed strong results compared to previous methods and was influential for future research on the problem of video prediction. [19]

#### 3.2.1.1 Method

Instead of predicting the future frame in the original resolution the multi-scale architecture, shown in figure 3.9, first generates the future frame at a lower resolution. In each stage, the prediction of the previous stage is upscaled and fed into the next stage, alongside n previous frames of the current stage resolution. This approach enables the lower resolution stages to learn large motions and higher resolution stages to capture small motions as well as to produce a sharp prediction. [19]

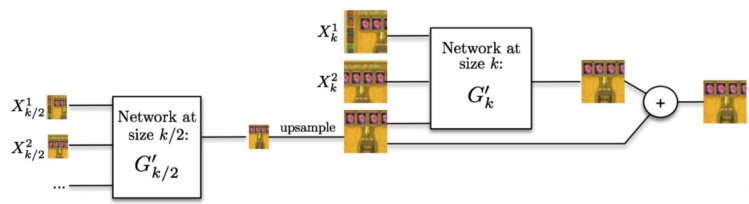

**Figure 3.9:** Multi-scale generator network architecture for future frame prediction. [19]

The introduced multi-scale architecture is trained in an unsupervised and adversarial fashion. To achieve the adversarial training setting for each stage a separate discriminator network is utilized. The whole model is trained on the classical adversarial loss [8], computed for each stage, and an image gradient differences loss function. This image gradient differences loss function penalizes the differences of the image gradients and should lead to sharper predictions. [19]

### 3.2.1.2 Results

The multi-scale architecture has been trained and tested on the Sport1m dataset (Fig. 3.10) and the UDF101 dataset. The authors evaluated the performance of the multi-scale architecture for different loss functions. [19]

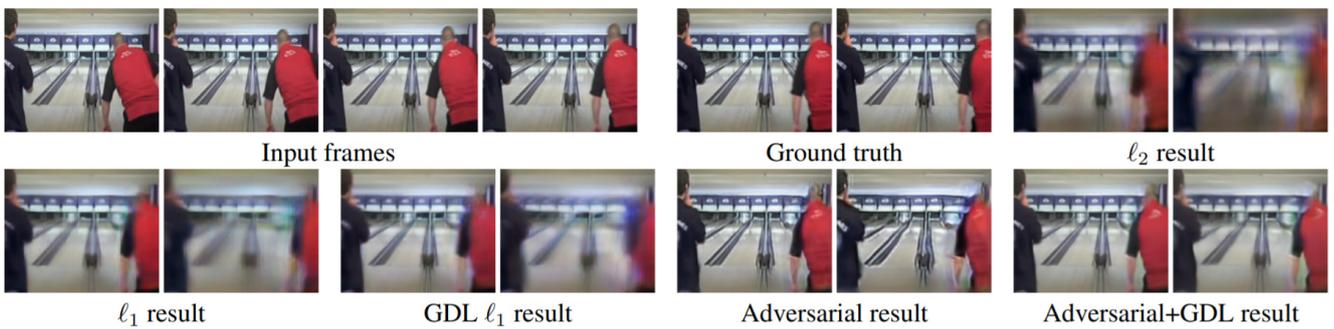

**Figure 3.10:** Results of the multi-scale generator architecture for different training settings. [19]

The quantitative results in the paper, as well as the qualitative results, are shown in figure 3.10, lead to the conclusion that training a direct video prediction model in an adversarial setting leads to much better results. In particular, the adversarial loss in combination with the gradient difference loss outperformed standard loss functions like the L1 loss or the MSE loss. [19]

### 3.2.2 SDC-Net

The paper "SDC-Net: Video prediction using spatially-displaced convolution " by Reda et al. [22] proposes a novel approach to the problem of future frame prediction. The introduced spatially-displaced convolution (SDC) combines the ideas of flow-based and kernel-based approaches for video prediction. Alongside the SDC the authors also propose a 3d U-Net based architecture for predicting the parameters of the SDC. The proposed framework achieved state-of-the-art performance on natural video sequences as well as on synthetic computer game sequences. [22]

### 3.2.2.1 Spatially-Displaced Convolution

The problem of future frame prediction given a sequence of past frames $\mathbf{I}_{1:t}$ can be formulated as a transformation learning problem [22]

$$\mathbf{I}_{t+1} = \mathscr{T}\left(\mathscr{G}\left(\mathbf{I}_{1:t}\right), \mathbf{I}_t\right). \tag{3.10}$$

Where $\mathcal{T}$ is the transformation function and $\mathcal{G}$ a learned function which predicts the parameters needed for the transformation $\mathcal{T}$.

Flow-based methods typically apply bilinear backward resampling for predicting $I_{t+1}$ [20, 22]

$$\mathbf{I}_{t+1}(x, y) = f\left(I_t(x + u, y + v)\right). \tag{3.11}$$

Where $(u, v)$ is the motion vector of the pixel $(x, y)$ in $\mathbf{I}_t$ and $f$ the bilinear interpolator. The motion vector is typically predicted by $\mathcal{G}$. [20, 22]

Kernel-based methods define the transformation $\mathcal{T}$ as a convolution. This operation learns combines resampling and displacement in a one operation and is defined as: [95, 22]

$$\mathbf{I}_{t+1}(x, y) = K(x, y) * \mathbf{P}_t(x, y), \tag{3.12}$$

where $\mathbf{P}_t(x, y)$ is the centered patch, of the shape $N \times N$, at $(x, y)$ in $\mathbf{I}_t$. Additionally, $K(x, y) \in \mathbb{R}^{N \times N}$ is a 2d kernel matrix typically predicted by the learned function $\mathcal{G}$ at the position $(x, y)$. [95, 22]

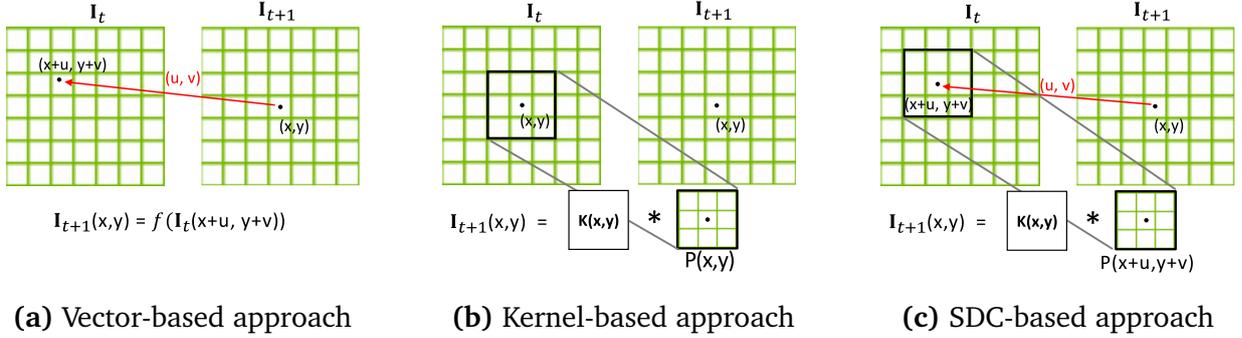

**(a)** Vector-based approach    **(b)** Kernel-based approach    **(c)** SDC-based approach

**Figure 3.11:** Illustration of transformations for future frame prediction. (a) A vector-based operation which utilizes bilinear interpolation to produce the future frame. (b) Kernel-based method which uses a convolution with a centered patch. (c) The spatially-displaced convolution which employs a displaced patch. [22]

The spatially-displaced convolution, as shown in figure 3.11c, combines the flow and kernel-based operation. SDC utilizes a motion vector $(u, v)$ as well as an adaptive kernel $K(x, y)$. The spatially-displaced convolution is mathematically defined as: [22]

$$\mathbf{I}_{t+1}(x, y) = K(x, y) * \mathbf{P}_t(x + u, y + v). \tag{3.13}$$

The kernel $K(x, y)$ gets convolved with a patch $\mathbf{P}$ displaced at the location $(x+u, y+v)$ by the motion vector $(u, v)$ from $\mathbf{I}_t$, to predict the pixel in the future frame $\mathbf{I}_{t+1}(x, y)$. To achieve the patch $\mathbf{P}_t$ bilinear sampling (backward resampling) is utilized since the motion vector contains real values. [22]

When the motion vector $(u, v)$ is all-zero the SDC is reduced to equation 3.12. In the case if the kernel $K(x, y)$ is all-zero except for the center element, with the value 1, equation 3.11 can be derived from the SDC. [22]

For a more efficient estimation of the kernel $K$ not the whole $N \times N$ kernel for each pixel is predicted. Instead of the whole kernel a pair of 1d kernels is predicted and the final 2d kernel is achieved by the outer-product of the 1d kernel pair. [22]

### 3.2.2.2 Architecture

The SDC-Net generator architecture $\mathscr{G}$ is based on a CNN that estimates the kernels pairs $\mathbf{K}_v$ and $\mathbf{K}_h$ as well as the motion vectors $U$ and $V$ for each pixel in $\mathbf{I}_t$. The whole video prediction model can be formulated as

$$\mathbf{I}_{t+1} = \mathscr{T}(\mathscr{G}(\mathbf{I}_{1:t}, \mathbf{F}_{2:t}), \mathbf{I}_t) \tag{3.14}$$

where $\mathscr{T}$ is the spatially-displaced convolution module. Additionally, the tensor $\mathbf{F}_{2:t}$ represents the backward optical flow between the previous input frames $I_{1:t}$. These flow maps gets fed into the generator model together with previous video frames $I_{1:t}$. The backward optical flow especially is utilized since backward resampling is utilized. used since The whole model described in formula 3.14 can be seen in figure 3.12.

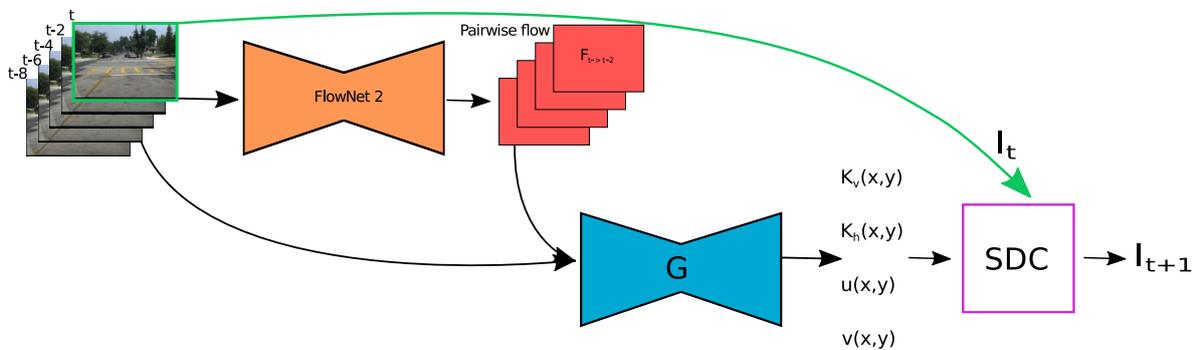

**Figure 3.12:** The full architecture of the SDC-Net including the FlowNet 2 for the backward optical flow estimation, the generator network $\mathscr{G}$, and the spatially-displaced convolution module. [22]

As illustrated in figure 3.12 the backward optical flow maps are computed by the deep convolutional neural network FlowNet 2 [15], which was pre-trained in a supervised setting on the Chairs and Things3d optical flow datasets. For a detailed description of the FlowNet 2 model see [15]. The CNN generator network $\mathscr{G}$ follows the architecture of the 3d U-net model and convolves over the spatial dimensions as well as the time dimension. The generator model, however, consist of three different encoder heads. One head predicts the motion vectors and the other two heads predicts the one of the separate 1d kernels each. [22]
A detailed description about the generator architecture and the chosen hyperparameters can be found here [22].
One important fact to mention is that the motion vectors $U$ and $v$ predicted from the generator are not equivalent to the backward optical flow. This is due to the fact that the pure backward optical flow is not defined. [22]

### 3.2.2.3 Training

The SDC-Net generator is trained in an unsupervised fashion on video sequences. The authors proposed a three stage training process. These three stage training process include a motion training, a kernel initialization training, and a fine tune training. [22]

The motion training only optimizes the encoder of the generator and the motion prediction head to predict the motion vector for the backward resampling. This backward resampling is equal to equation 3.11 or to equation 3.13 if all kernels are one valued in the middle element and zero valued elsewhere. This training step is done to learn large displacements. In this training stage a simple L1 loss is utilized which is defined for the predicted frame $\hat{\mathbf{I}}_{t+1} \in \mathbb{R}^{c \times h \times w}$ and the true video frame $\mathbf{I}_{t+1} \in \mathbb{R}^{c \times h \times w}$ as [22]

$$L_1 = \frac{1}{c\,h\,w} \left\| \mathbf{I}_{t+1} - \hat{\mathbf{I}}_{t+1} \right\|_1. \tag{3.15}$$

To have a smooth transition between the motion training and the fine tune training a kernel initialization training is utilized where only the kernel prediction heads are optimized. To force the 1d kernel predictions to always result in a middle-one-hot vector a L2 loss between the kernel prediction and the desired fixed middle-one-hot vector is used. [22]
The final fine-tune training step optimizes the whole generator network $\mathcal{G}$. Here a combined loss function between the L1 loss, the perceptual loss [96], and the style loss [96], based on a pre-trained VGG-16 [97], is utilized. [22]

### 3.2.2.4 Results

The SDC-Net was trained on the CaltechPed video dataset including complex real world scenes and on a subset of the YouTube-8M dataset. The results on the CaltechPed dataset showed that the SDC-Net outperforms previous methods like the BeyondMSE model and reached state-of-the-art performance in the task of future frame prediction. [22]

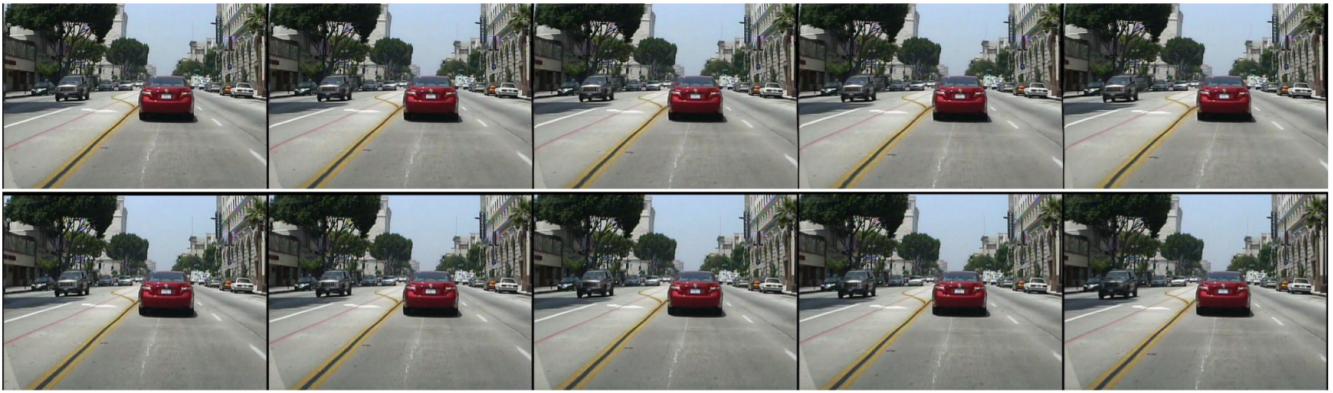

**Figure 3.13:** Results for five frame prediction of the SDC-Net in the top row and the corresponding ground truth frames in the bottom row. [22]

When predicting multiple future frames in an autoregressive way the SDC-Net also showed strong results as shown qualitative in figure 3.13. Despited that the SDC-Net is only trained to predict one frame even multiple predcted frames are sharp and detailed. [22]

## 3.3 Additional Related Work

This section briefly introduces multiple recent methods related to the tasks of future frame prediction, optical flow estimation, and generative adversarial networks. Especially, methods regarding the StyleGAN [12, 13] architecture are introduced.

### 3.3.1 Image2StyleGAN

The StyleGAN architectures, described in sections 3.1.4 & 3.1.3, are able to perform style transformation of two or even multiple images. This style transfer process is shown in figure 3.6. However, to perform style transfer, the latent code of the images to be utilized, in this process, has to be known. Thus if style transfer should be performed on real images the corresponding latent vector has to be estimated. The paper "Image2StyleGAN: How to Embed Images Into the StyleGAN Latent Space?" by Abdal et al. [98] proposes an algorithm to derive the latent vector from real images. However, the images get not transformed into the latent space $\mathbb{W}$ but into an extended latent space $\mathbb{W}^+$. This extended latent space includes one latent vector for one styled stage in the StyleGAN architecture. [98]

**Algorithm 4** Latent space embedding ($\mathbb{W}^+$) algorithm for the StyleGAN [12, 13] architecture. [98]

**Require:** An image $\mathbf{I} \in \mathbb{R}^{c \times h \times w}$; a pre-trained Style GAN generator $G$ (only the synthesis network $g$).
1: Initialize the latent code $w^*$
2: **while** not converged **do**
3: $\quad L \leftarrow L_{\text{perceptual}}(G(w^*, \mathbf{I})) + L_{\text{MSE}}(G(w^*, \mathbf{I}))$
4: $\quad w^* \leftarrow w^* - \eta \mathbf{\nabla}_{w^*} L$
5: **end while**
6: **return** The embedded latent code $w^+ \in \mathbb{W}^+$ and the embedded image $G(w^+)$.

Performing this algorithm, which is based on the perceptual loss $L_{\text{perceptual}}$ [96] and the mean-squared-error $L_{\text{MSE}}$, however, takes between six and seven minutes on a high-end workstation GPU. This long runtime makes the algorithm impractical to use in real-world applications. [98]

### 3.3.2 $R_1$ Regularization

One common regularization of the generative adversarial training is the $R_1$ regularizer. This regularizer penalizes large gradients of the discriminator for real data samples. This should force the discriminator from deviating from the Nash-equilibrium. The $R_1$ regularizer is defined as:

$$R_1 = \frac{\gamma}{2} \mathbb{E}_{\mathbf{x} \sim P_{\text{data}}} \left[ \|\mathbf{\nabla} D(\mathbf{x})\|_2^2 \right] \quad (3.16)$$

where $\gamma \in \mathbb{R}$ is a weights factor to handle the strength of the regularization, $\mathbf{x}$ a real sample and $D$ the discriminator network. [69]
Corresponding to the $R_1$ regularization which uses real samples for the regularization, the $R_2$ regularizer utilizes fake samples to commute the regularization term. [69]

### 3.3.3 PWC-Net for Optical Flow Estimation

The PWC-Net is a CNN based state-of-the-art model for optical flow estimation. Compared to the previous state-of-the-art FlowNet 2 model the PWC-Net achieve better optical flow prediction and reduces the number of learnable parameters by a factor of 17 resulting in a much faster runtime. These improvements are achieved by incorporating domain knowledge of traditional optical flow estimation methods into deep convolutional neural networks. [16]

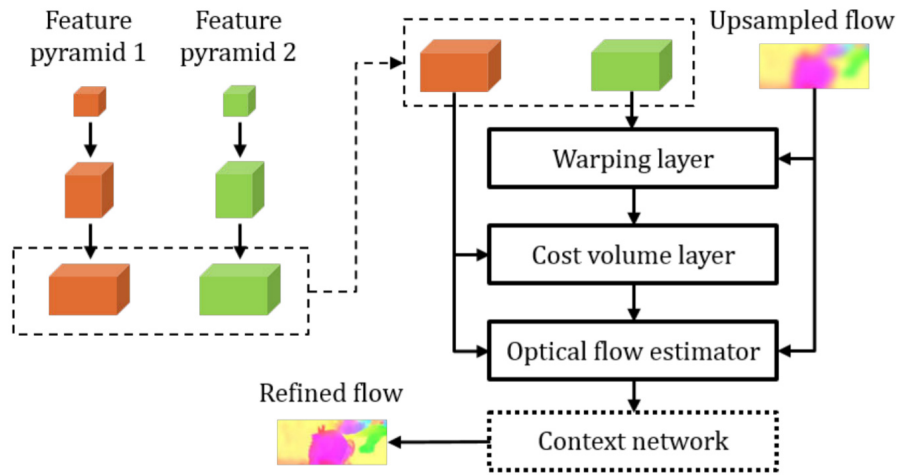

**Figure 3.14:** Architecture of the PWC-Net for optical flow estimation. [16]

As illustrated in figure 3.14 the PWC-Net architecture first predicts feature pyramids for each of the two consecutive input frames. In each stage of the feature pyramids, except for the lowest stage, the features of the second frame are backward wrapped by the upsampled predicted flow of the lower previous stage. With the backward wrapped features and the features of the first frame, a cost volume is produced. Based on this cost volume as well as the upsampled previous optical flow and the feature of the first video frame a CNN predicts the optical flow for the current stage which is upsampled and passed to the next higher stage. [16]
The PWC-Net is trained in a supervised fashion with ground truth optical flow labels. The network is optimized on a multi-scale L2 loss. To be precise each stage of the PWC-Net is supervised with a downsampled version of the ground truth optical flow. This multi-scale loss showed in previous work great improvements for optical flow estimation [15]. The authors also introduced a fine-tune training where the PWC-Net model is optimized on a multi-scale L1 loss. [16]

### 3.3.4 DeepFovea: Fovea Sampled Video Reconstruction

The paper "DeepFovea: Neural Reconstruction for Foveated Rendering and Video Compression using Learned Statistics of Natural Videos" by Kaplanyan et al. [99] introduces the task of fovea sample video reconstruction by a CNN. Fovea sampling is the process of sampling an image or video frame with a focus region where many pixel information are preserved and an out-of-focus region where less pixel information are preserved. This sampling process can be seen in figure 3.15 (left) is inspired by the focus of the human eye and has many used cases especially in rendering virtual reality environments. [99]

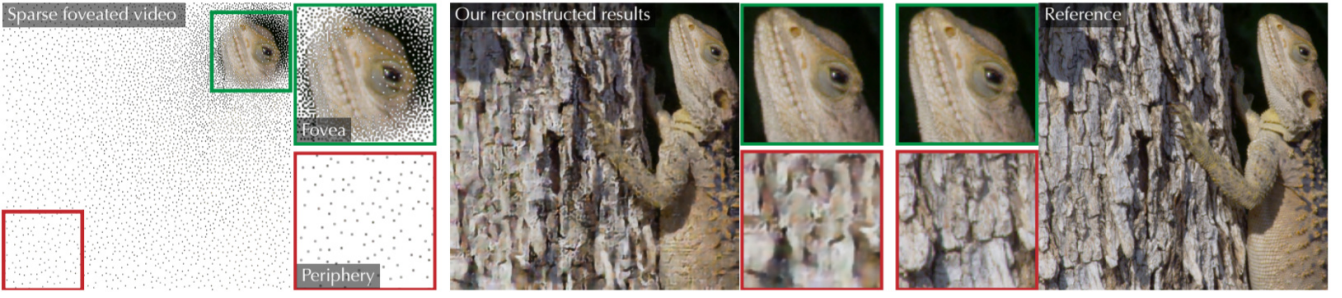

**Figure 3.15:** Results of the partly adversarial based training process for fovea sampled video sequence reconstruction. [99]

The DeepFovea framework utilizes a residual and recurrent 2d U-Net as the reconstruction model. The reconstruction model is trained to learn the reconstruction of the fovea sampled input video sequence in the most plausible way. A detailed visualization of the reconstruction network can be seen in figure 3.16. [99]

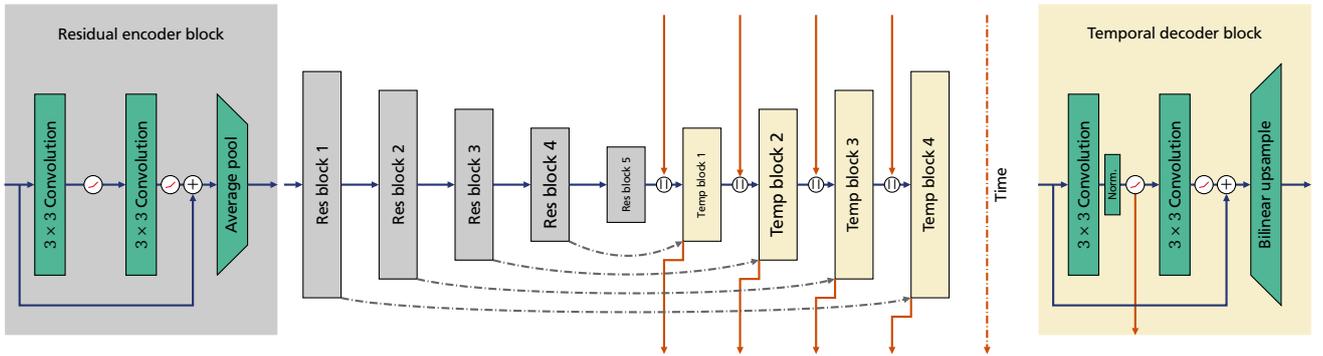

**Figure 3.16:** Architecture of the DeepFovea reconstruction network based on a residual and recurrent U-Net architecture. [99, 45, 80]

The reconstruction network is trained in an unsupervised setting on a combined loss function

$$L = w_{\text{adv}} L_{\text{adv}} + w_{\text{adv fft}} L_{\text{adv fft}} + w_{\text{LPIPS}} L_{\text{LPIPS}} + w_{\text{flow}} L_{\text{flow}}. \quad (3.17)$$

Where $L_{\text{LPIPS}}$ is the perceptual loss [96] between the reconstructed and the original video sequence. Additionally, $L_{\text{flow}}$ indicates the flow loss which comprises the process of backward wrapping each video frame to the previous predicted video frame with the optical flow estimated by a pre-trained FlowNet 2 model and the commutation of the L1 loss between the wrapped and the original predicted frame. $L_{\text{adv}}$ and $L_{\text{adv fft}}$ represents the adversarial losses where $L_{\text{adv}}$ is produced by a standard ResNet-like [43] discriminator which takes the whole reconstructed as an input. The adversarial loss $L_{\text{adv fft}}$ is, however, predicted by a separate ResNet-like discriminator network which however takes the 3d FFT spectrum of the whole video sequence as an input. This FFT adversarial loss in utilized because natural video frames have a characteristic statistics of a vanishing Fourier spectrum. [99]

### 3.3.5 General and Adaptive Robust Loss Function

Many computer vision learning tasks like monocular scene flow [100] estimation or optical flow [14, 15, 16, 18] estimation rely on convex regression losses. However, choosing the right loss function is a non-trivial task. Barron proposed a general and adaptive robust loss (GARLoss) function. This GARLoss is a generalization the Cauchy, Geman-McClure, Welsch, generalized Charbonnier, smooth L1, and L2 (MSE) loss functions. It future introduces the robustness as a continuous parameter. The general and adaptive robust loss function is defined as: [32]

$$\rho : \mathbb{R} \times \mathbb{R} \times \mathbb{R}^+ \to \mathbb{R}, \ \rho(x, \alpha, c) = \frac{|\alpha - 2|}{\alpha} \left( \left( \frac{(x/c)^2}{|\alpha - 2|} + 1 \right)^{\alpha/2} - 1 \right), \tag{3.18}$$

where $x \in \mathbb{R}$ represents the difference between some label and some prediction. The shape parameter $\alpha \in \mathbb{R}$ controls the shape of the loss function and the parameter $c > 0$ is a scale parameter that controls the quadratic bowl size of the loss $\rho$ near $x = 0$. [32]

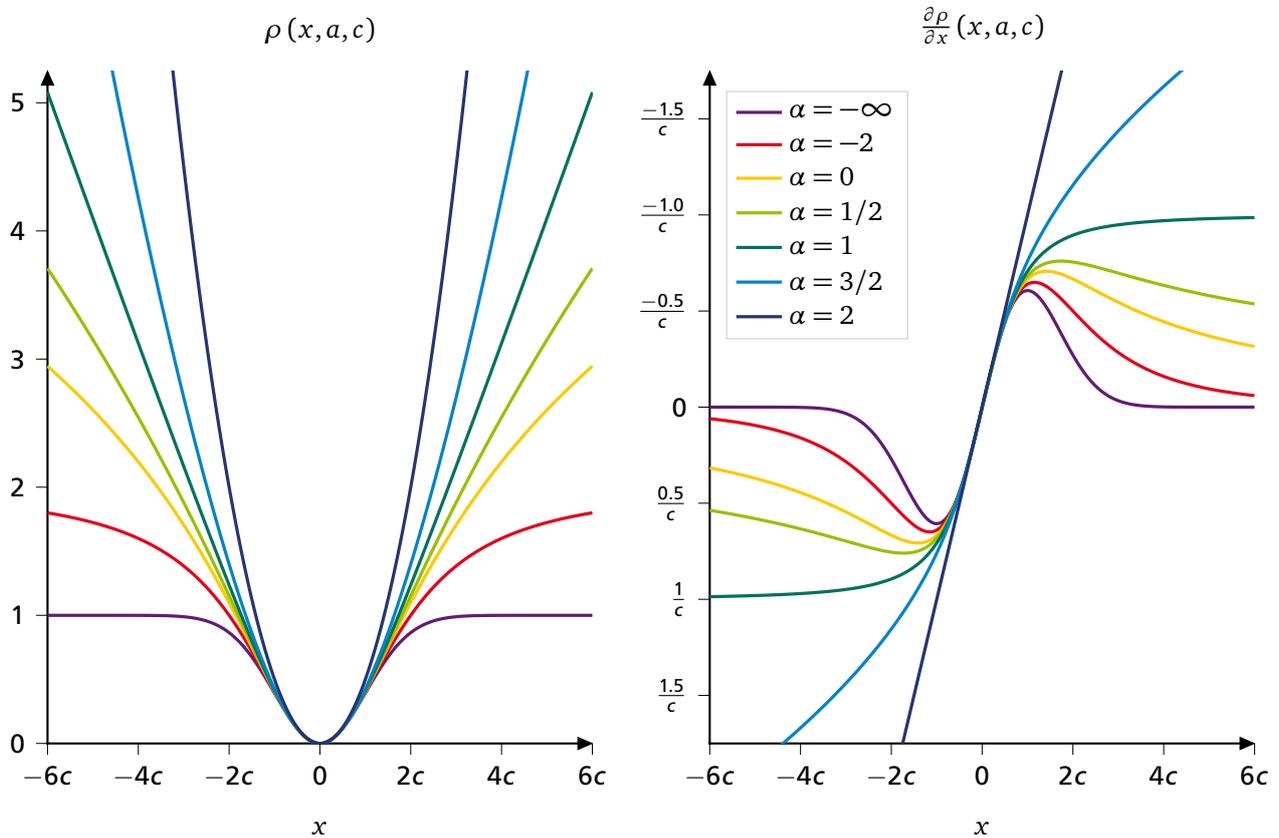

**Figure 3.17:** General loss function on the left and its gradient on the right for different values of the shape parameter $\alpha$ (in different colors). Some $\alpha$ values reproduce exiting loss functions: $\alpha = 2$ the L2 loss ▆, $\alpha = 1$ the Charbonnier loss ▆, $\alpha = 0$ the Cauchy loss ▆, $\alpha = -2$ the Geman-McClure loss ▆, and $\alpha = -\infty$ the Welsch loss ▆. [32]

As shown in figure 3.17 the GARLoss can reproduce existing loss functions. This can also be derived mathematically. For a detailed mathematical derivation see [32].
Choosing the right loss function for a specific learning problem is non-trivial since the loss function type is a hyperparameter which can only be optimized by computational expansive hyperparameter optimization. The GARLoss introduces robustness as a continuous shape parameter $\alpha$. This shape parameter, however, can not be trained directly by backpropagation since the training would end in a trivial solution. [32]

To overcome this limitation Barron proposed the approach of interpreting the GARLoss as the negative log of a univariate density. The probability density function is defined as:

$$p(x|\alpha,c) = \frac{1}{c\,Z(\alpha)} \exp(-\rho(x,\alpha,c)) \qquad (3.19)$$

$$Z(\alpha) = \int_{-\infty}^{\infty} \exp(-\rho(x,\alpha,1)). \qquad (3.20)$$

The general probability distribution $p(x|\alpha,c)$ is only defined if $\alpha \geq 0$, since $Z(\alpha)$ diverges $\forall \alpha < 0$. Estimating and differentiating $Z(\alpha)$ is non-trivial, however, can be approximated with a cubic hermite spline. [32]

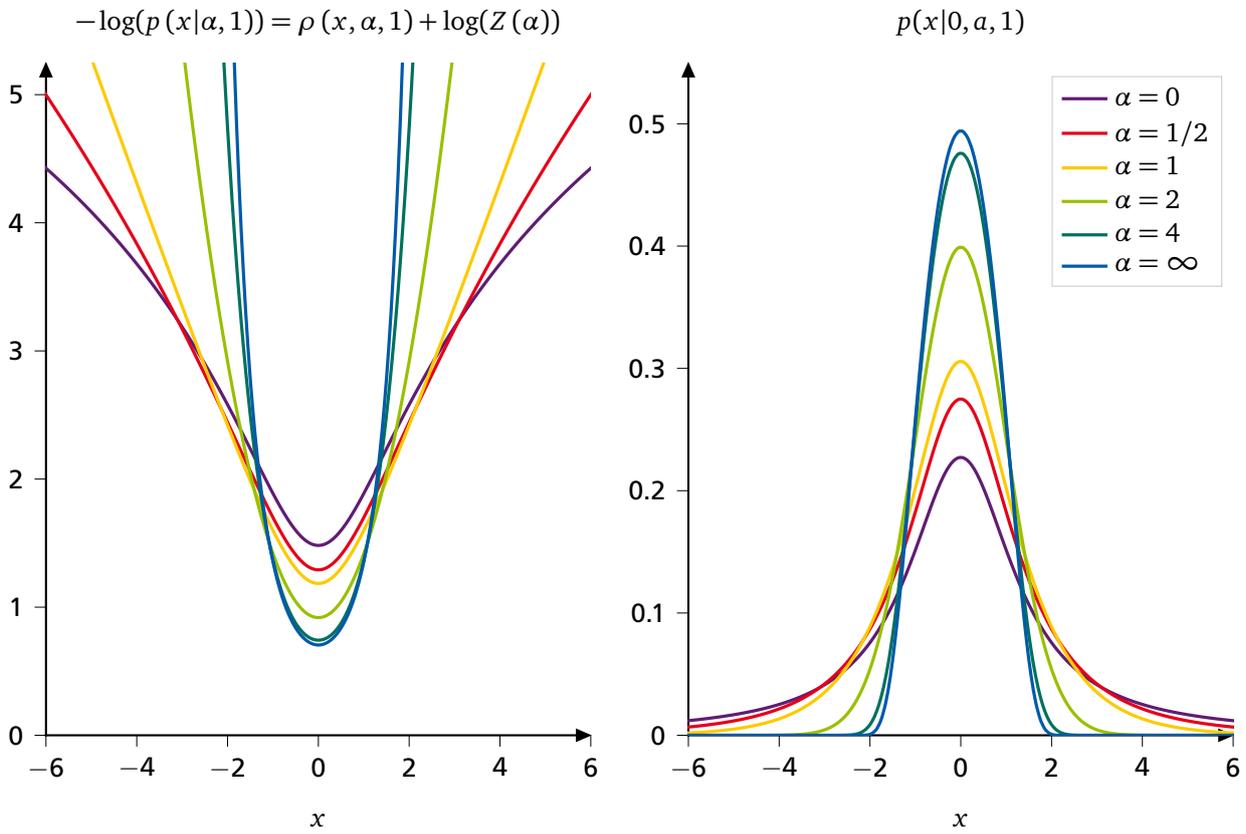

**Figure 3.18:** Plot of the negative log-likelihood (left) and the probability density (right) function of the General loss function. When $\alpha = 2$ ■ the PDF (eq. 3.19) reproduces a Gaussian normal distribution $\mathcal{N}(0,1)$. [32]

By utilizing the negative log-likelihood (NLL) $-\log(p(x|\alpha,c))$ for training enables to optimize a neural network as well as the shape parameter $\alpha$ and the scale parameter $c$. The important property of the NLL to not end up in a trivial solution when optimizing is that the NLL puts a penalty on inliers when giving a discount on outliers. This property can be observed in figure 3.18. [32]

### 3.3.6 Padé Activation Unit

Non-linear activation functions are an essential part of deep neural networks of all kinds. The activation function often influences the performance of a deep model. And the choice of the optimal activation function is non-trivial. In the past, multiple activations like, for example, Rectified Linear Unit (ReLU) [101], LeakyReLU [52], Swish [102], or the ELU [103] has been proposed. Furthermore, learnable activation functions like PReLU [53], SReLU [104], or Maxout [105] has been introduced. All proposed activation functions are fixed in some way. The Padé Activation Unit (PAU) [106] solves this issue by utilizing flexible parametric rational functions resulting in a learnable activation which is able to reproduce multiple previous activation as shown in figure 3.19. [106]

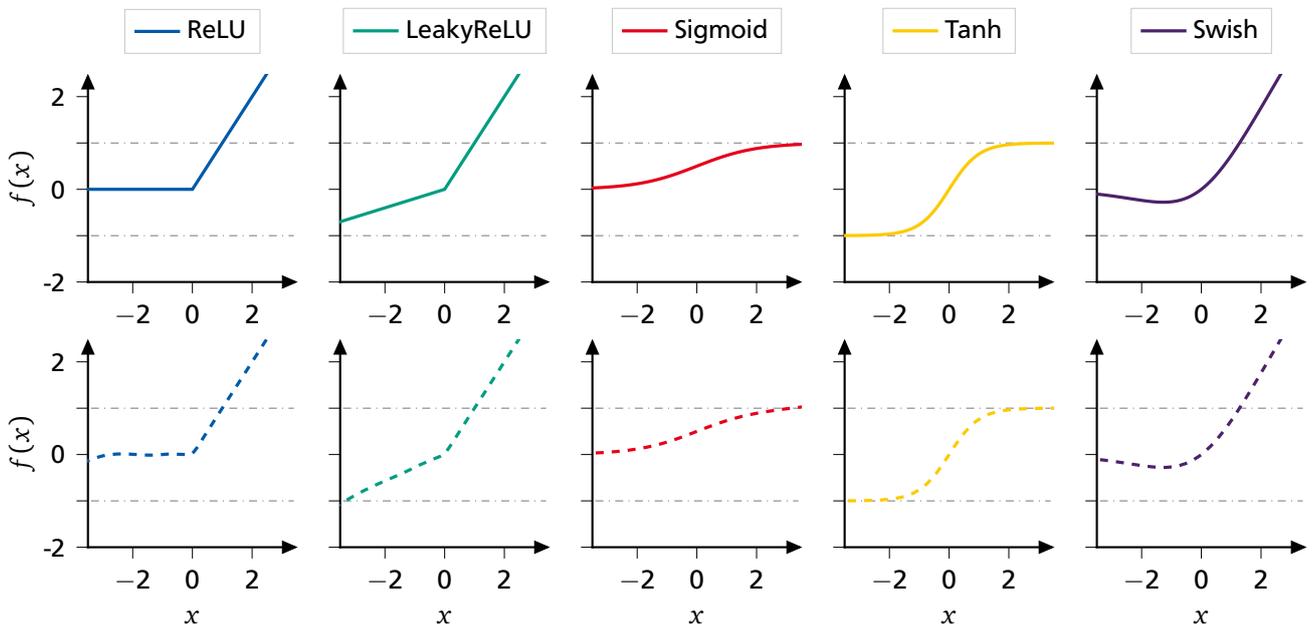

**Figure 3.19:** Common activation functions (top) approximated by the Padé Activation Unit (bottom). [106]

The Padé Activation Unit is the fraction of two polynomials $P(x), Q(x)$ of order $m, n$ and is mathematically defined as: [106]

$$\text{PAU} : \mathbb{R} \to \mathbb{R}, \text{PAU}(x; \boldsymbol{a}, \boldsymbol{b}) = \frac{P(x; \boldsymbol{a})}{Q(x; \boldsymbol{b})} = \frac{\sum_{j=0}^{m} a_j x^j}{1 + \left|\sum_{k=1}^{n} b_k x^k\right|} = \frac{a_0 + a_1 x + a_2 x^2 + \ldots a_m x^m}{1 + |b_1 x + b_2 x^2 + \ldots + b_n x^n|}. \tag{3.21}$$

The polynomial $Q$ is constrained to be $> 1$ to prevent PAU from having poles. [106]
The PAU showed, by setting $m = 5$ and $n = 4$, great improvements over traditional non-parameterized activation functions and also over more recent parameterized activation functions. However, the performance of the PAU was only evaluated in the task of image classification. A trained PAU can be seen in section 9.2 of the appendix. [106]

# 4 Trapped Yeast Cell Time-Series Dataset

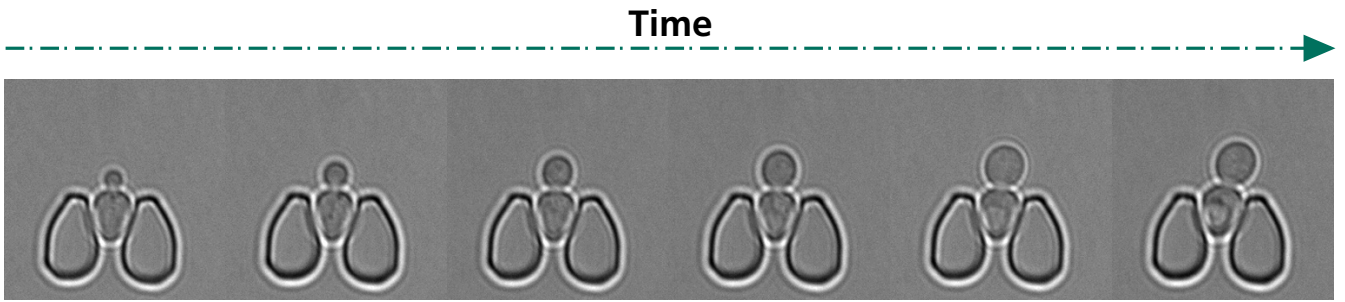

**Figure 4.1:** Example sequence of the yeast cell time-series dataset including six time-steps. Time-step $t = 0$ on the left, increasing from left to right, to the final time-step $t = 5$ on the right (image histograms adopted for better visibility).

For this thesis, a time-series dataset of bright field microscopy images including trapped yeast cells from one experiment was collected. The dataset consists of 4748 images ordered in 641 time-series including at least 3 microscopy images of consecutive time steps. An example time-series of six time-steps including one mother cell and one growing daughter cell can be seen in figure 4.1.

## 4.1 Data Acquisition

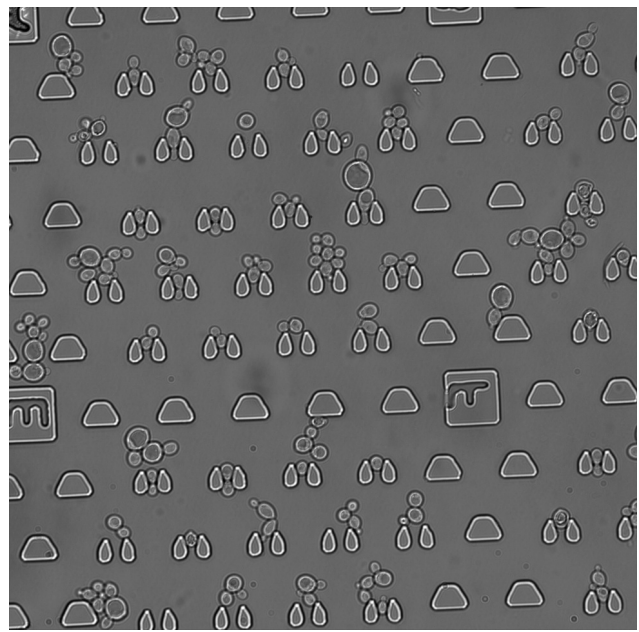

**Figure 4.2:** Example of a big brightfield microscopy image from one position (image histogram adopted for better visibility).

The dataset was acquired from one time-lapse fluorescence microscopy experiment (*60x_10BF_ 200GFP_200RFP20_3Z_10min_3*) of Jascha Diemer. This experiment includes 31 positions, 70 time-steps with a $\Delta t$ of 10min, and 3 $z$ slices for the brightfield channel.

To get clean images with healthy cells, first, positions with no-overcrowded traps were selected. The selected positions include the position 18, 20, 21, 22, 24, 25, 26, 27, 28, 29, and 30. Secondly, a time interval with no-overcrowded and sharp images was selected for each position. The interval was chosen to be as large as possible. Thirdly, the big images (shown in figure 4.2) of each position over time were cropped to a resolution of 256 × 256 by pattern matching to produce small images of each trap. Finally, each produced time-sequence of the cropped images was visited and selected to not include too many cells ($\gtrapprox$ 8) or no cells and highly random cell movements. To be precise if many cells get flush in or out of the image sequence the sequence was not selected.

## 4.2 Data Variation

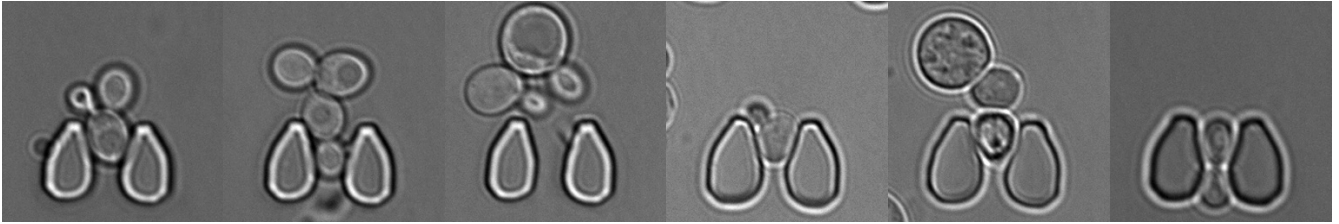

**Figure 4.3:** Gray-scale brightfield microscopy image samples from different sequences of the yeast cell dataset (image histograms adopted for better visibility).

The trapped yeast cell time-series dataset includes a high variance of different cell formations as can be seen in figure 4.3. It is important to note that each image includes at least one mother cell. Over time at least one daughter cell is occurring in the image. The dataset includes samples where a daughter cell is growing to the top of the image as well as samples where the daughter cell is growing to the bottom. Also washed out daughter cells are included in the dataset.

## 4.3 Dataset Versions

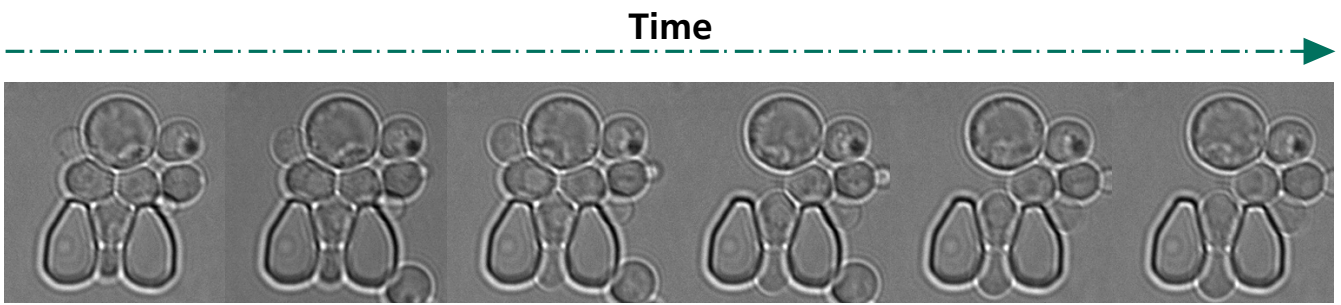

**Figure 4.4:** Example sequence of the yeast cell time-series dataset with randomness including six time-steps. Time-step $t = 0$ on the left, increasing from left to right, to the final time-step $t = 5$ on the right (image histograms adopted for better visibility).

From the original trapped yeast cell time-series dataset also a smaller subset is extracted. In this smaller dataset, as much randomness as possible is eliminated. This means that the occurrence of flushed out cells is reduced to a minimum, as can be observed in figure 4.4. The resulting subset of the original dataset includes 2739 images structured in 392 sequences. This dataset is especially used for future frame prediction to explicitly model the movements and the growth of the cells. Further in this thesis, this dataset is called trapped yeast cell time-series dataset without randomness.

## 4.4 Dataset Splits

The following splits shown in table 4.1 are applied to the datasets. It is important to notice that the datasets are split sequence wise with respect to the z position. Which means that one splitted set contains z positions of one sequence.

Table 4.1: Different splits of the trapped yeast cell time-series dataset with and without randomness.

| Yeast Dataset | Use-case | Train sequences/images | Validation sequences/images | Test sequences/images |
|---|---|---|---|---|
| Randomness | Generation | 601/4489 | - | 40/250 |
| No Randomness | Generation | 367/2531 | - | 25/208 |
| Randomness | Simulation | 593/4297 | 24/259 | 24/192 |
| No Randomness | Simulation | 356/2478 | 18/138 | 18/123 |

The dataset splits for generation does not include a validation dataset since the network architecture is trained in an unsupervised setting and is not optimized on the test set.

## 4.5 Preprocessing and Augmentation

The preprocessing of the yeast cell dataset is reduced to two steps. In the first step, data augmentation is performed. To keep the geometric structure and original information as much as possible only random flipping on the vertical axis is utilized. In the second and final step normalization is performed. For the task of generation the images are normalized to a mean of one and a standard deviation of one. In the case of future frame prediction, the images are normalized to a pixel value range of zero to one.

# 5 Cell-GAN: Adversarial-Based Conditionalized and Unconditionalized Image Generation for Microscopy Imagery

This chapter proposes the novel Cell-GAN architecture for unconditionalized and guided (conditionalized) image generation. The Cell-GAN includes a style-bases generator network $\mathcal{G}$, a U-Net discriminator $\mathcal{D}$, a style mapping network $f$, and a guidance encoder $\mathcal{E}$. The style-based generator as well as the style mapping network is based on the StyeGAN 2 [13] architecture. The discriminator is inspired by the U-Net based discriminator network proposed in the CVPR 2020 by Schonfeld et al. [107].

## 5.1 Method

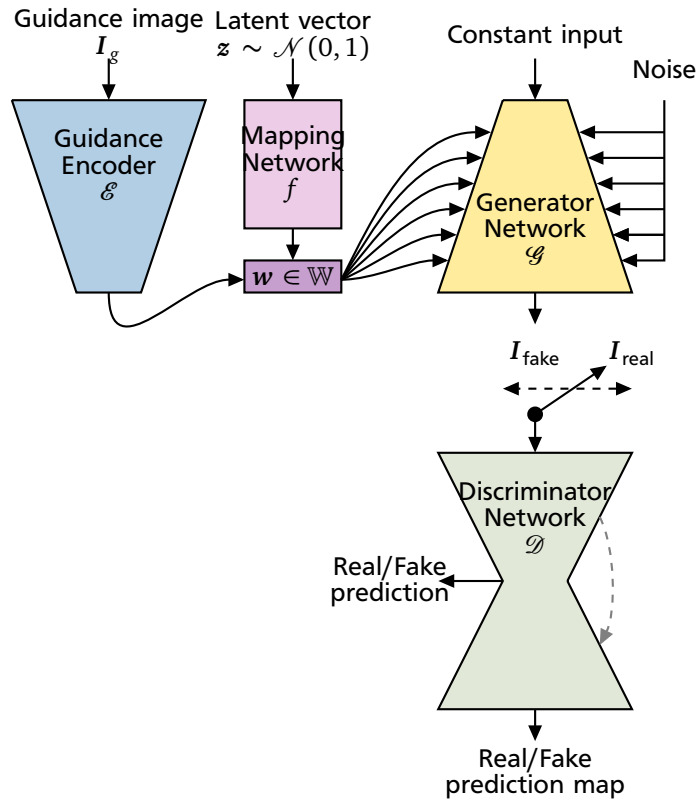

**Figure 5.1:** Schematic of the Cell-GAN architecture.

The Cell-GAN architecture, shown in figure 5.1, can be utilized for unconditionalized and guided image generation.

In the case of unconditionalized image generation, a random latent vector $z$ is produced and fed into the mapping network $f$ which produces an intermediate latent vector $w \in \mathbb{W}$. This intermediate latent vector can also be seen as an embedded latent vector where the essence of the image to be generated is captured. To produce the final image sample $I_{\text{fake}} \sim p_{\text{data}}$ a constant but learnable input is fed into the style CNN generator network $\mathcal{G}$. Additionally, in each stage of the generator the intermediate latent vector $w \in \mathbb{W}$ is processed by a separate linear transformation to achieve a styled vector. This style vector is utilized in each styled convolution of the current generator stage to finally produce the image sample. Additionally in each stage, a random noise bias is added.

For guided image generation first, a guidance image $I_g$ is fed into the guidance encoder $\mathcal{E}$. In the encoder, the guidance image is mapped into the intermediated latent vector $w \stackrel{\approx}{\in} \mathbb{W}$. With the produced intermediated latent vector $w$ the generator network then samples an image $I_{\text{fake}} \sim I_g$ approximately similar to the guidance image.

### 5.1.1 Generator Network

The generator network $\mathcal{G}$ follows the architecture of the StyleGAN 2 [13] generator. However, compared to the original StyleGAN 2 generator architecture the Cell-GAN generator has a reduced number of parameters for faster training and utilizes padé activation units [106] instead of LeakyReLU [52, 13] activation function.

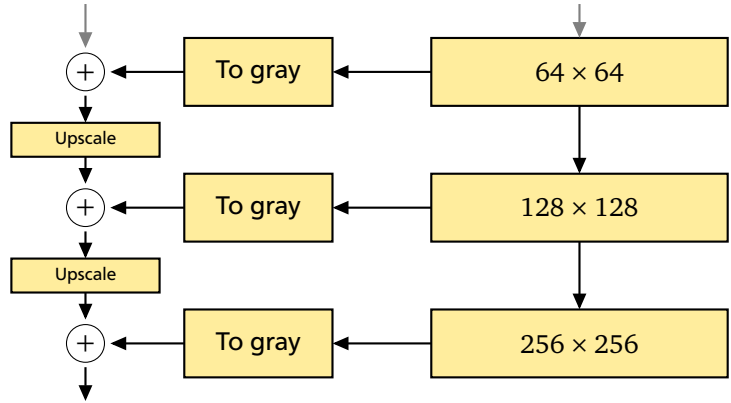

**Figure 5.2:** Generator architecture with output skip connections. In each resolution stage of the output of the stage is fed to the next resolution stage but also mapped by a 1 × 1 styled convolution (To gray) to a gray-scale output image. This output gray-scale image is added to the bilinear upsampled image of the previous. This process is repeated until in the last stage the final high-resolution output image is produced.

The original StyleGAN 2 showed experiments on different generator architectures. The best performing generator architecture thought multiple tests was a generator with skip connections to the output domain. The Cell-GAN generator follows this output skip connections architecture, which can be seen in figure 5.2.

Each resolution stage, shown in figure 5.2, is composed of two styled convolutions. These styled convolutions are taken from the StyleGAN 2 [13] architecture. Each styled convolution first transformed the incoming intermediated latent vector $z$ by an linear layer into a style-vector. This style vector is then utilized to modulate weights (formula 3.7) of the 3 × 3 convolution. After the modulation, a demodulation (formula 3.8) step is performed. After the modulation

5.1 Method

and demodulation of the convolution weights is performed, the resulting weights are used in the convolution to transform the incoming feature tensor. The resulting feature maps are then transformed by a padé activation unit [106]. Then a bias and a random noise are added channel-wise to the feature tensor. The second styled convolution utilizes the same operations as the first style convolution, however, additionally performs an upsampling of the incoming feature tensor. This upsampling is implemented as a bilinear upsampling followed by an FIR filter operation. The whole styled convolution block can be seen in figure 5.3.

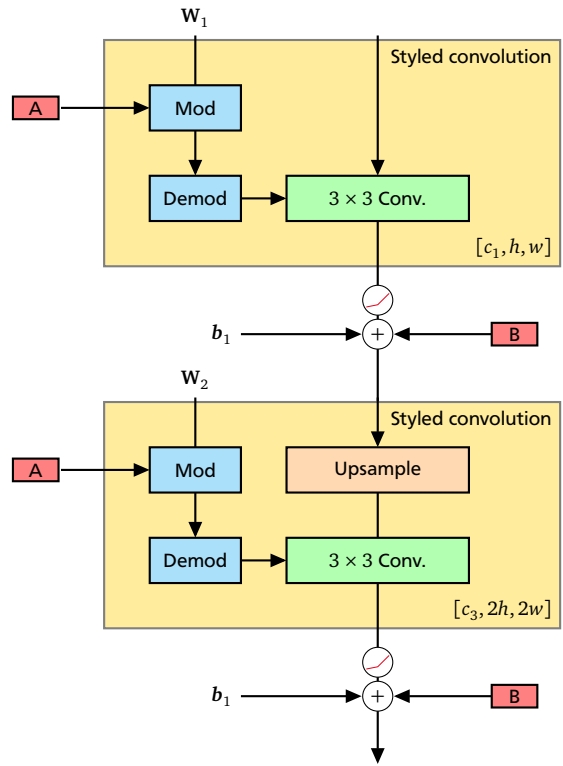

**Figure 5.3:** Schematic the styled convolutional block including two styled convolution. The second convolution utilizes in addition to the raw styled convolution also an upsampling operation. A indicates the linear transformation of the incoming intermediated latent vector to the style vector used in the modulation step. Additionally, B indicates a noise broadcast operation.

The Cell-GAN generator utilizes one starting styled convolution and six styled convolution blocks (figure 5.3) to transform the constant but learnable input of a resolution of $4 \times 4$ to the desired high-resolution ($256 \times 256$) output image. In the first four styled convolutional blocks 512 convolutional filters are utilized. In the next two blocks, 256 and 128 are used, respectively. For each block also an additional style convolution is utilized to perform the mapping to output skip path.

### 5.1.2 Mapping Network

The mapping network $f$ based on the StyleGAN 2 [13] architecture is utilized to map the random latent vector $z$, sampled from a normal distribution $\mathcal{N}(0,1)$, to the intermediated latent vector $w \in \mathbb{W}$. This mapping is used since the space of the input vector is non-optimal to capture the distribution of the image features corresponding to $p_{\text{data}}$. The mapping network learns during training the transformation of the random input vector to the intermediate latent vector and each styled convolution the mapping form the intermediate latent vector to the corresponding image feature.

The mapping network $f$ is implemented as an eight-layer feed-forward neural network with equalized linear layers [11] and PAU [106] non-linear activation function.

### 5.1.3 Discriminator Network

For the discriminator network $\mathscr{D}$, a U-Net [45, 44] is utilized following the proposed architecture of Schonfeld et al. [107]. The Cell-GAN U-Net discriminator, however, adopts the original U-Net discriminator architecture by using learnable padé activation units and applying a mini-batch standard deviation layer [11] to the bottleneck stage (last encoder block) of the U-Net. This mini-batch standard deviation layer computes mini-batch-wise statistics of the feature maps which enables the generator to capture the whole variance of the target data distribution $p_{\text{data}}$.

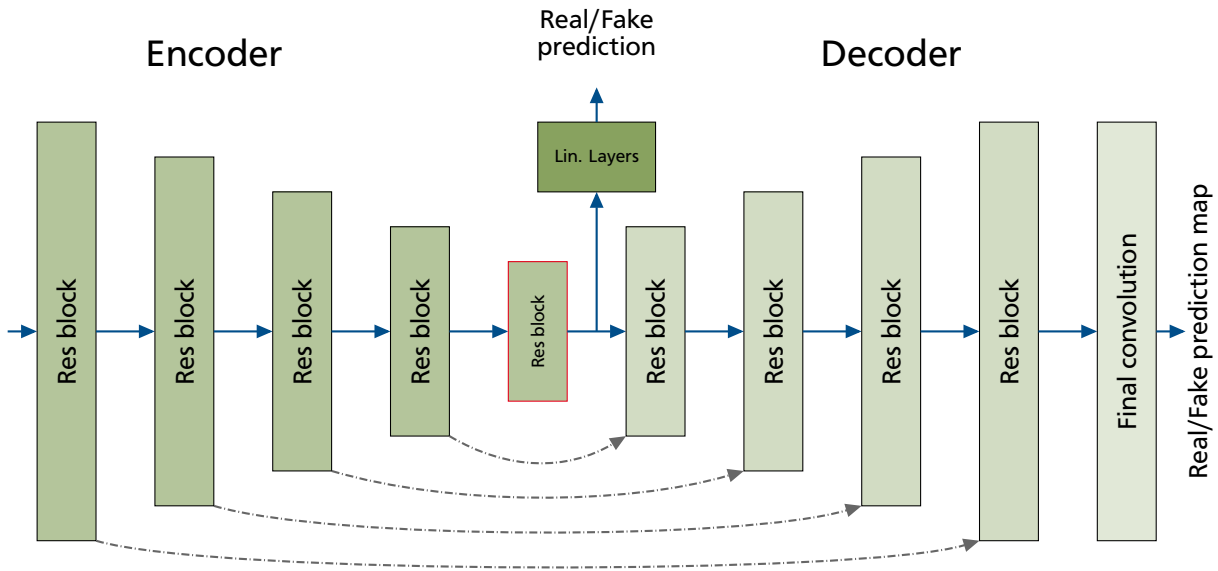

**Figure 5.4:** U-Net discriminator architecture with five residual encoder blocks ▬, four residual decoder blocks ▭ and skip connections between the encoder and the decoder. However, the not downsampled maps are passed by the skip connections to the decoder. Bottleneck encoder block with mini-batch standard deviation layer outlined in ▬. Final convolution layer to predict the final pixel-wise real/fake prediction in ▭. Linear layers to produce the scalar real/fake prediction in ▬.

The U-Net discriminator architecture is based on a standard residual U-Net [80]. The standard residual U-Net is extended by two linear layers that produce a scaler real/fake prediction. Standard GANs often suffer from local inconsistencies the U-Net discriminator [107] network solves this issue by also producing a pixel-wise real/fake prediction.

As can be observed from figure 5.4 the Cell-GAN U-Net discriminator utilizes five residual encoder blocks, four residual decoder blocks, two equalized linear layers, and a final convolution layer.

In the five encoder residual blocks 48, 96, 192, 384, and 768 convolutional filters are utilized, respectively. Each residual encoder block, as can be seen in figure 5.5, consists of two equalized $3 \times 3$ convolutional layers followed each by a padé activation unit, an average pooling layer with a kernel size of 2 × 2, and a residual mapping. The residual mapping is implemented as an equalized 1 × 1 convolution and a factor of $1/\sqrt{2}$ is multiplied to the tensor after the residual addition to prevent too large feature magnitudes, which can harm the adversarial training process.

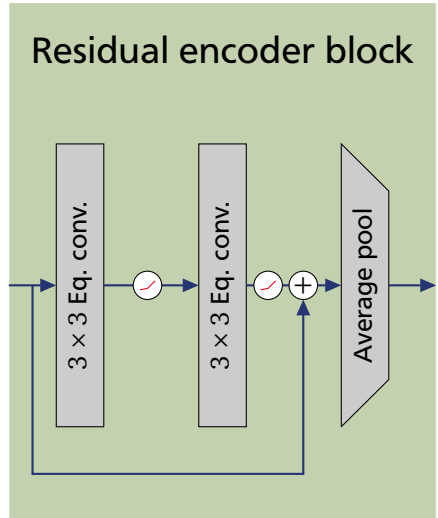

**Figure 5.5:** Residual U-Net encoder block with two equalized 3×3 convolutions, two activation functions, a residual mapping, and an average pooling layer.

In the five decoder residual blocks 48, 96, 192, and 384 convolutional filters are utilized, respectively. Each decoder block, as can be seen in figure 1, consists of a bilinear upsampling layer, a concatenation, two equalized 3 × 3 convolutional layers followed each by a padé activation unit, and a residual mapping. The residual mapping is implemented as in the U-Net residual encoder block.

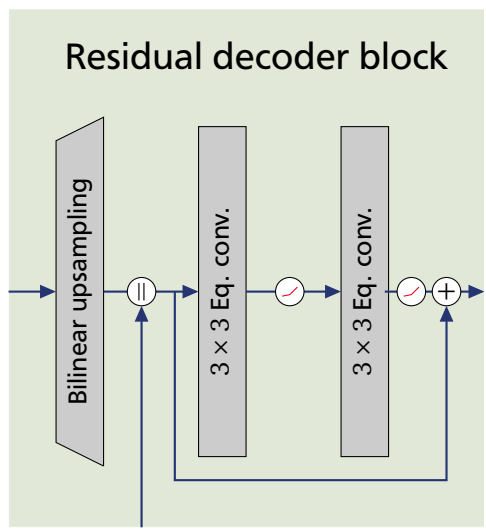

**Figure 5.6:** Residual U-Net decoder block with a bilinear upsampling layer, two equalized 3 × 3 convolutions, two activation functions, a residual mapping.

### 5.1.4 Encoder Network for Guidance

The standard StyleGAN 2 [13] architecture can not be conditionalized directly and even the intermediate latent space $\mathbb{W}$ can not be interpreted. To efficiently produce an image similar to a guidance image $I_g$ an encoder network is trained to map the guidance image into the intermediate latent space $\mathbb{W}$. The generator then can produce an image with the produced intermediate latent vector $w \in \mathbb{W}$. The produced image $I_{\text{fake}} \sim I_g$ should be approximately similar to the guidance image. This means the image, produced by the generator, should include the same key features, like the number of cells or the z position, as the guidance image. With this approach the Cell-GAN generator can be conditionalized in an very efficient way compared to other methods [98, 108].

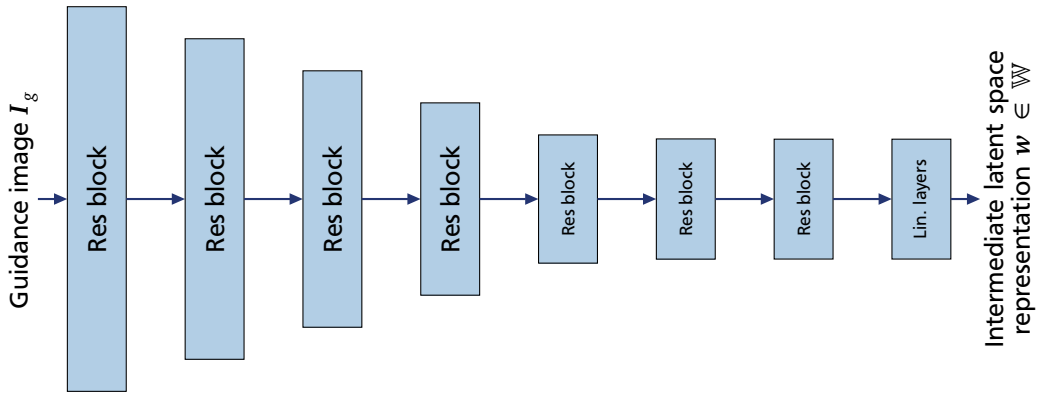

**Figure 5.7:** Cell-GAN encoder architecture.

The utilized encoder follows a ResNet-like [43] architecture and is composed of seven residual blocks and a linear layer to transform the guidance image to the intermediate latent space $\mathbb{W}$.

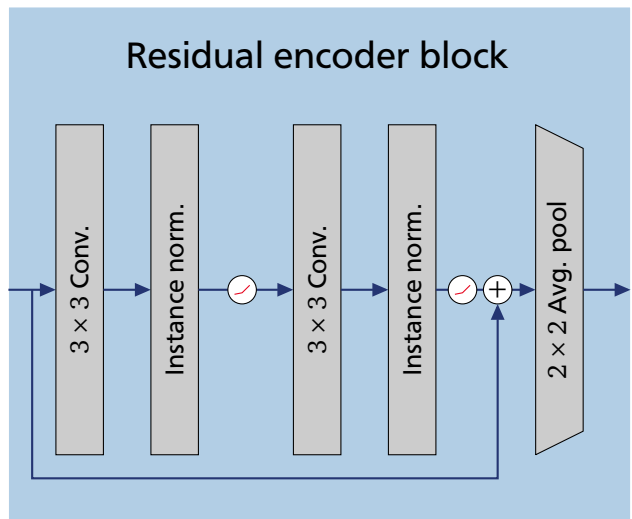

**Figure 5.8:** Residual convolution of the Cell-GAN encoder with two $3 \times 3$ convolutional layers, two instance normalization layer [54], two activation functions, an residual mapping, and a final $2 \times 2$ average pooling layer to downscale the spatial dimensions.

The seven blocks include 32, 64, 128, 256, 256, 256, and 128 convolutional filters, restively. A detailed illustration of the used ResNet-like block can be seen in figure 5.8.

### 5.1.5 Training Approach

The Cell-GAN model is trained in a unsupervised two-step training process. In the first training process, the mapping network $f$, the generator $\mathcal{G}$, and the discriminator $\mathcal{D}$ is trained in an adversarial setting. In the second training step the guidance encoder $\mathcal{E}$ to embed an image into the intermediated latent space $\mathbb{W}$.
The first training process optimizes the generator, the mapping network, and the discriminator network in an adversarial min-max game as introduced in 3. However, not the original GAN loss (eq. 2.6) is utilized but the non-saturating GAN loss [8] (NS GAN loss), which is defined as:

$$L_{\mathcal{D}} = -\mathbb{E}\left[\log \mathcal{D}\left(I_{\text{real}}\right)\right] - \mathbb{E}\left[\log(1 - \mathcal{D}\left(\mathcal{G}\left(f\left(z\right)\right)\right))\right], \tag{5.1}$$
$$L_{\mathcal{G}} = -\mathbb{E}\left[\log \mathcal{D}\left(\mathcal{G}\left(f\left(z\right)\right)\right)\right]. \tag{5.2}$$

Where $\mathcal{D}$ aims to minimize $L_{\mathcal{D}}$ and $\mathcal{G}$ on the other hand aims to minimize $L_{\mathcal{G}}$. This means the generator $\mathcal{G}$ and the mapping network $f$ learns to map a latent variable $z \sim \mathcal{N}(0,1)$ to a real-looking image which is in $p_{\text{data}}$, while the discriminator $\mathcal{D}$ aims to distinguish between a fake sample, generated by $\mathcal{G}$, and real samples form $p_{\text{data}}$. However, since a U-Net discriminator [107] is utilized, which produces a scalar real/fake prediction and a pixel-wise real/fake prediction the losses changes to:

$$\begin{aligned} L_{\mathcal{D}} = &-\mathbb{E}\left[\log \mathcal{D}_{\text{scalar}}\left(I_{\text{real}}\right)\right] - \mathbb{E}\left[\log \mathcal{D}_{\text{pixel}}\left(I_{\text{real}}\right)\right] \\ &- \mathbb{E}\left[\log(1 - \mathcal{D}_{\text{scalar}}\left(\mathcal{G}\left(f\left(z\right)\right)\right))\right] - \mathbb{E}\left[\log\left(1 - \mathcal{D}_{\text{pixel}}\left(\mathcal{G}\left(f\left(z\right)\right)\right)\right)\right], \end{aligned} \tag{5.3}$$

$$L_{\mathcal{G}} = -\mathbb{E}\left[\log \mathcal{D}_{\text{scalar}}\left(\mathcal{G}\left(f\left(z\right)\right)\right)\right] - \mathbb{E}\left[\log \mathcal{D}_{\text{pixel}}\left(\mathcal{G}\left(f\left(z\right)\right)\right)\right]. \tag{5.4}$$

Where $\mathcal{D}_{\text{scalar}}$ indicates the scaler prediction of the U-Net discriminator and $\mathcal{D}_{\text{pixel}}$ the pixel-wise prediction.
Additionally, lazy $R_1$ regularization of the discriminator is utilized. This means the $R_1$ regularization term

$$L_{\mathcal{D}}^{reg} = \frac{\gamma_{R_1}}{2}\mathbb{E}\left[\|\boldsymbol{\nabla}\mathcal{D}\left(I_{\text{real}}\right)\|_2^2\right] \tag{5.5}$$

is only computed every 16 training iterations. The weights factor $\gamma_{R_1}$ was set to 160. For more information regarding the $R_1$ regularizer see section 3.3.2. In addition to the discriminator regularization also a regularization of the generator is utilized. Therefore, a path length regularization $P_{lr}$ term

$$L_{\mathcal{G}}^{reg} = \gamma_{R_{pl}}\mathbb{E}_{\boldsymbol{w},\boldsymbol{z}\sim\mathcal{N}(0,I)}\left[\left\|\boldsymbol{J}_{\boldsymbol{w}}^{\mathsf{T}}\boldsymbol{y}\right\|_2 - a\right]^2 \tag{5.6}$$

with the weights parameter $\gamma_{R_{pl}}$ set to 8, is utilized. For a detailed description on path length regularization see section 3.1.4. This training process follows some key points of the StyleGAN 2 [13] training.

For an additional pixel-wise consistency regularization, CutMix consistency regularization [107] is utilized. In CutMix consistency regularization first a mini-batch of real images $I_{\text{real}}$ and a mini-batch of fake images $I_{\text{fake}}$ is predicted by the U-Net discriminator to achieve the pixel-wise real/fake predictions $\hat{P}_{\text{real}}$ and $\hat{P}_{\text{fake}}$. Then a map is produced with a randomly placed patch which indicates the combination of the real and fake image mini-batch. Then an combined real and fake input is constructed and predicted by the discriminator. Finally, a L2 loss is applied to the prediction of the mixed input image and to the mixed predictions $\hat{P}_{\text{real}}$ and $\hat{P}_{\text{fake}}$. This CutMix consistency regularization loss can be described as:

$$L_{\mathcal{D}}^{mix} = ||\text{Mix}(\mathcal{D}(I_{\text{real}}), \mathcal{D}(I_{\text{fake}})) - \mathcal{D}(\text{Mix}(I_{\text{real}}, I_{\text{fake}}))||_2. \quad (5.7)$$

This regularization term is also computed in a lazy way, every 16 training iterations.

In the second stage of the training, the guidance encoder $\mathcal{E}$ is trained in an unsupervised autoencoder fashion to map a given guidance image into the intermediated latent space $\mathbb{W}$. First, a mini-batch of real images is fed into the guidance encoder, which then produces the intermediated latent vector $w$. The vector $w$ is then used by the fixed generator network $\mathcal{G}$ to produce images similar to the input images. To train the guidance encoder an L1 loss, averaged over all pixels, is then utilized between the predicted images $\hat{I}$ and the real input images is utilized.

$$L_{\mathcal{E}} = \frac{1}{hw}||\hat{I} - I_{\text{real}}||_1 \quad (5.8)$$

To regularize the guidance encoder, pairs of fake images $I_{\text{fake}}$ and there corresponding intermediated latent vectors $w_{\text{fake}}$ are produced. The guidance encoder is then trained by a L1 loss, averaged over the latent vector dimension $l_{\text{dim}}$

$$L_{\mathcal{E}}^{reg} = \frac{1}{l_{\text{dim}}}||\mathcal{E}(I_{\text{fake}}) - w_{\text{fake}}||_1 \quad (5.9)$$

to regress the latent vector from the fake image. This regularization is performed in each training step.

## 5.2 Experiments

This section demonstrates the performance of the Cell-GAN model in unconditionalized and guided image generation. For training and evaluation, the yeast cell time-series dataset (generation) with randomness is used. For a more detailed description regarding this dataset see chapter 4.

### 5.2.1 Technical Details

The whole Cell-GAN architecture, as well as the necessary dataset and data-loader classes, is implemented in the PyTorch [109] framework. Additionally, the custom CUDA [110] upsampling/-downsampling layer of the original StyleGAN 2 [13] implementation is utilized. Furthermore, the official CUDA-based padé activation unit [106] implementation of the authors is employed. To train the Cell-GAN model the AdamP [64] (algorithm 2) with a learning rate of $10^{-3}$ is utilized. The learning rate for the mapping network was set to $10^{-5}$. The first and second-order momentum moving average factors were 0.1 and 0.99 respectively. Additionally, a weight decay of $10^{-4}$ is employed. The additional parameter of the AdamP [64] optimizer are set to the default value of the original implementation of the authors.

At training time a batch size of 32 was utilized. Furthermore, 30 epochs of adversarial training (fist training step) and 30 epochs of encoder training (second training step) are performed. The whole training took about 8 to 10 hours on a first-generation DGX station with four Nvidia Tesla V100 (16GB).

### 5.2.2 Results

The Cell-GAN architecture is able to generate real looking brightfield microscopy images. This can be observed from figure 5.9, which shows hand-picked unconditional samples from the Cell-GAN generator. However, in case if multiple cells are generated the quality sometimes falls short compared to real images or samples with fewer cells.

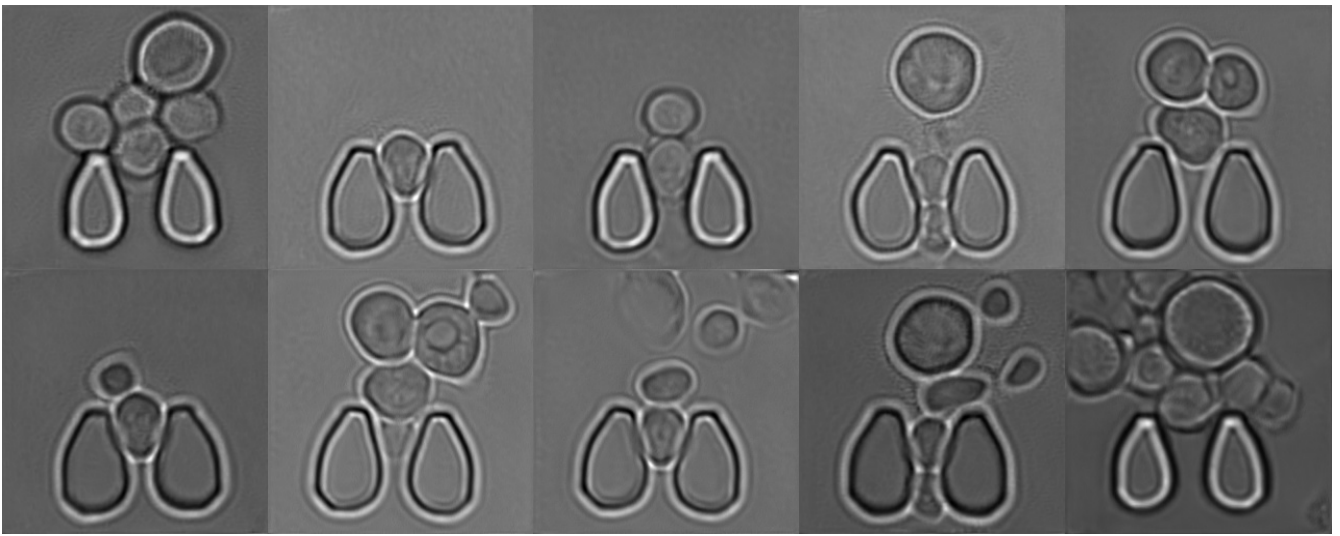

**Figure 5.9:** Random unconditionalized samples generated by the Cell-GAN model.

The quantitative results, shown in table 5.1, confirm the strong performance observed in figure 5.9. The Cell-GAN is able to generate samples with an inspection score of 1.73, which nearly matches the dataset's inception score of 1.799. This implies a strong performance.

The Fréchet Inception Distance shows future the strength of the Cell-GAN model for unconditionalized image generation. But also shows the weeker performance for conditionalized image generation. The drop from an FID of 29.588 for unconditionalized generation to 60.067 for conditionalized generation not only implies a weaker performance in conditionalized generation but it is also an indicator that the generator has learned to generate new microscopy images instead of remembering the images from the training set. This claim can be further reinforced by the qualitative reconstruction results shown in figures 5.10, 5.11 and 5.12. Since the Cell-GAN is able to generate conditionalized images, based on fake unconditionalized samples, with high quality and preserved key features. This is qualitatively shown in figure 5.10.

**Table 5.1:** Quantitative results of the Cell-GAN architecture on the trapped yeast cell time-series dataset. FID (lower is better) represents the Fréchet Inception Distance (sec. 2.4.3.2) and IS (higher is better) the Inception Score (sec. 2.4.3.1). The Inception Score of the dataset is 1.799. The evaluation for the conditionalized setting was done with guidance images from the test set, since the test set provides not enough images.

| Setting | FID ↓ | IS ↑ | Parameters $\times 10^6$ |
| --- | --- | --- | --- |
| Unconditionalized | 29.588 | 1.730 | 24 ($\mathcal{G}$) & 2 ($f$) |
| Conditionalized | 60.067 | 1.557 | 24 ($\mathcal{G}$) & 17 ($\mathcal{E}$) |

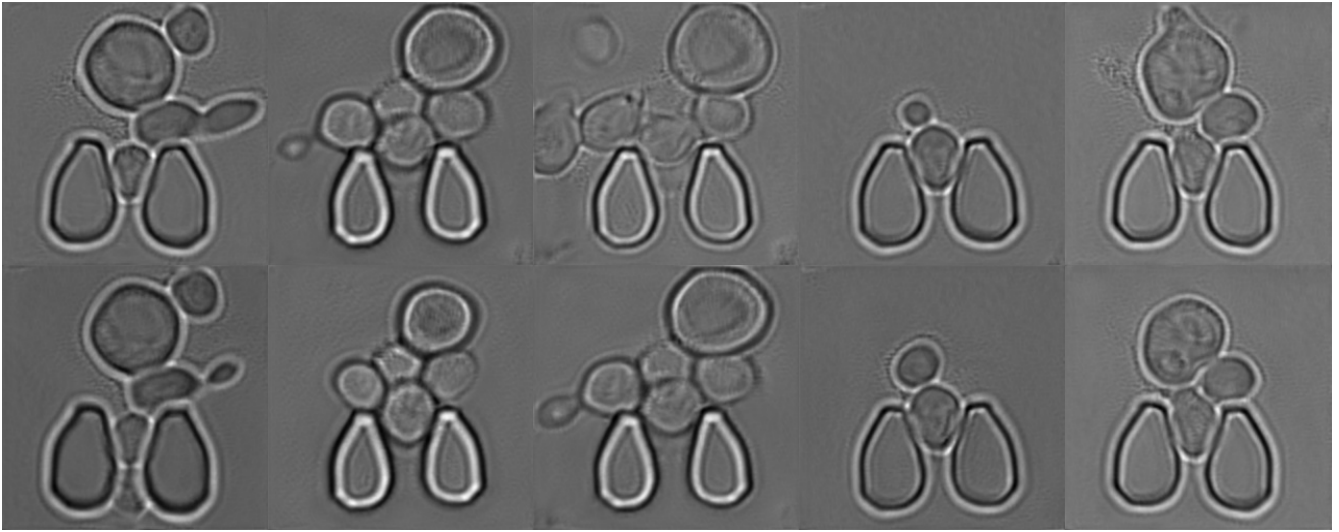

**Figure 5.10:** Conditionalized samples generated by the Cell-GAN model for fake guidance images. Fake guidance samples generated from the Cell-GAN generator on the top and the corresponding reconstructed image by the encoder and generator on the bottom.

However, the quality of the conditionalized generated images, with guidance images from the training dataset, is weaker. Since the Cell-GAN is only partly able to preserve the key features of the guidance image. But the produced conditionalized samples still have good visual qualities. Which implies the latent space of the Cell-GAN seems to be smooth.

The qualitative results for conditionalized image generation with real image from the never-seen test dataset, shown in figure 1 lead to similar observations in terms of the generation quality. The trained Cell-GAN can generate images with roughly the same key features as the guidance image if a simple guidance image is utilized. But is the image to complicated the conditionalized generation leads to images with less preserved key features.

The adversarial training process on small datasets remains difficult as can be seen from the loss curve shown in figure 5.13. Often the training does not converge to a good equilibrium resulting in bad image samples. And sometimes the training seems to stick in a bad equilibrium as can be observed in the loss curve between training step 3200 and 5500. But surprisingly the generator and discriminator corresponding to the loss curve does recover and converged to a good performance.

Additional plots, including conditionalized and unconditionalized image samples, for different training steps, can be seen in section 9.4 of the appendix .

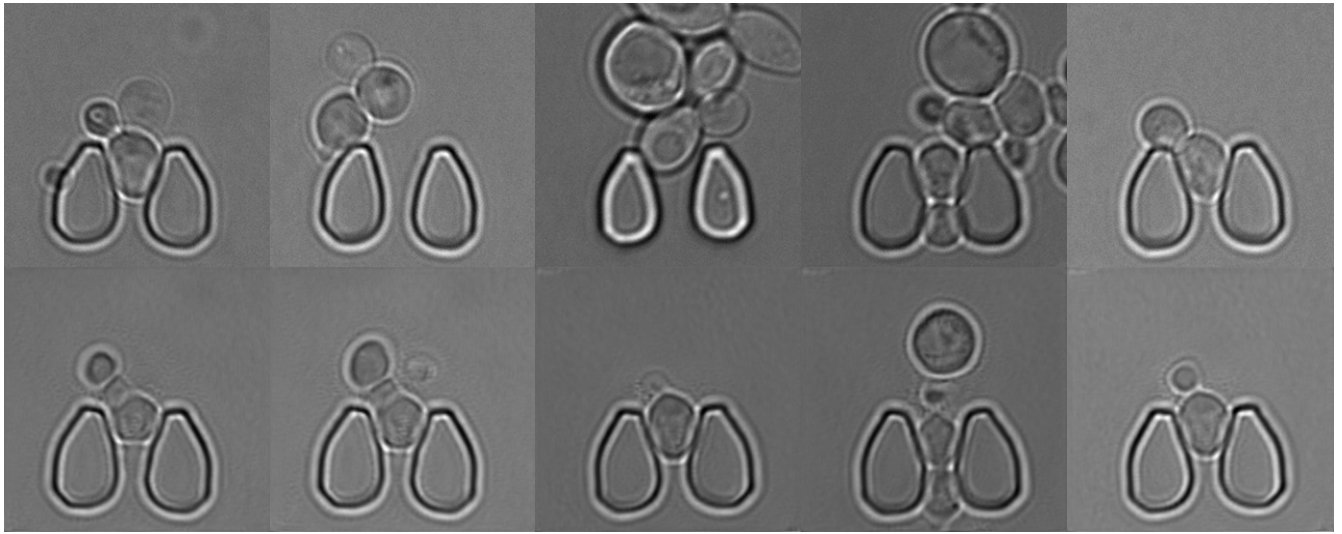

**Figure 5.11:** Conditionalized samples generated by the Cell-GAN model for training guidance images. Real guidance images from the training set on the top and the corresponding reconstructed image by the encoder and generator on the bottom.

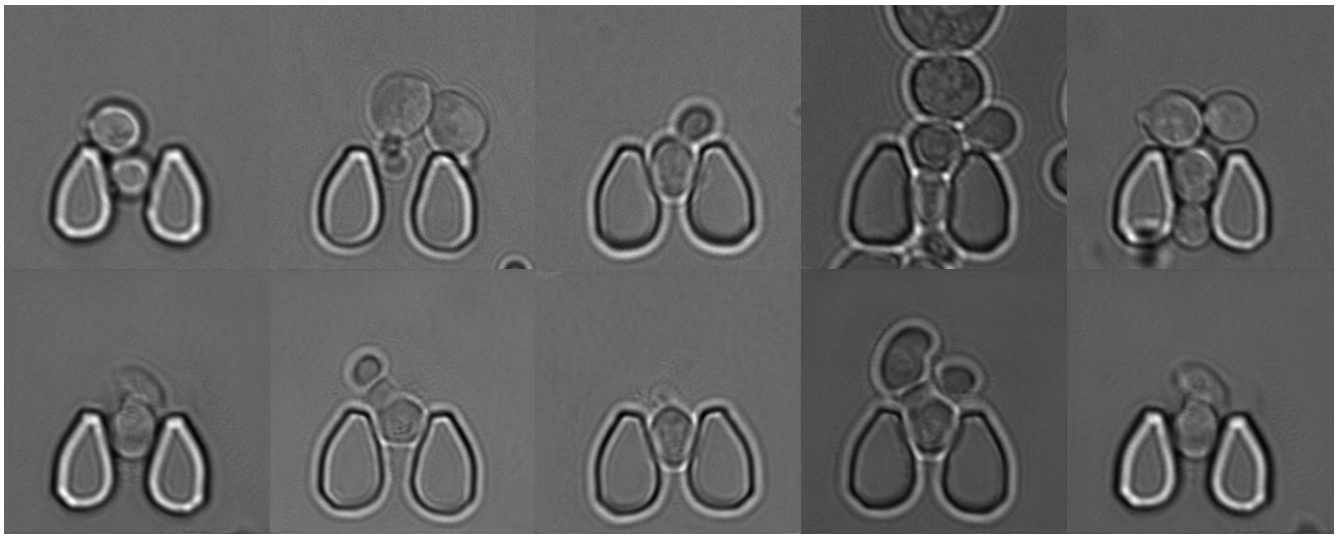

**Figure 5.12:** Conditionalized samples generated by the Cell-GAN model for test guidance images. Real guidance images from the test set on the top and the corresponding reconstructed image by the encoder and generator on the bottom.

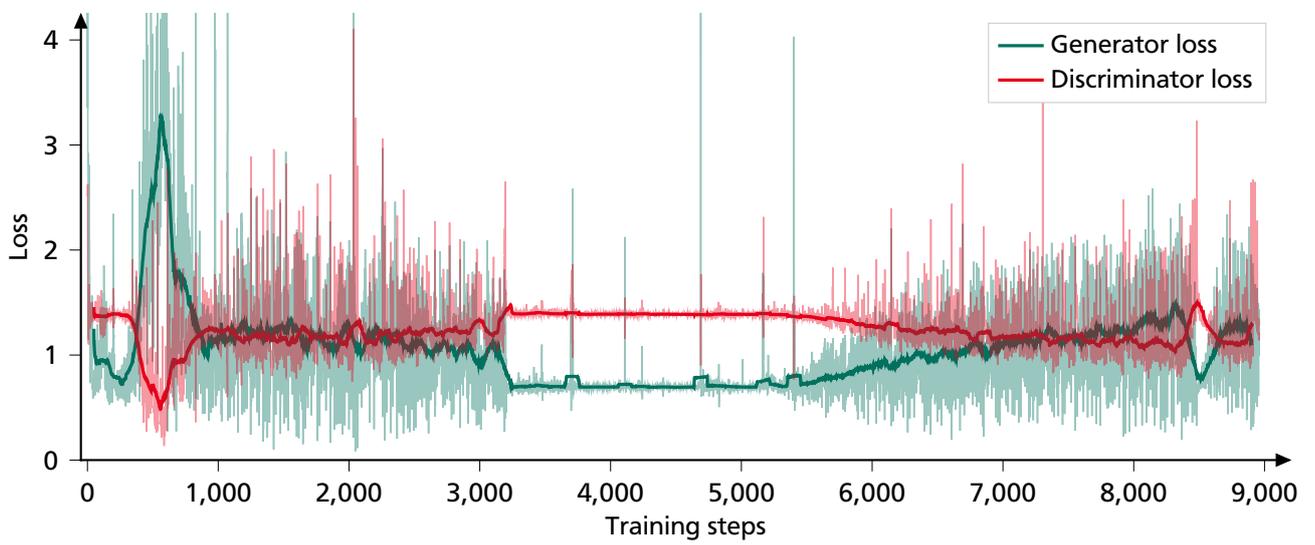

**Figure 5.13:** Cell-Gan loss curve with the running averages of the generator loss in green ▬ and the discriminator loss in red ▬. Running averages computed with a window size of 100.

# 6 SDC-Net++: Multiple Future Frame Prediction of Microscopy Image Sequences

This chapter proposes the novel SDC-Net++ architecture for unsupervised future frame prediction of brightfield microscopy image sequences. The SDC-Net++ model is composed of a pre-trained PWC-Net [16] for optical flow estimation, a pre-trained U-Net for semantic segmentation [6], the 3d U-Net generator $\mathcal{G}$ to be trained, and a spatially-displaced convolution [22] to produce the final future. The whole SDC-Net++ model is trained in an unsupervised setting and inspired by the original SDC-Net [22] model. However, the SDC-Net++ proposed and advanced adversarial multiple step prediction training as well as an advanced parameter prediction architecture.

## 6.1 Method

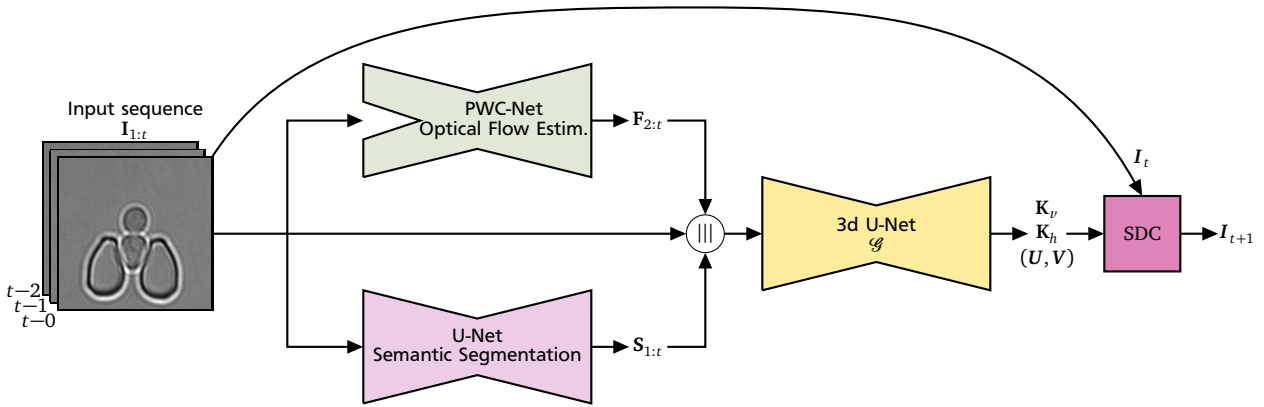

**Figure 6.1:** Schematic of the SDC-Net++ architecture. The pre-trained PWC-Net [16] take the past frames as an input and estimates the privious backward optical flows. The pre-trained U-Net also takes the past frames as an input and predicts semantic segmentation maps. The past frames, the backward optical flows, and the semantic segmentation maps gets concatenated form the input of the 3d U-Net generator network, which predicts the parameters of the spatially-displaced convolution [22], to finally produce the future frame.

The SDC-Net++ follows the idea of previous work in future frame prediction [22, 20, 95] (see 3.2.2) to formulate the future frame prediction problem as a transformation learning task. This utilized learnable transformation can be described as

$$I_{t+1} = \text{SDC}(\mathcal{G}(\mathbf{I}_{1:t}, \mathbf{F}_{2:t}, \mathbf{S}_{1:t}), I_t), \tag{6.1}$$

where the transformation is utilized as a spatially-displaced convolution SDC with the predicted transformation parameters by the 3d U-Net generator model $\mathcal{G}$ and the current bright-field microscopy image $I_t$. The generator network predicts the transformation parameter based on the past frames $I_{1:t}$, the backward optical flow maps $F_{2:t}$, predicted by a pre-trained PWC-Net [16], and the semantic segmentation maps $S_{1:t}$, produced by a pre-trained U-Net [6]. Both the PWC-Net and the U-Net are fixed during training. Providing the generator with additional information like the optical flow or the semantic segmentation showed great improvements, however, makes the generator reliant on the accuracy of the additional information [22]. For more information regarding semantic segmentation see section 2.6 and for a detailed description on optical flow estimation and the PWC-Net [16] see sections 3.3.3 and 2.7

One important detail to notice is that the flow maps $F_{2:t}$ include the backward optical flow, because the spatially-displaced convolution utilizes backward resampling. This means the sampling location in $I_t$ is predicted for each sampling location in $I_{t+1}$ [22]. Multiple future frames can be predicted with the SDC-Net++ by feeding the previously predicted frame and the last $n-1$ input frames back to the model.

### 6.1.1 3d U-Net Generator Network

The 3d U-Net generator network $\mathcal{G}$, shown in figure 6.2, follows the architecture of the original SDC-Net generator. A 3d CNN is utilized to convolve over the spatial dimensions as well as the time dimension to learn to interpolate the motions and occlusions of the objects in the past frames. As mentioned earlier the input to the generator network are the past frames, the backward optical flow maps (2 feature dimensions), and the semantic segmentation maps (4 feature dimensions). All feature maps are concatenated at the feature dimension resulting in 7 feature dimensions.

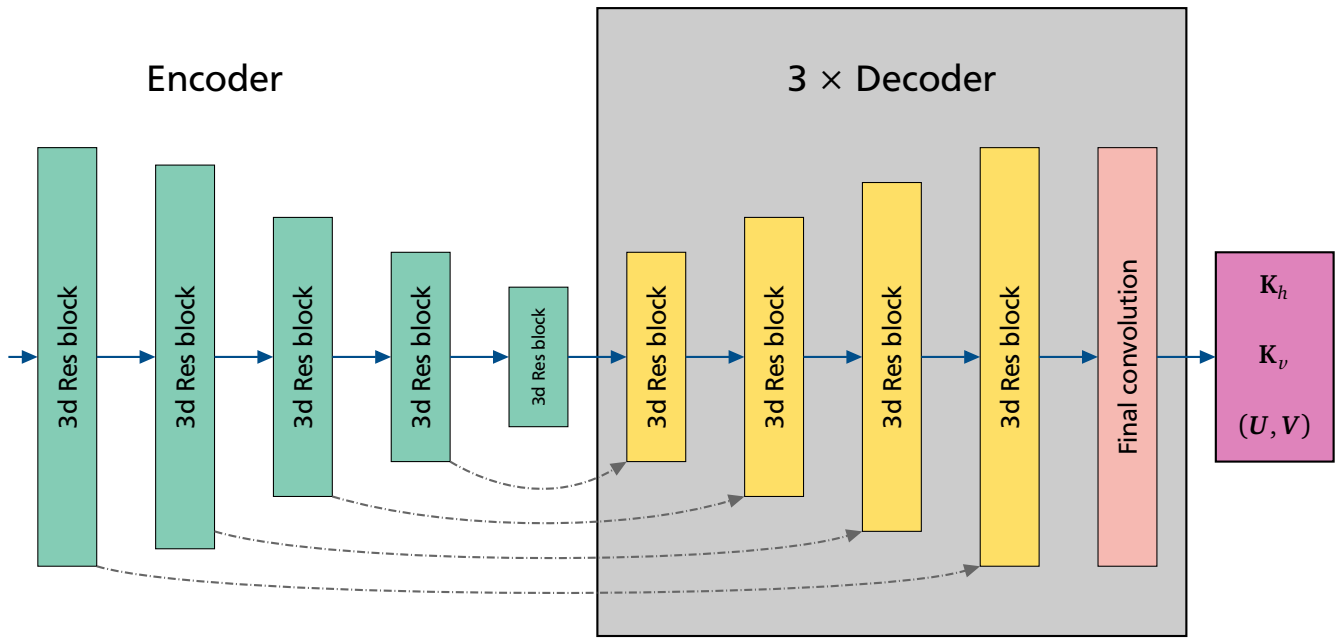

**Figure 6.2:** 3d U-Net generator architecture with five 3d residual encoder blocks ■ and three decoder heads ■ with each four 3d residual decoder blocks ■ and a final convolution ■. One decoder predicts the flow vectors $(U, V)$ for each pixel and the other two decoders predict the 1d kernels $K_h$ and $K_v$. Additionally skip connections bridge from the encoder to the decoder. However, the not downsampled maps are passed by the skip connections to the decoder.

The 3d U-Net generator network utilizes a ResNet-like [43] encoder and also three ResNet-Like decoders to predict the flow vectors ($U, V$) and the 1d kernels $\mathbf{K}_h$ and $\mathbf{K}_v$. Recent work has shown that multiple encoder heads result in a better overall performance throughout multiple like for example panoptic [111] segmentation and also future frame prediction [22, 112].

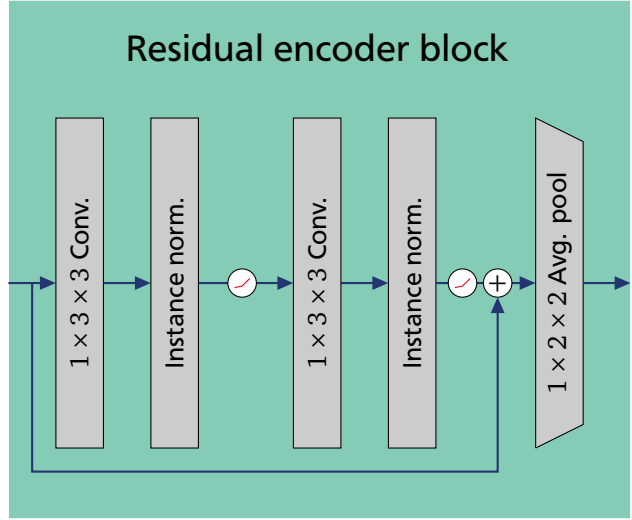

**Figure 6.3:** 3d residual encoder block composed of two $1\times3\times3$ convolutions, two instance normalization [54] layers, two padé activation units [106], a residual mapping, and an $1 \times 2 \times 2$ average pooling layer which downsamples the spatial dimensions by a factor to two.

The encoder block, shown in figure 6.3, and the decoder block, shown in figure 6.4, are utilizing two 3d $1 \times 3 \times 3$ convolutions to convolve over the time dimension as well as over the spatial dimensions. However, to avoid a too large model, which also results in a large increase in the computational cost, only a 2d convolutional kernel (first dim. one) is utilized.

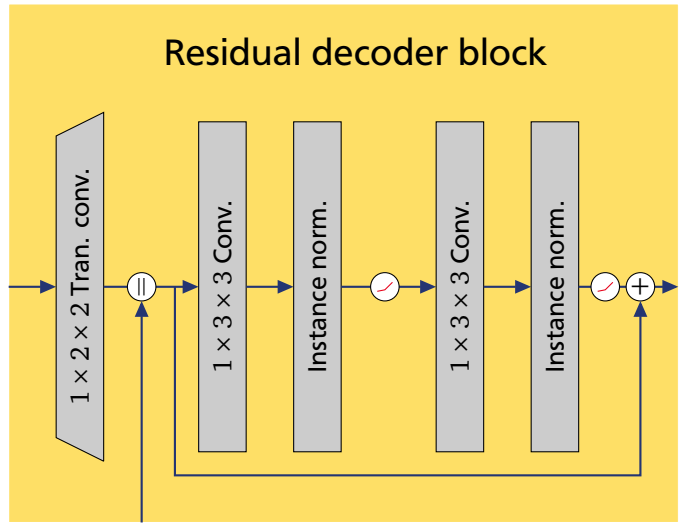

**Figure 6.4:** 3d residual decoder block including a $1 \times 2 \times 2$ transposed convolution with a stride of two to upsample the spatial dimensions of the incoming feature tensor. Additionally, two $1 \times 3 \times 3$ convolutions, two padé activation units [106], and two instance normalization layers [54] are included.

The decoder block, furthermore, performs a 3d transposed convolution to upsample the spatial dimensions of the incoming feature tensor, by a factor of two. However, in the last two decoder blocks a trilinear upsampling layer followed by a 1 × 1 × 1 convolution is used instead of the transposed convolution. This is done to prevent the predictions of suffering from checkerboard artifacts [59].

The final convolution is employed to reduce the feature dimensions to the desired output features. In case of the flow/motion encoder, two feature dimensions are predicted for each pixel, corresponding the $(U, V)$ motion vectors. The two kernel encoders on the other hand predict $n$ feature dimensions for each pixel. Where $n \times n$ is the size of the kernel predicted for each pixel.

### 6.1.2 Training Approach

The SDC-Net++ is trained in an unsupervised setting. The overall training process is composed of four stages. First, a flow training stage, second a kernel initialization training stage, third a fine-tune training stage, and a final partly adversarial multi-frame prediction training.
In the flow training stage only the generator encoder and the flow decoder head is trained. The flow training is done by predicting the parameters of the SDC with standard kernels (overall zero expect for the middle element). In this case the SDC is equal to standard bilinear backward resampling operation (see 3.2.2). In this training stage large displacements should be learned by the predicted pixel-wise motion/flow vectors $(U, V)$. The generator is trained to minimize the following loss

$$L_{\text{flow}} = L_g \left[ \bm{I}_{t+1} - \hat{\bm{I}}_{t+1} \right], \tag{6.2}$$

where $L_g$ is the general and adaptive robust loss function [32] introduced in section 3.3.5, $\bm{I}$ is the real target image and $\hat{\bm{I}}$ the predicted image.
In the kernel initialization training, followed form the original SDC-Net [22], the following loss is minimized by the kernel decoder heads

$$L_{\text{kernel}} = \frac{1}{hw} \sum_{x=1}^{h} \sum_{y=1}^{h} \left( \left\| K_u(x,y) - 1^{<\frac{N}{2}>} \right\|_2^2 + \left\| K_v(x,y) - 1^{<\frac{N}{2}>} \right\|_2^2 \right) \tag{6.3}$$

all other parameters are fixed during this training stage. To be precise, the kernel loss $L_{\text{kernel}}$ computes the mean squared error between each kernel, predicted for each pixel in the spatial dimensions $h$ and $w$, and a middle-one-hot vector $1^{<\frac{N}{2}>}$. This training stage has the purpose to initialize the kernels to have a smooth transition between the flow training and the fine time training, since the flow decoder head is trained with the standard bilinear backward resampling operation.
In the fine-tune training stage the whole generator network $\mathcal{G}$ is optimized. The future frame is predicted by the SDC and the following loss is applied

$$L_{\text{fine-tune}} = L_g \left[ \bm{I}_{t+1} - \hat{\bm{I}}_{t+1} \right]. \tag{6.4}$$

Where $L_g$ is again the general and adaptive robust loss function [32] (GARLoss). It is important to notice that the same parameters for the GARLoss are utilized as achieved during the flow training stage.

The final multi-frame prediction training the network predicts $n$ multiple future frames in an autoregressive way. The loss function utilized in this training steps can be decried as

$$L_{\text{multi}} = \frac{\lambda_g}{n} \sum_{i=1}^{n} L_g \left[ I_{t+n} - \hat{I}_{t+n} \right] - \lambda_{adv} \mathbb{E} \left[ \log \mathcal{D} \left( \hat{\mathbf{I}}_{t-(m-1):t+n} \right) \right]. \quad (6.5)$$

Where $L_g$ is again the GARLoss with the initial parameters of the previous training stage and the scalar weights factor $\lambda_g$ applied to each $n$ predicted frame and the corresponding real frame. The second term is the non-saturating GAN generator loss with the scalar weights factor $\lambda_{adv}$, where $\mathcal{D}$ is an discriminator network which takes the past $m$ initial input frames as well as the $n$ predicted frames as an input. The discriminator itself is trained the the corresponding GAN discriminator loss $-\mathbb{E}\left[\log \mathcal{D}\left(\mathbf{I}_{t-(m-1):t+n}\right)\right] - \mathbb{E}\left[\log\left(1 - \mathcal{D}\left(\hat{\mathbf{I}}_{t-(m-1):t+n}\right)\right)\right]$. This training process is lousily inspired by the DeepFovea [99] (section 3.3.4) training.

At the flow training stage, the fine-tune training, and the multi-frame prediction an optional weights map $W_{sv} \in \mathbb{R}h \times w$, in the supervised loss is employed. This weights map is applied to the difference between the predicted image $\hat{I}$ and the true image $I$ before computing the GARLoss. To compute the weights map $W_{sv}$, the semantic segmentation map of the last input frame and the labels, produced by the pre-trained U-Net is used. To be precise, a binary map indicating the occurrence of a cell is extracted from the semantic segmentation for both of the images. Then a map that models the areas where each cell grows is computed by the difference between the binary map of the image to be predicted and the last input image. These areas are further enlarged by simple dilation. If now a pixel belongs to the computed area the $W(i,j)$ is set to 1.5 if not $W(i,j)$ is set to 1.0.

### 6.1.3 Sequence Discriminator Network

Recent work has shown that adversarial training can improve the prediction of video frames [99]. Inspired by this results a sequence discriminator is utilized to train the SDC-Net++ generator at the multi-frame prediction training stage.

The sequence discriminator $\mathcal{D}$, shown in figure 1, takes the whole brightfield microscopy image sequence including the initial real images as well as the predicted images as an input. In the first step, the input sequence is processed by five 3d residual convolution blocks. Then the resulting features are classified as real or fake by a linear layer.

To stabilize the adversarial training, spectral normalization [113] is applied to all weights included in the discriminator network. Each residual block follows the architecture shown in figure 6.3. However, no normalization layers are utilized.

## 6.2 Experiments

This section introduces the empirical results of the SDC-Net++ on the yeast time-series dataset (simulation). Furthermore, this section also describes the technical details regarding the implementation and introduces the an evaluation approach of the SDC-Net++.

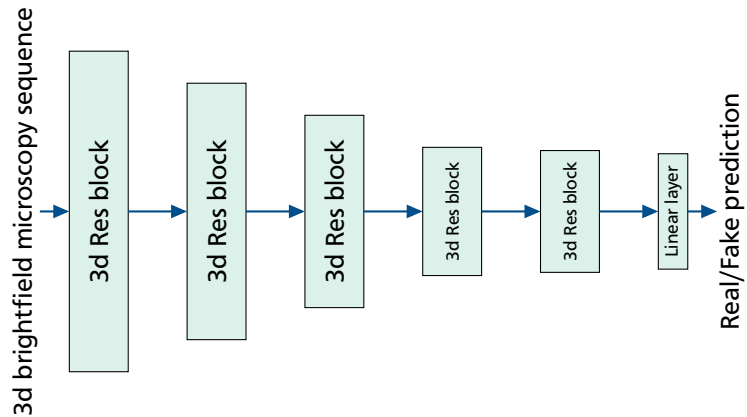

**Figure 6.5:** 3d discriminator architecture composed of five 3d residual convolution blocks, and a final linear layer.

### 6.2.1 Technical Details

As well as the Cell-GAN, the SDC-Net++ is also implemented in PyTorch [109]. Additional utilities, like dataset classes, are also implemented in native PyTorch. The spatially-displaced convolution CUDA/PyTorch [110] implementation is taken from the official SDC-Net [22] repository. For the bilinear backward resampling operation, used in the first training stage, the CUDA/PyTorch implementation included in the NVIDIA FlowNet 2 repository [114] is employed. The official GARLoss PyTorch implementation of the author is future utilized.

For predicting the backward flow maps, inputed to the generator network $\mathscr{G}$, the official pre-trained PWC-Net [16] is utilized. The pre-training was done on multiple datasets including the FlyingChairs dataset [14], the FlyingThings3 [115], the Sintel dataset [82], and the KITTI dataset [116]. For a detailed description on the PWC-Net training process see [16]. Some backward optical flow predictions for the pre-trained PWC-Net can be seen in figure 6.6. In can be observed that the PWC-Net is able to produce accurate flow maps for the yeast cell time-series dataset in the most cases. But it can also be observed that, in some cases, the PWC-Net fails to produce a precise flow prediction.

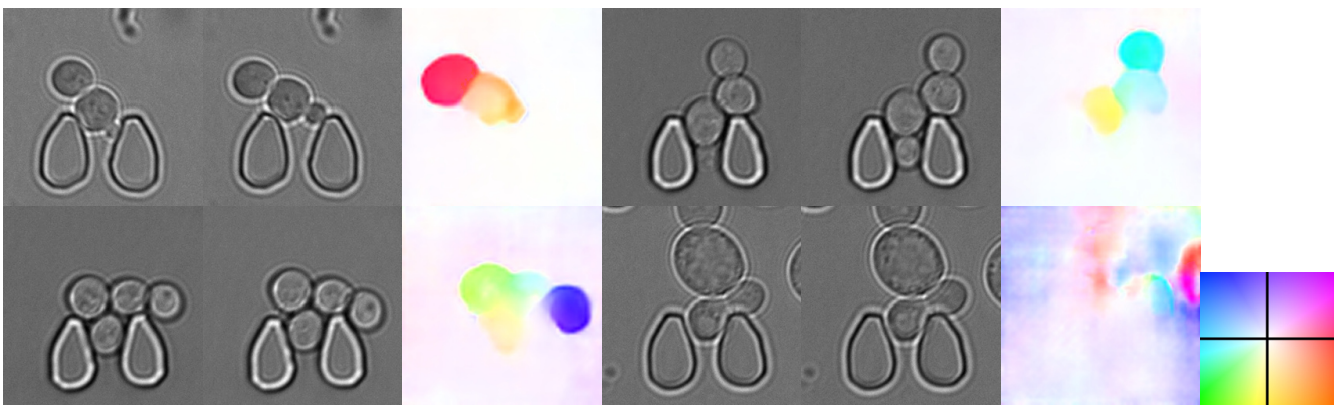

**Figure 6.6:** Backward optical flow prediction by the pre-trained PWC-Net [16] for the trapped yeast cell time-series dataset. First frame on the left, second frame in the middle, and the predicted backward optical flow on the right. Optical flow encoded in the color wheel proposed (top) in [83].

To produce the semantic segmentation input maps for the generator network $\mathcal{G}$ a pre-trained U-Net [6], trained supervised on a related trapped yeast cell dataset is employed. This network was converted from the original Tensorflow [117] implementation to PyTorch [109].

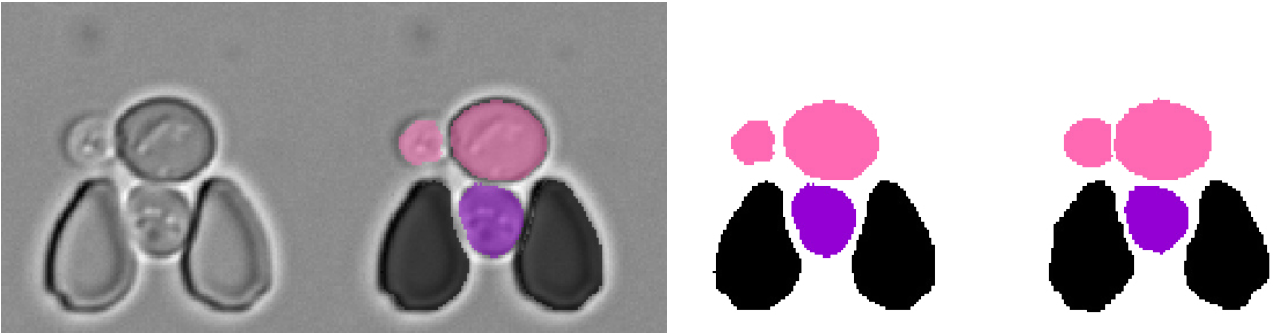

**Figure 6.7:** Semantic segmentation prediction from the pre-trained U-Net [6]. Input brightfield microscopy image on the left, alongside the prediction overlay and the prediction map. Segmentation label on the right. Background in white/transparent, traps in black ■, cell of interest in violet ■, and auxiliary cells in light violet ■. [6]

To optimize the SDC-Net++ while training the AdamP [64] (algorithm 2) optimizer is utilized. The learning rate is set to $3 \cdot 10^{-4}$ for the prediction heads. The generator encoder and the discriminator is trained with a slightly lower learning rate of $3 \cdot 10^{-4}$. Further weight decay of $10^{-2}$ is utilized. The additional hyperparameters of the AdamP optimizer are set to the corresponding default values. Furthermore, a kernel size $7 \times 7$ is utilized in the SDC and predicted by the kernel encoders. The number of input frames to the SDC-Net++ is set to three.

The generator $\mathcal{G}$ is further utilized with 16, 32, 64, 128, and 256 convolutional filters in the encoder blocks. The decoder blocks, for each of the tree decoders, employ 128, 64, 32, and 16 convolutional filter.

The wights parameters of the multi-frame prediction loss $L_{\text{multi}}$ are set to $\lambda_g = 1$ and $\lambda_{adv} = 0.01$.

The training is composed of 5 epochs of flow training, 1 epoch of kernel initialization training, 20 epochs fine-tune training and 5 epochs of multi-frame prediction training. While training a batch size of 64 is utilized except for the multi-frame prediction training, where a batch size of 1 is used. This is because future frames for sequences with different lengths are predicted in an autoregressive way. Furthermore, the autoregressive prediction process is additionally very GPU memory intensive, since the activation of all predicted frames have to be stored until the adversarial generator training steps with the whole sequence is performed.

Training the SDC-Net++ on the yeast cell time-series dataset without randomness took about 2 hours on a first-generation DGX station with four Nvidia Tesla V100 (16GB).

### 6.2.2 Evaluation Approach

The quality of the frames predicted by the SDC-Net++ is evaluated on the L1 metric, the Mean-Squared-Error (MSE/L2) [118], the Peak-Signal-To-Noise (PSNR), and the Structural-Similarity-Image-Metric (SSIM). The L1 metric is defined as

$$\text{L1} = \frac{1}{hw}\left|\left|\mathbf{I} - \hat{\mathbf{I}}\right|\right|_1, \quad \mathbf{I}, \hat{\mathbf{I}} \in \mathbb{R}^{h \times w} \tag{6.6}$$

where $I$ is the target image and $\hat{I}$ the predicted image. The Mean-Squared-Error (MSE/L2) is further defined as

$$\text{MSE} = \frac{1}{hw}||I - \hat{I}||_2^2, \quad I, \hat{I} \in \mathbb{R}^{h \times w}. \tag{6.7}$$

The Peak-Signal-To-Noise (PSNR) and the Structural-Similarity-Image-Metric (SSIM) for two image are defined as

$$\text{PSNR} = 10 \log_{10}\left(\frac{\max\{\hat{I}\}^2}{\text{MSE}(\hat{I}, I)}\right) \tag{6.8}$$

$$\text{SSIM} = \frac{4 \mathbb{E}[\hat{I}] \mathbb{E}[I] \text{Cov}[\hat{I}, I]}{\left(\mathbb{E}[\hat{I}]^2 + \mathbb{E}[I]^2\right)\left(\text{Var}[\hat{I}] + \text{Var}[I]\right)}. \tag{6.9}$$

where a higher SSIM $\in [-1, 1]$ indicates a better similarity between the true image $I$ and the predicted image $\hat{I}$. A higher PSNR also indicates a better prediction. One important detail to notice is that the SSIM is evaluated on the images normalized to a range of zero to one which is common practice.

### 6.2.3 Results

The SDC-Net++ architecture is able to predict the future microscopy image accurately, as can be seen in figure 6.8. But the predicted frame dose not include the detailed as the original frame and appear a bit blurry.

From the visualized motion prediction in figure 6.8 it can also be observed that the SDC-Net++ is able to precisely estimate the regions of growing and moving cells. Additionally, it can be seen that the predicted kernels, for each pixel, especially apply in regions with growing cells.

**Table 6.1:** Quantitative results of the SDC-Net++ architecture on the trapped yeast cell time-series dataset for **one** predicted future frame.

| Training stage | $W_{sv}$ | MSE ↓ | L1 ↓ | SSMI ↑ | PSNR ↑ |
|---|---|---|---|---|---|
| After flow training | ✗ | 0.3445 | 0.3128 | 0.8202 | 26.2748 |
| After flow training | ✓ | 0.3471 | 0.3125 | 0.8201 | **26.2995** |
| After fine-tune training | ✗ | **0.2797** | **0.2828** | **0.8445** | 25.5198 |
| After fine-tune training | ✓ | 0.2811 | 0.2880 | 0.8372 | 23.8990 |
| After multi pred. training | ✗ | 0.3159 | 0.3080 | 0.8031 | 19.3601 |
| After multi pred. training | ✓ | 0.3074 | 0.2992 | 0.8178 | 22.8368 |
| After multi pred. training (no adv. training) | ✗ | 0.3034 | 0.2997 | 0.8228 | 23.8532 |

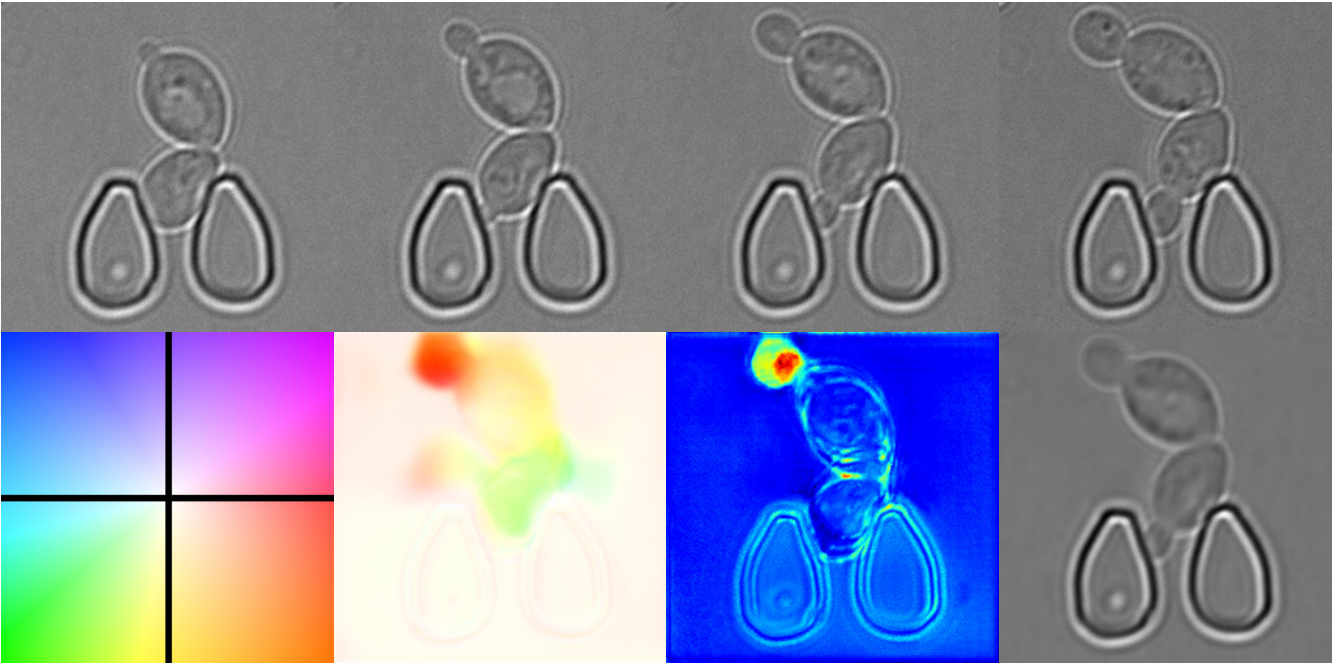

**Figure 6.8:** Motion vectors, kernels, and image prediction of the SDC-Net++ on never seen test data. Three input images in the top row on the left. Target image in the top row on the right. In the bottom row from left to right the classical optical flow color wheel [83], the encoded predicted motion (flow) vectors, the kernel prediction, and the predicted future frame. The kernel prediction is encoded as the L1 metric between the default kernel (all zero except for the middle element which is one) for each pixel. Regions in red ■ indicate kernels which are less similar to the default kernel, regions in blue ■ indicate kernels very similar to the default kernel.

**Table 6.2:** Quantitative results of the SDC-Net++ architecture on the trapped yeast cell time-series dataset for **five** predicted future frame.

| Training stage | $W_{sv}$ | MSE ↓ | L1 ↓ | SSMI ↑ | PSNR ↑ |
|---|---|---|---|---|---|
| After flow training | ✗ | 0.5758 | 0.4139 | 0.6791 | 22.0579 |
| After flow training | ✓ | 0.6250 | 0.4247 | 0.6664 | 22.7817 |
| After fine-tune training | ✗ | 0.46234 | 0.4245 | 0.7432 | **23.3211** |
| After fine-tune training | ✓ | 0.42634 | 0.3640 | 0.7364 | 22.1782 |
| After multi pred. training | ✗ | **0.4215** | **0.3525** | **0.7484** | 22.1277 |
| After multi pred. training | ✓ | 0.4764 | 0.3667 | 0.7276 | 22.8194 |
| After multi pred. training (no adv. training) | ✗ | 0.4445 | 0.3542 | 0.7422 | 22.7901 |

The tables 6.1, 6.2, and 6.3 include the quantitative evaluation of the SDC-Net++ after different training stages and for the prediction of one, five, and seven predicted future frames. For predicting one future frame the SDC-Net++ performs best after the fine-tune training. For predicting five and seven future frames the SDC-Net++ after the multi-frame prediction training stage outperforms the other models. Surprisingly, if the SDC-Net++ is trained with a weighted loss ($W_{sv}$) the performance drops compared to the model trained without the weighed loss.

Furthermore, the adversarial training in the multi-frame prediction training boost the performance slightly, compared to the multi-frame prediction training where no adversarial training is utilized.

**Table 6.3:** Quantitative results of the SDC-Net++ architecture on the trapped yeast cell time-series dataset for **seven** predicted future frame.

| Training stage | $W_{sv}$ | MSE ↓ | L1 ↓ | SSMI ↑ | PSNR ↑ |
|---|---|---|---|---|---|
| After flow training | × | 0.5766 | 0.4513 | 0.6802 | 22.4218 |
| After flow training | ✓ | 0.5794 | 0.4407 | 0.6864 | 22.3579 |
| After fine-tune training | × | 0.4245 | 0.3864 | 0.7583 | **24.2843** |
| After fine-tune training | ✓ | 0.4197 | 0.3971 | 0.7347 | 21.2110 |
| After multi pred. training | × | **0.3989** | **0.3743** | **0.7701** | 23.0549 |
| After multi pred. training | ✓ | 0.4877 | 0.3958 | 0.7273 | 22.3779 |
| After multi pred. training (no adv. training) | × | 0.4045 | 0.3708 | 0.7652 | 22.7487 |

Qualitative results for multiple predicted future frames can be seen in figure 6.9. The SDC-Net++ is able to accurately predict the first frame, however, loses texture in the further predicted frame. This results in blurry images. But form the predicted motion it can be observed that the generator network is still able to capture the motion of the growing cell. Additionally, the network tries to catch more texture by predicting more complex kernels, however, the network is not able to work against the loss in texture.

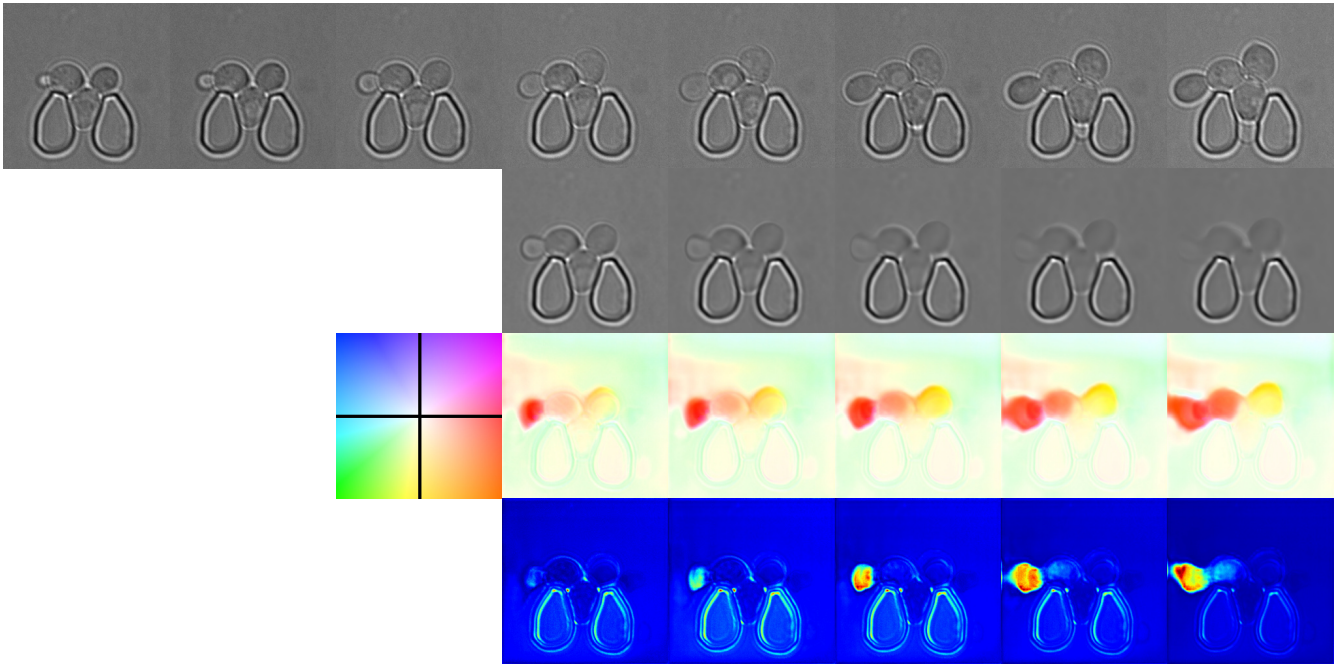

**Figure 6.9:** Multi future frame prediction of the SDC-Net++ with motion vectors and kernels. In the top row, from left to right, first the three input frames, beside the ground truth future frames are visualized. In the second row, the five autoregressively predicted future frames are shown. In the third and fourth row, the predicted motion vectors and the kernels are visualized. The motion vectors are encoded in the classical color wheel [83] also, shown in the third row. The kernel predictions are visualized, as described in the caption of the figure 6.8.

Besides the raw performance results of the SDC-Net++, an interesting observation regarding the learned parameters of the GARLoss can be made. The GARLoss namely learns the regions of the image where typically cells occur and utilize a loss function, which penalizes the outliers stronger in these regions. The GARLoss, also, employs, in the cell regions, a loss which puts less penalty on inliers. These effects can be seen in figure 6.10.

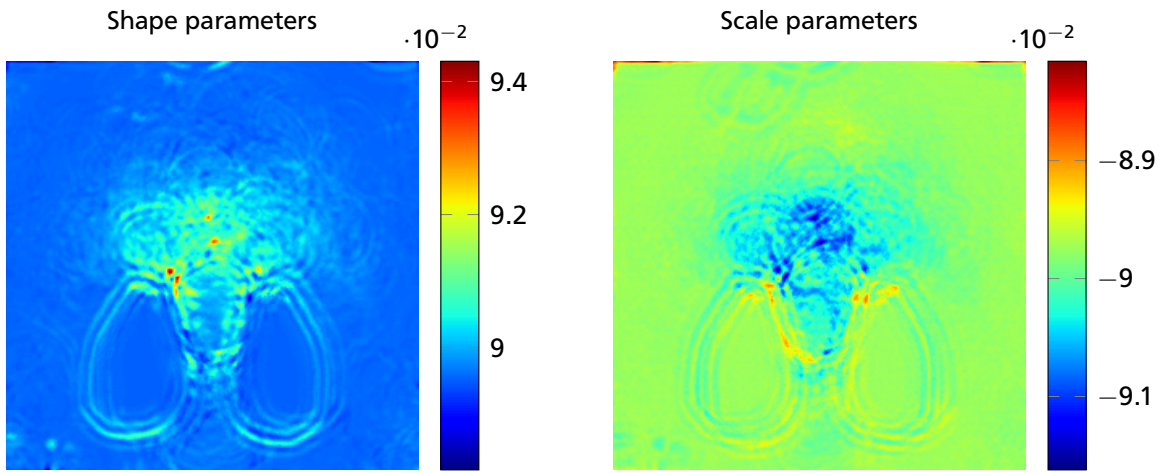

**Figure 6.10:** Trained parameters of the GARLoss [32] after the fine-tune training stage. On the left the shape parameter $\alpha$ and on the right the scale parameter $c$, for each pixel.

Additional plots of the SDC-Net++ predictions as well as loss curves of the training process can be found, in section 9.5 of the appendix.

# 7 Discussion & Outlook

This thesis presents the first step towards a deep learning-based image-level modeling of biological experiments. The modeling task was split into two main problems. First, the generation of synthetic brightfield microscopy images based on real images. And second, the simulation of brightfield microscopy images for future time-steps. For the generation of brightfield microscopy images, the Cell-GAN model was introduced. The novel Cell-GAN model can produce conditionalized and unconditionalized synthetic images. To further tackle the problem of simulation, the SDC-Net++ model was proposed, which is able to predict microscopy images for future time-steps based on images from the past. Both models are not yet capable of operating together to model biological experiments despite the strong initial results. To achieve this, both the Cell-GAN model and the SDC-Net++ model have to be improved further. This, however, remains difficult, since the main problem lies in the availability of clean data.

The Cell-GAN model, based on a StyleGAN 2 [13] generator network, a U-Net discriminator [107], and a ResNet-like encoder, is able to produce conditionalized and unconditionalized brightfield microscopy image samples. For unconditionalized generation, the Cell-GAN achieved an Inception score [71] of 1.730, which is close to the dataset's Inception score of 1.799, indicating a strong performance in terms of sample quality and variety. For the conditionalized generation of microscopy images, with the goal of generating an image with the key feature of a guidance image, the Inception score drops from 1.730 to 1.557, indicating a drop in performance. This performance drop engaged by the increase of the Fréchet Inception Distance [72] from 29.588, in the unconditionalized setting, to 60.067 for the conditionalized setting. Both, the Inception score and the Fréchet Inception Distance rely on the InceptionNet [73, 74] pre-trained on ImageNet [65]. Since the yeast cell dataset used to train the Cell-GAN is far different from the ImageNet the Inception score and the Fréchet Inception Distance can be artificially bad. The qualitative results reveal that the conditionalized generation based on synthetically generated images is functioning better than if real images are utilized as guidance. This implies that the generator does not overfit to the training dataset and is able to generate new samples.

To close the gap between simulation and generation, the Cell-GAN model has to be improved in conditionalized generation based on real images. Since in practice, a real sequence of images could be taken to generate a new synthetic sequence, similar to the real sequence, which could then be simulated by the SDC-Net++. Another approach would be to train the Cell-GAN to generate not one microscopy image but to generate a synthetic sequence of images instead. These generated sequences could further be simulated by the SDC-Net++. However, first tests on generating a sequence of three microscopy images failed, resulting in a non-convergence to an equilibrium of the adversarial training process.

Possible improvements to the Cell-GAN could be to enlarge the intermediated latent space in which the guidance encoder maps the guidance image, similar to the extended intermediated latent space proposed the Imge2StyleGAN paper [98, 108]. This could enable the Cell-GAN to improve in conditionalized generation. Another promising approach that could be adopted is an adaptive discriminator augmentation mechanism [119]. This mechanism augments the input to the discriminator network resulting in overall improved performance on small datasets. This could possibly enable the Cell-GAN to predict a sequence of images, since utilizing sequences of three images instead of a single image at training time effectively reduces the number of training samples by a factor of three. The Hessian Penalty [120] is a novel and simple regularization of the adversarial latent space. Applying such a weak prior could possibly improve the overall performance of the Cell-GAN model but could also improve conditionalize generation since the intermediated latent space could be more disentanglement.

The SDC-Net++ model, based on the original SDC-Net [22] can predict multiple future brightfield microscopy frames based on a given input sequence of past frames. To train the SDC-Net++, multiple different training stages are employed, also including a partly adversarial stage. The results show that the model is able to accurately predict the future frame of a given input sequence. However, it lacks in terms of the image texture when predicting multiple frames in advance. The qualitative results have shown that the SDC-Net++ generator is able to extrapolate the object motion for multiple future frames. Nevertheless, the predicted frames for multiple steps are blurry and texture is lost. This could possibly be caused by inaccurately predicted kernels but needs further investigation.

To use the SDC-Net++ in practice, the network should be able to accurately predict at least five future brightfield microscopy frames. The first step to achieve improved performance is by studying why and where the model fails when predicting multiple frames. However, also recent improvements in the field of optical flow estimation could be employed to boost the performance of the SDC-Net++. Since predicting an extrapolated motion field is slightly related to optical flow estimation.

Feasible changes to the SDC-Net++ which can improve the performance could be for example to employ a feature pyramid network [121, 16, 18] for each input frame, instead of the 3d U-Net [50, 79, 22]. Additionally, multi-stage predictions [15, 16], for different resolutions, could be employed. These approaches showed great improvements in supervised and unsupervised optical flow estimation [18, 16, 15]. Recent work on future frame prediction also utilized instance segmentation maps as additional guidance to the optical flow maps [23]. But using additional information from pre-trained models makes the future prediction model more complex and highly reliant on the additional model. Moreover, when employing a future frame prediction model with an additional instance segmentation network, domain labels have to be available. This is not always given. Also, the SDC-Net++ relies on a pre-trained semantic segmentation model, which should possibly be omitted in the future, depending on the use-case. However, to compute the necessary optical flow maps of the past frames, a separate pre-trained model is employed. This deep optical flow estimation model can be fine-tuned, in an unsupervised setting, with the same dataset as used to learn future frame prediction. The SDC-Net might also benefit from self-supervised training or smoothness regularization of the predicted motion vector, as employed in unsupervised optical flow estimation [18]. It is also further imaginable to apply a full adversarial training in the final multi-frame prediction training stage since the network should learn a plausible prediction of the cell behavior for future frames. This plausible prediction can differ from the true future frames but can still be correct [99]. But also recent more developments like axial-attentions [122, 123, 124] should be considered in future research.

# 8 Conclusion

In conclusion, this thesis introduced two novel methods towards a deep learning-based image-level modeling of time-lapse-fluorescence-microscopy experiments. The proposed Cell-GAN model, for conditionalized and unconditionalized microscopy image generation, showed strong performance in the conditionalized setting with an inception score of 1.730 (dataset IS 1.799). However, in the unconditionalized setting, the performance drops slightly to an inception score of 1.557. The introduced simulation model SDC-Net++ can accurately predict one future frame of a given input sequence. When predicting multiple future frames the resulting images suffer from blurriness. In general, the showcased results demonstrated that addressing the modeling of TLFM experiments with separate deep learning-based models, for generation and simulation, is a promising approach for future research.

# Acknowledgments

First, I would like to thank my supervisors **Tim Prangemeier**, **Christian Wildner** and **Prof. Koeppl** for the great support, through my thesis and additional projects. Furthermore, I want to acknowledge the work of **Tim Kircher**, **Tizian Dege**, and **Florian Schwald** to prepare the necessary data with me. Moreover, I thank **Marius Memmel**, **Nikita Araslanov** as well as **Martin Schlotthauer** for the useful discussions. Finally, I especially thank **Markus Baier** and **Bastian Alt** for providing the computational infrastructure.

# List of Algorithms

# List of Figures

# List of Tables

# 9 Appendix

## 9.1 Residual Variational Autoencoder

For a comparison to the adversarial approaches for microscopy image generation also a residual variation Autoencoder (VAE) has been trained in this thesis. The residual variational Autoencoder follows a ResNet-like [43] architecture and can be seen in figure 9.1.

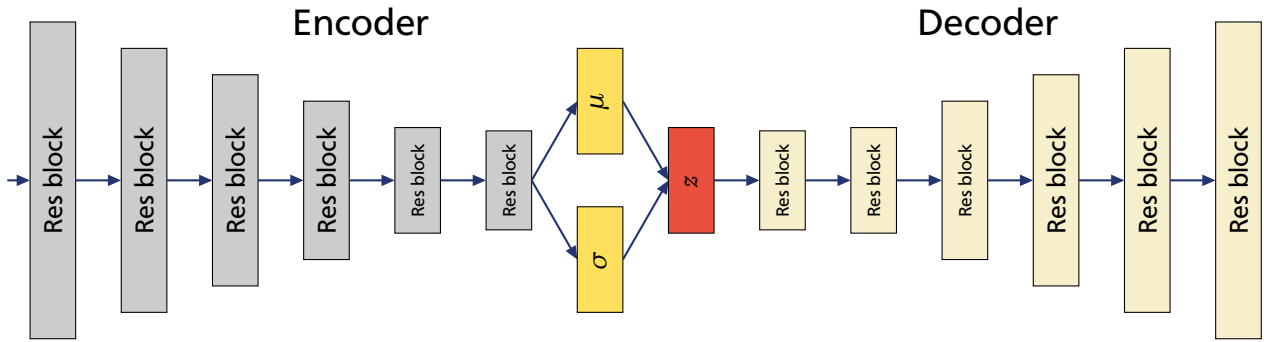

**Figure 9.1:** Architecture of the Res-Net like variational autoencoder with reparametrization between the encoder end decoder.

The encoder of the VAE aims encoder the input image $\mathbf{I}$ into a latent feature vector $\boldsymbol{z}$. From this tensor the vector $\boldsymbol{z}$ the parameters $\boldsymbol{\mu}$ and $\boldsymbol{\sigma}$ are learned. These parameters are used in the reparametrization step to construct the input to the decoder path $\boldsymbol{z}$. The decoder then aims to reproduce the original input image $\mathbf{I}$. Each residual blocks used in the residual VAE consists of two $3 \times 3$ convolutions followed by an instance normalization [54] and a leaky ReLU [52] activation, a residual mapping, and an up/downsampling operation.

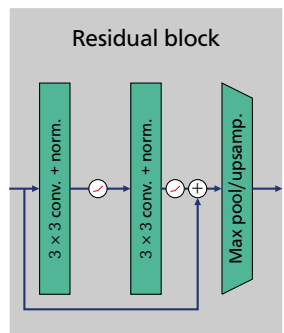

**Figure 9.2:** Architecture of the residual block used in the architecture of the residual variational autoencoder. Dependent if the block is utilized in the encoder path or the decoder path max pooling or bilinear upsampling is used.

For a more detailed description on the architecture of VAEs and the reparametrization step see Kingma and Welling [75] or Goodfellow et al. [1] (section 20.10.3).

### 9.1.1 Results

The residual variational autoencoder was trained for 100 epochs on the standard variational autoencoder loss [75]. For optimizing the network parameters the Adam optimizer [60] with a learning rate of $10^{-4}$ was utilized.

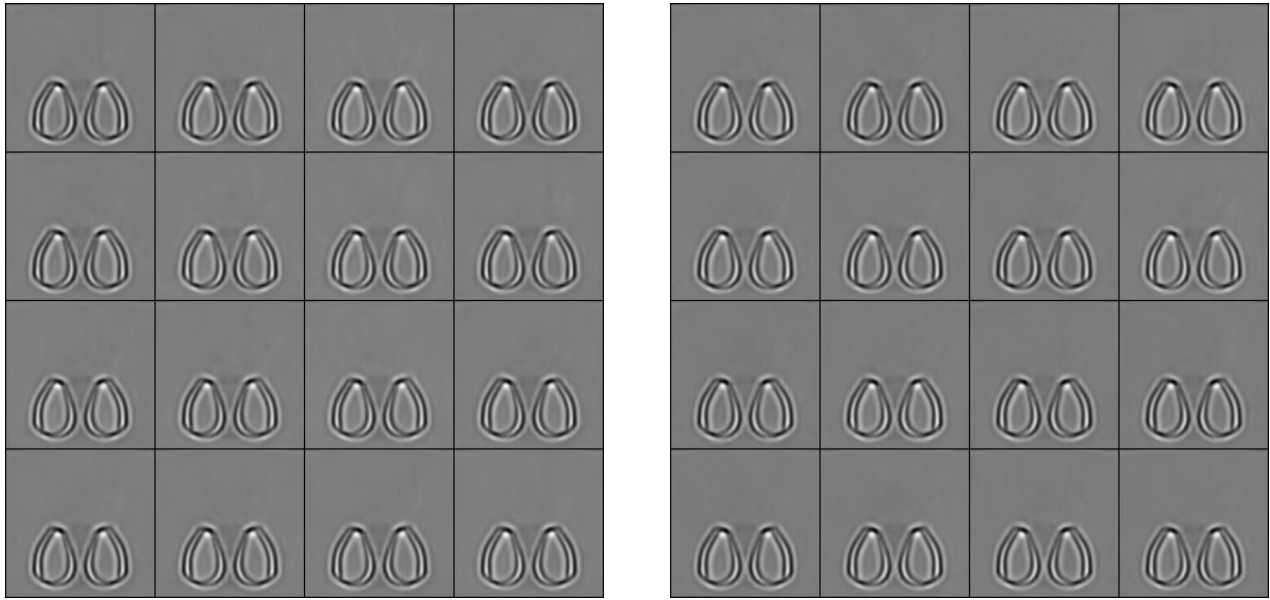

**Figure 9.3:** Reconstruction training results of the VAE on the left and random samples from the VAE on the right.

The qualitative results of the trained residual VAE (fig. 9.3) showed week performance compared to the results of the adversarial approaches. The VAE was only able to learn to produce traps without cells or traps with cell-like artifacts. For this reason, a quantitative analysis of the residual VAE has been omitted.
However, more advanced variational autoencoders like [31] have been published. It is leave up for future work to investigate the performance of these advanced VAE approaches on generating microscopy imagery.

## 9.2 Trained Padé Activation Unit

The figure below shows a PAU [106] trained (initialized as a LeakyReLU [52]) on the task of instance segmentation with a moderate learning rate of $10^{-4}$. After training the following parameter values for $\boldsymbol{a} \in \mathbb{R}^6$ and $\boldsymbol{b} \in \mathbb{R}^5$ of $\text{PAU}_{\text{trained}}$ have been reached:

$$\boldsymbol{a} = [-0.0174, 0.5433, 1.6947, 2.0711, 1.0022, 0.2311]$$
$$\boldsymbol{b} = [-1.7421 \cdot 10^{-5}, 3.9152, 3.0160 \cdot 10^{-5}, 0.21971]$$

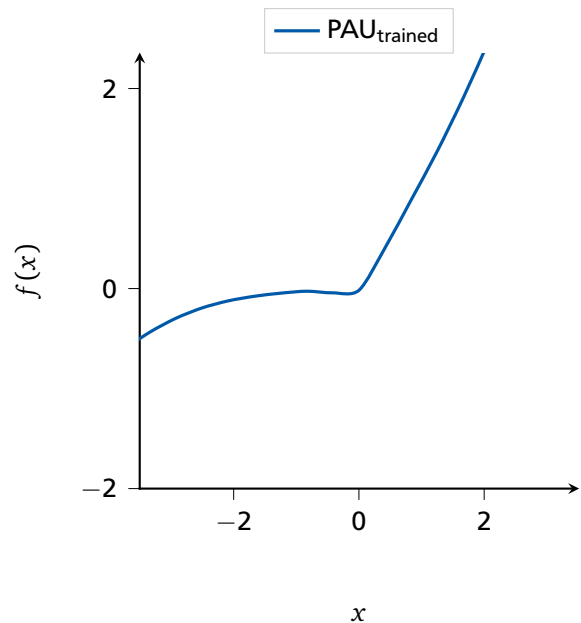

**Figure 9.4:** Example of a trained Padé Activation Unit. [106]

From figure 9.4 can be observed that the trained Padé Activation Unit does not differ much from the initial LeakyReLU activation. The reason for this behavior is unclear and requires future investigation. Potential reasons, however, could be a bad gradient flow to the parameters of the Padé Activation Unit for a small learning rate or the fact that the LeakyReLU is a too good choice as an initial activation function.

## 9.3 Mode Collapse Example

The mode collapse test example trains a small multi-layer perceptron generator and discriminator in an adversarial setting to produce the data distribution $p_{\text{data}}$. The data distribution $p_{\text{data}}$ is produced by 8 Gaussian distributions $\mathcal{N}(\mu, 0.05)$ ordered in a circle with radius one. The generator network is implemented as a three-layer feed-forward neural network with 10 hidden features and a random noise input feature vector sampled from $U(0, 1)$ with the dimension $z \in \mathbb{R}^2$ is used as the input to the model. The discriminator is also implemented as the generator network as a three-layer feed-forward neural network, however, produces one scalar output instead of the 2d output of the generator network. For the adversarial training, the standard GAN loss 2.6 is utilized. To optimize the generator and discriminator network standard stochastic gradient descent is used as in 3. The whole code for reproducing the plots is available at: https://github.com/ChristophReich1996/Mode_Collapse.

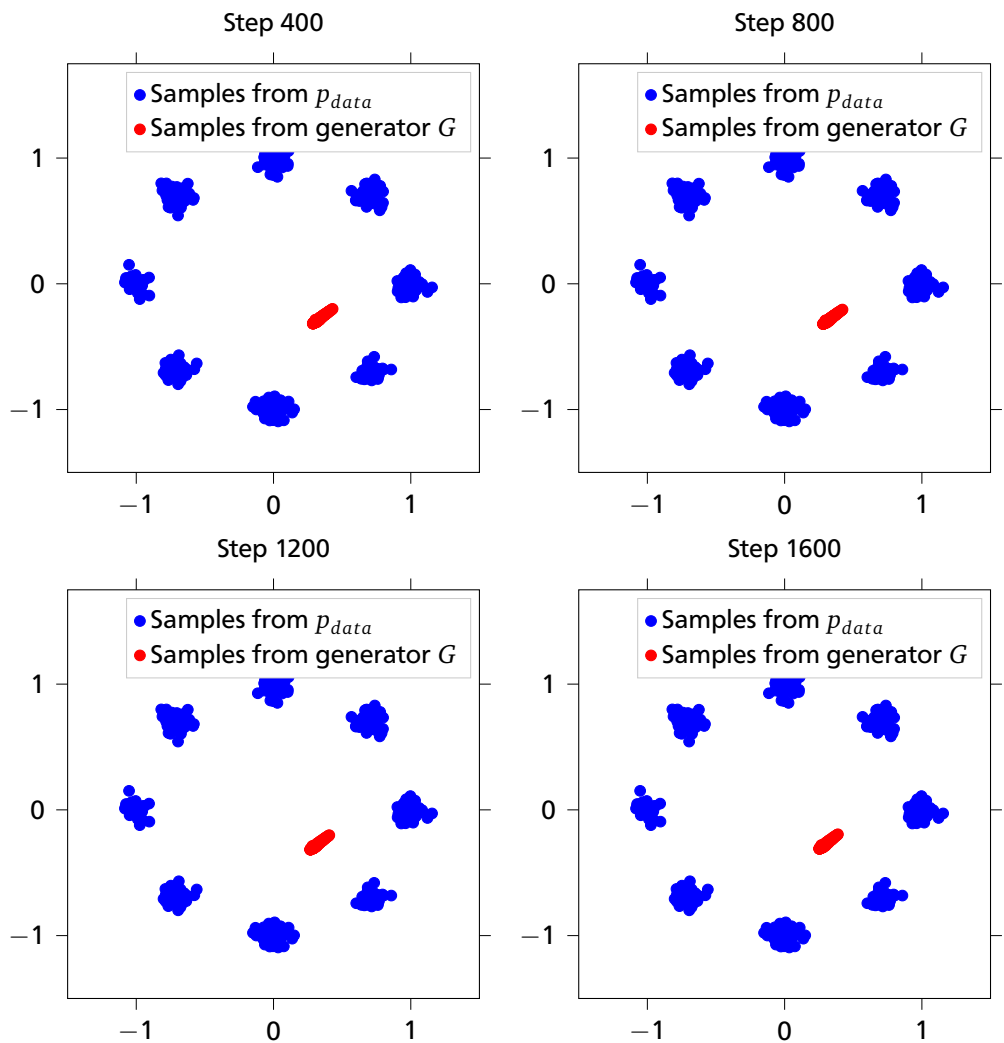

**Figure 9.5:** Mode collapse example in $\mathbb{R}^2$ after epoch 1 to 4.

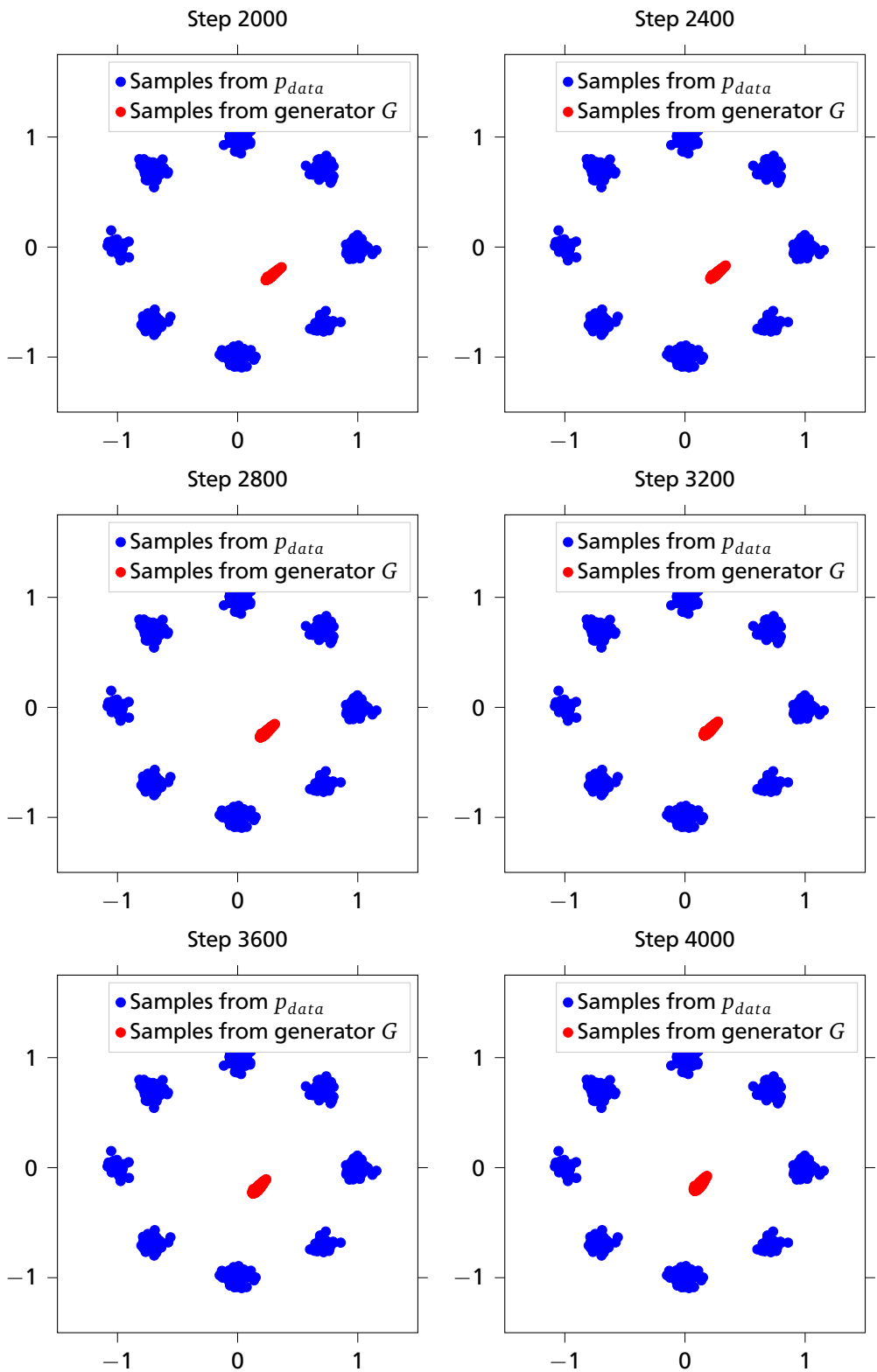

**Figure 9.6:** Mode collapse example in $\mathbb{R}^2$ after epoch 5 to 10.

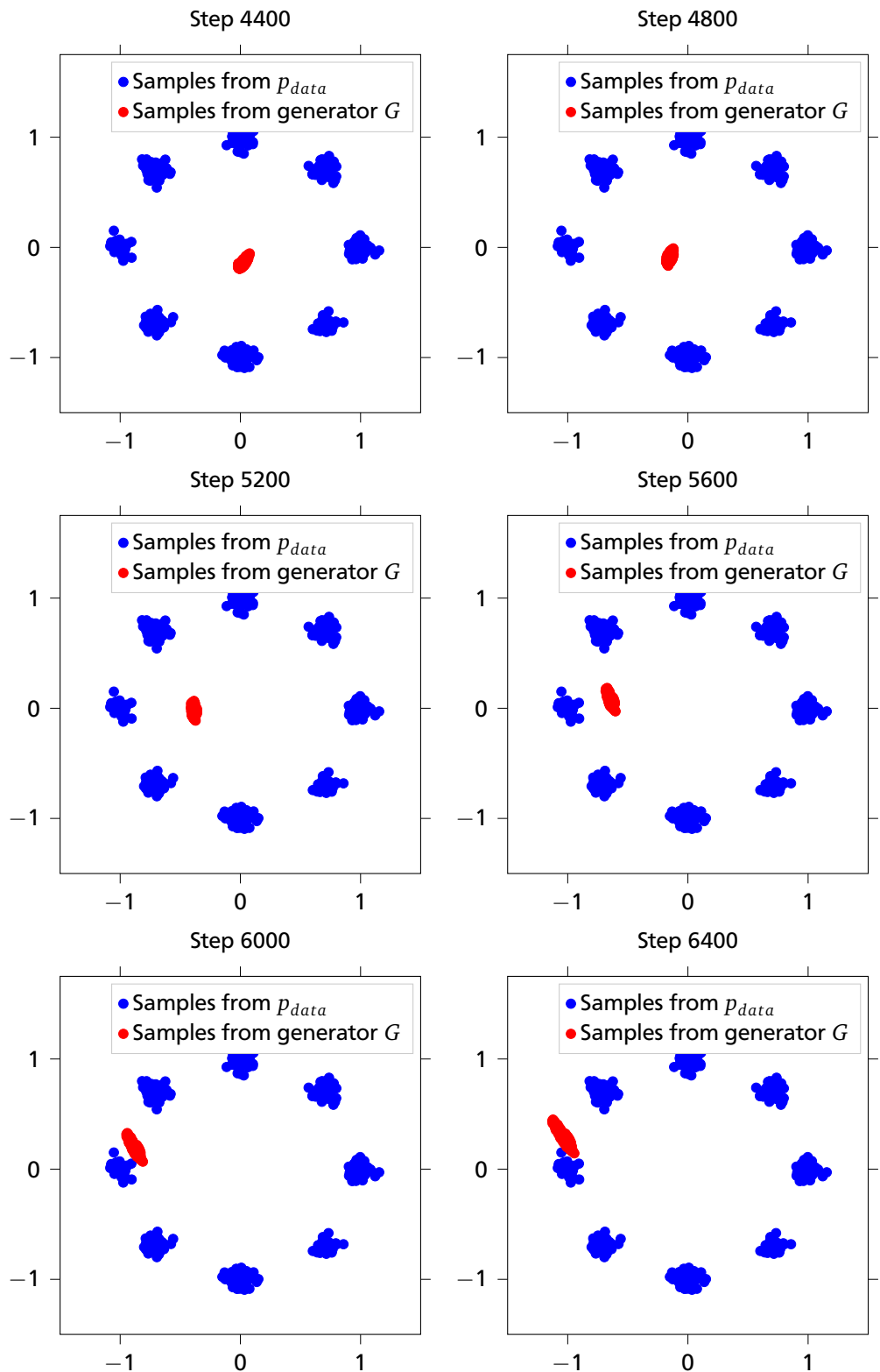

**Figure 9.7:** Mode collapse example in $\mathbb{R}^2$ after epoch 11 to 16.

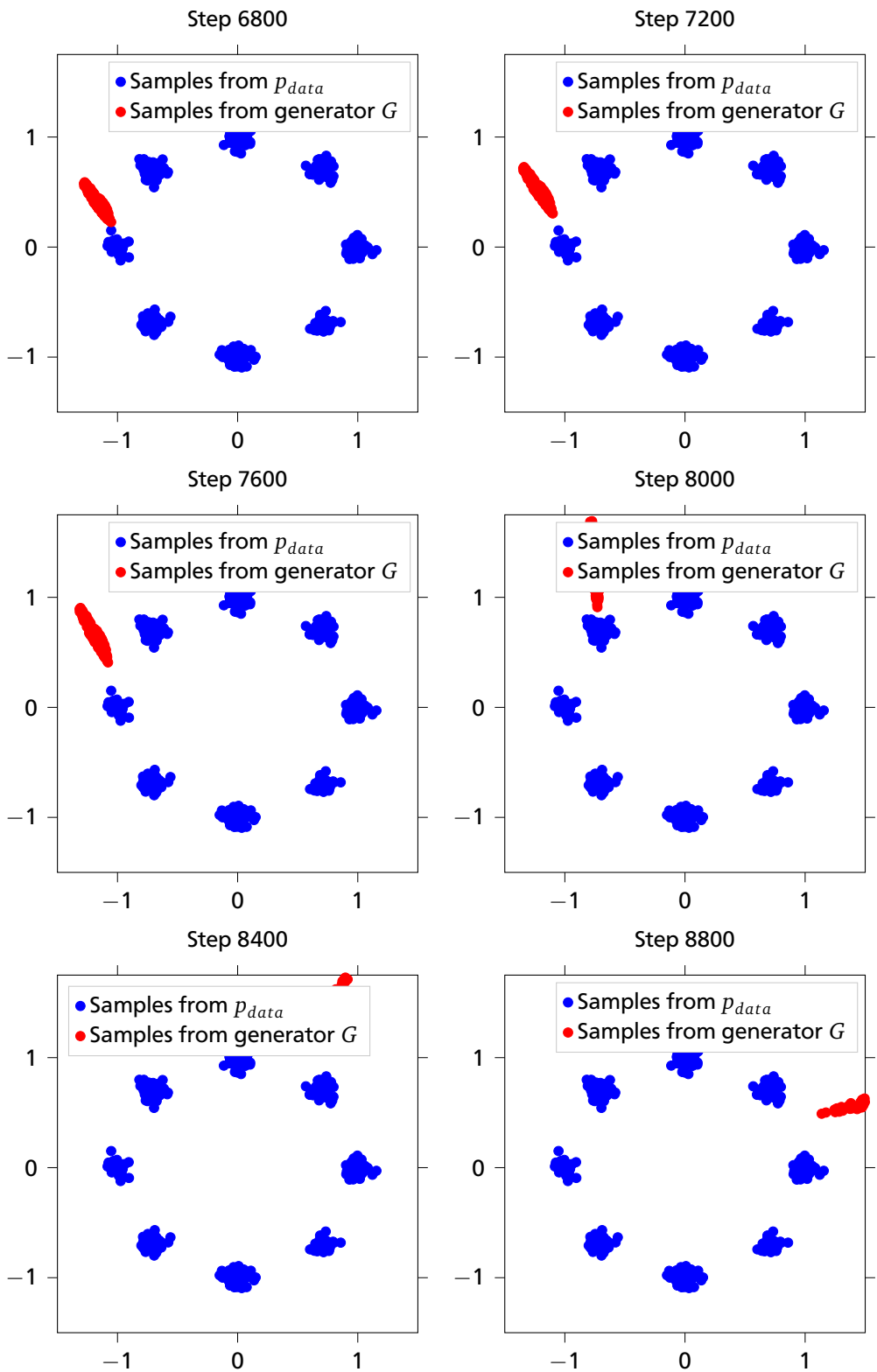

**Figure 9.8:** Mode collapse example in $\mathbb{R}^2$ after epoch 17 to 22.

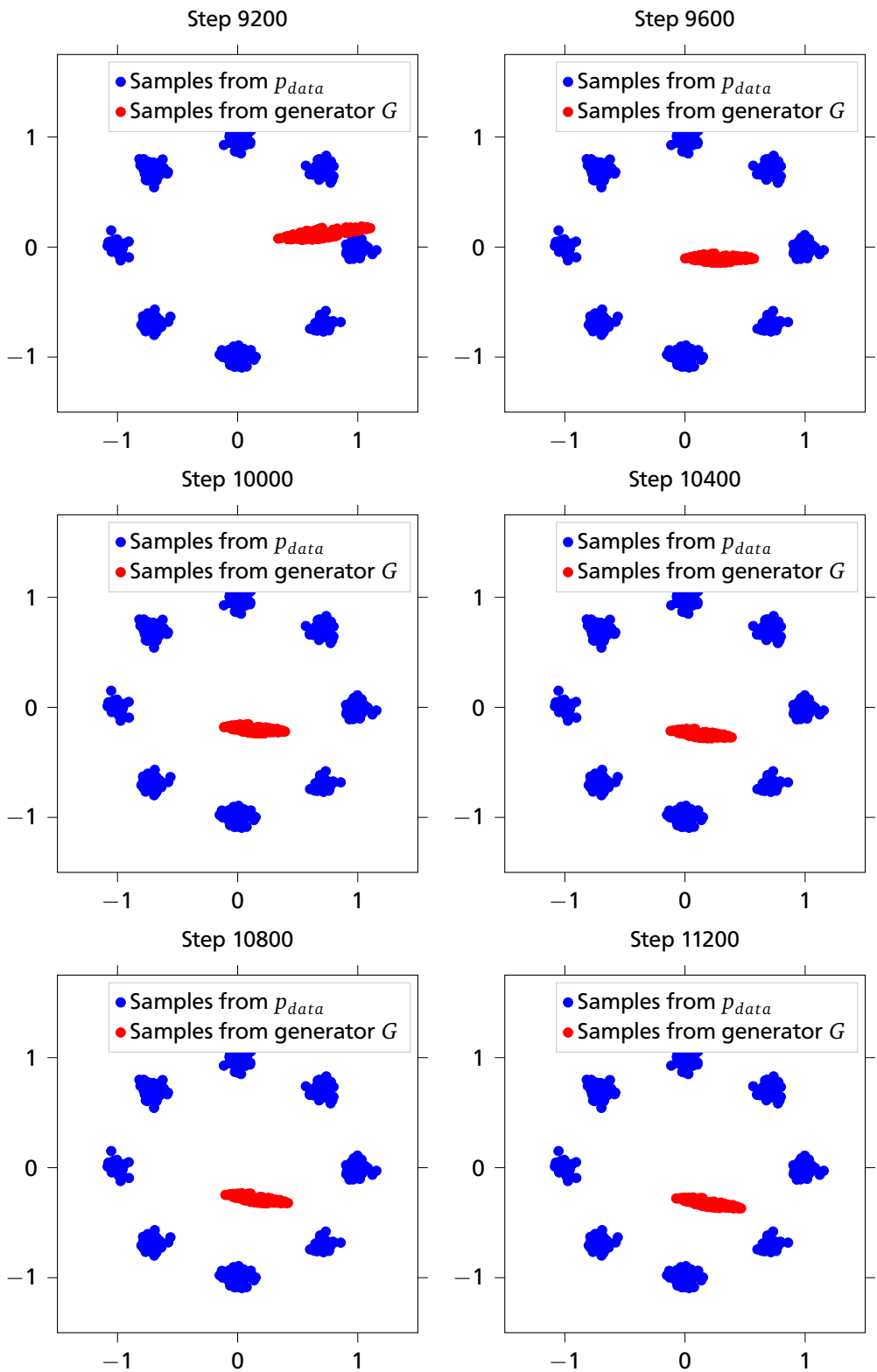

**Figure 9.9:** Mode collapse example in $\mathbb{R}^2$ after epoch 23 to 28.

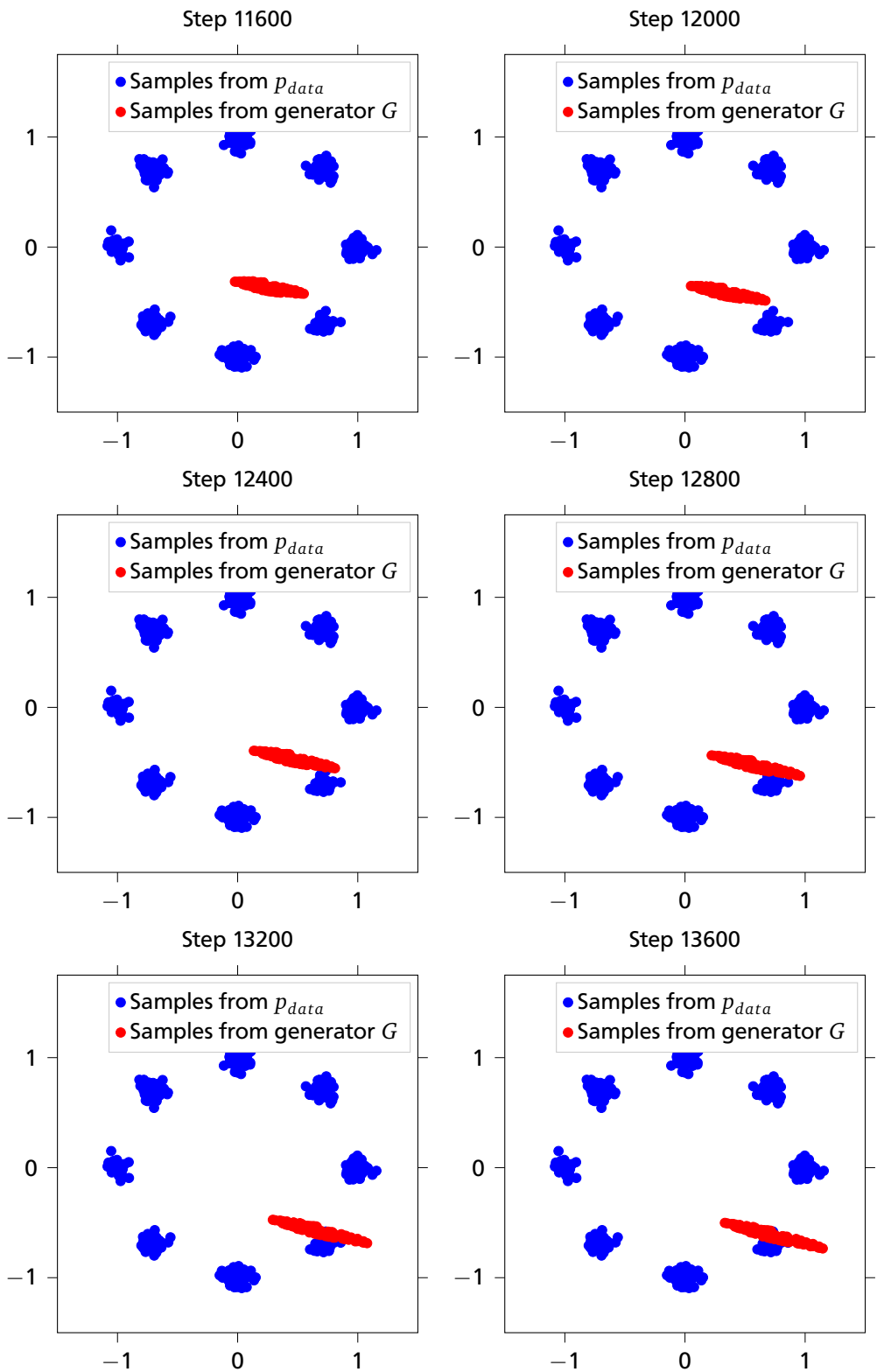

**Figure 9.10:** Mode collapse example in $\mathbb{R}^2$ after epoch 29 to 34.

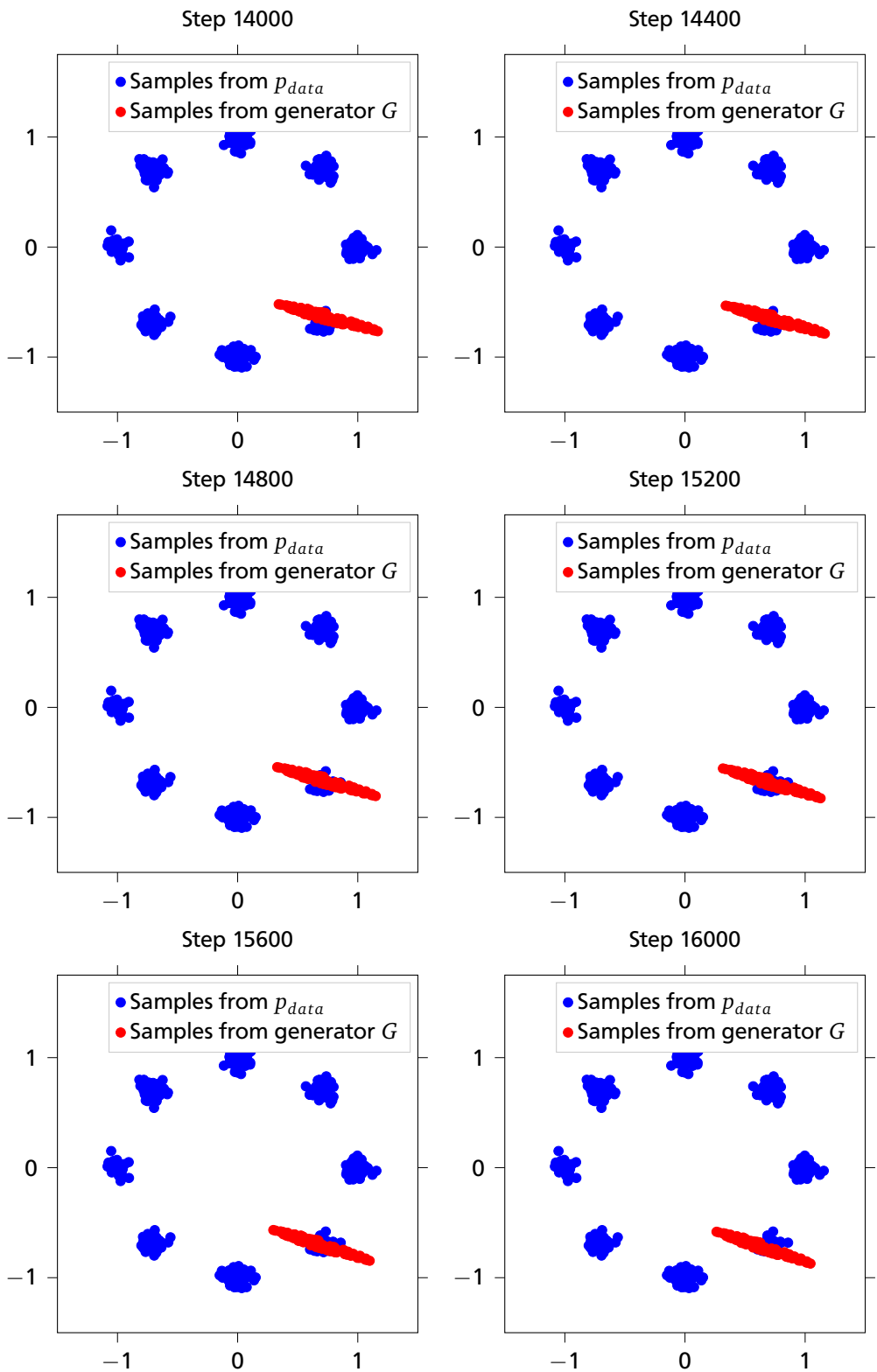

**Figure 9.11:** Mode collapse example in $\mathbb{R}^2$ after epoch 35 to 40.

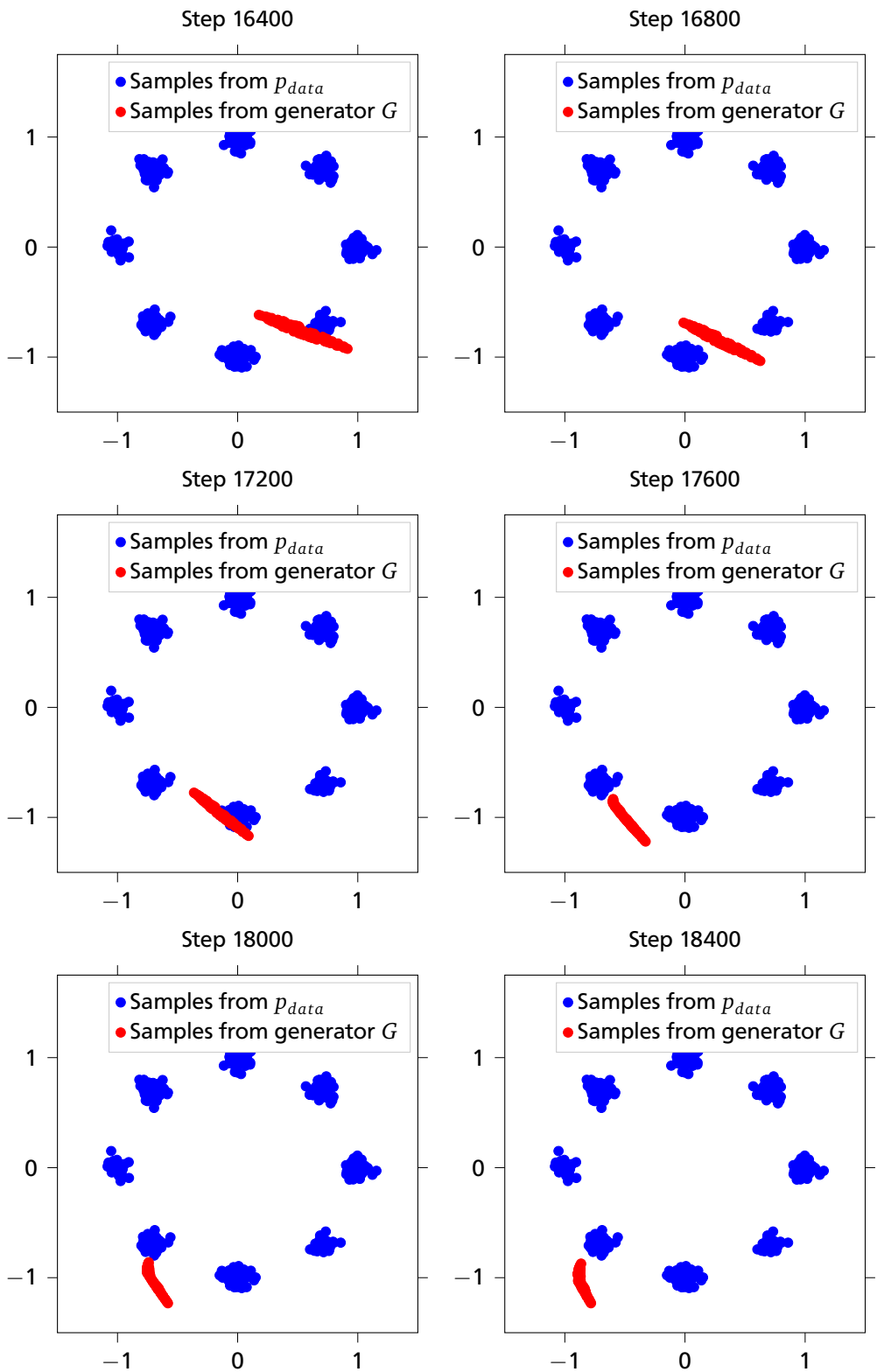

**Figure 9.12:** Mode collapse example in $\mathbb{R}^2$ after epoch 41 to 46.

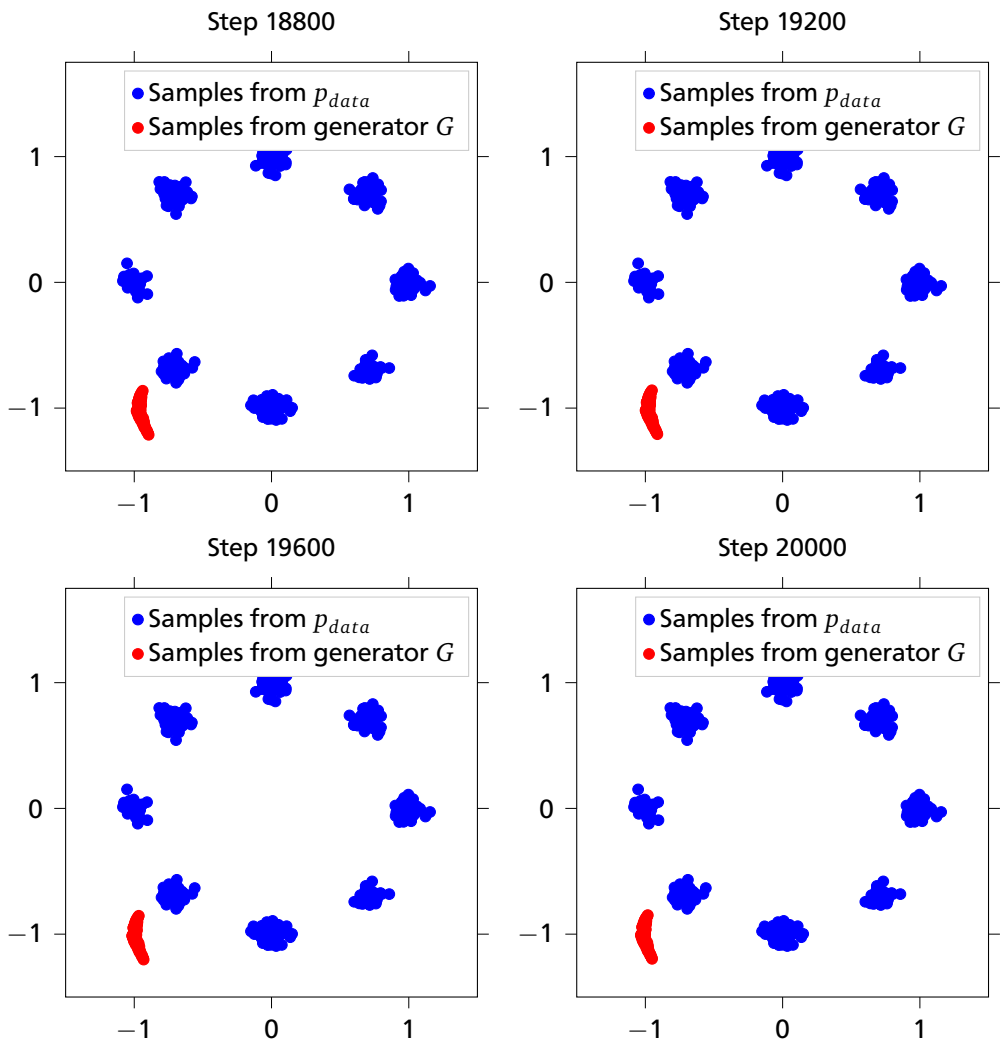

**Figure 9.13:** Mode collapse example in $\mathbb{R}^2$ after epoch 47 to 50.

## 9.4 Cell-GAN

This section shows additional conditionalized and unconditionalized samples from the Cell-GAN model after different training steps.

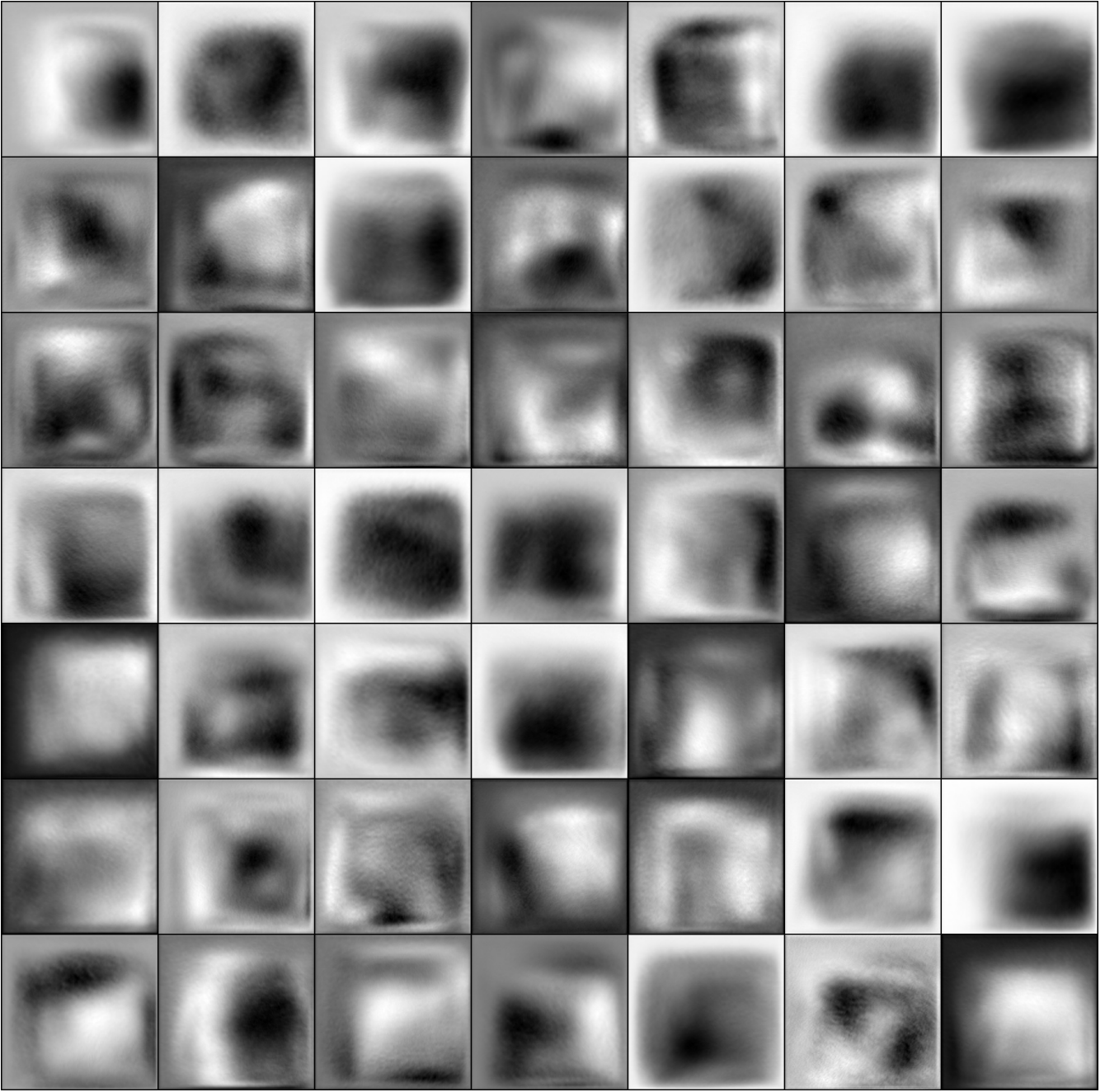

**Figure 9.14:** Unconditionalized generated samples from the Cell-GAN generator after 1 epoch.

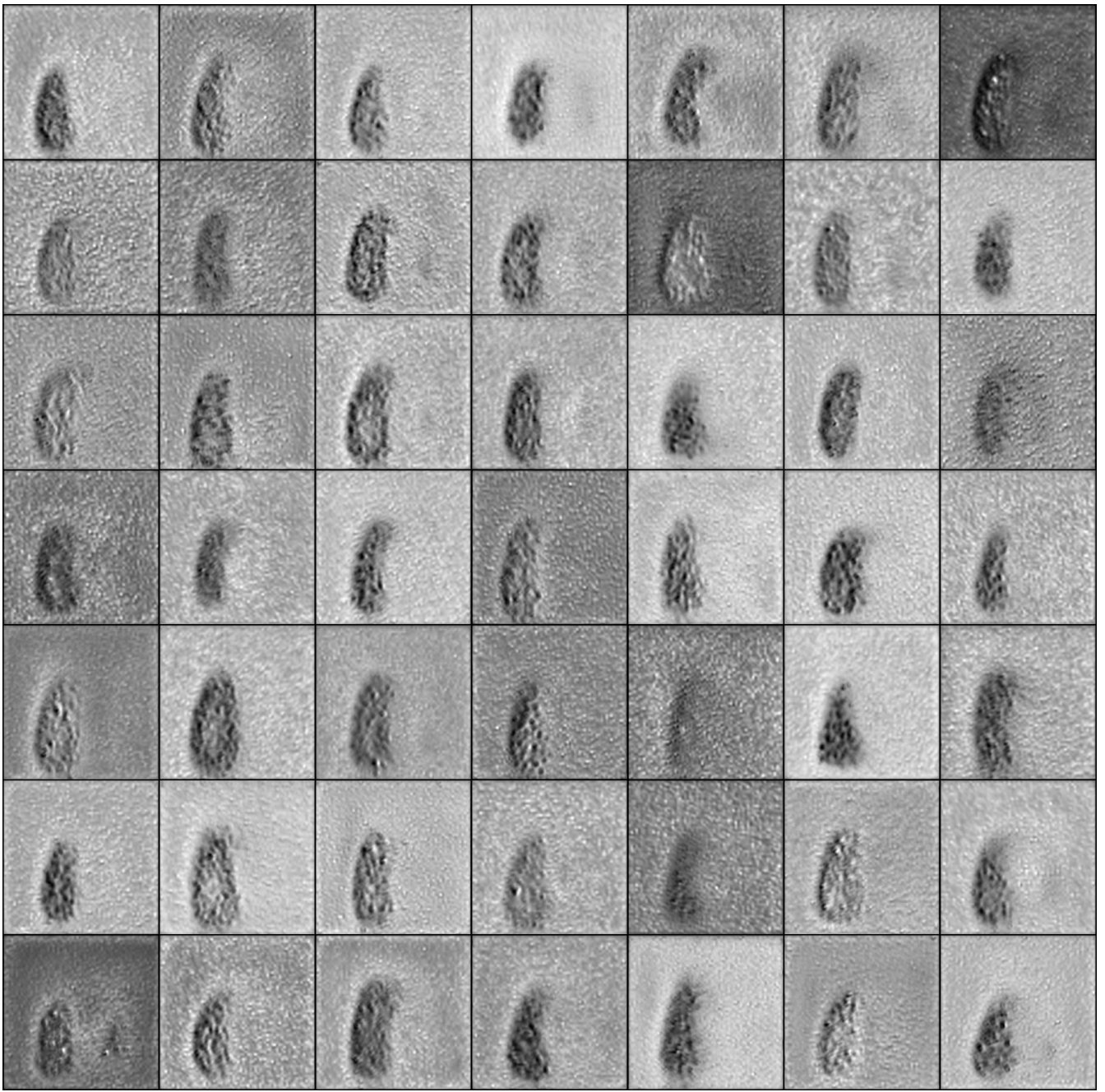

**Figure 9.15:** Unconditionalized generated samples from the Cell-GAN generator after 5 epochs.

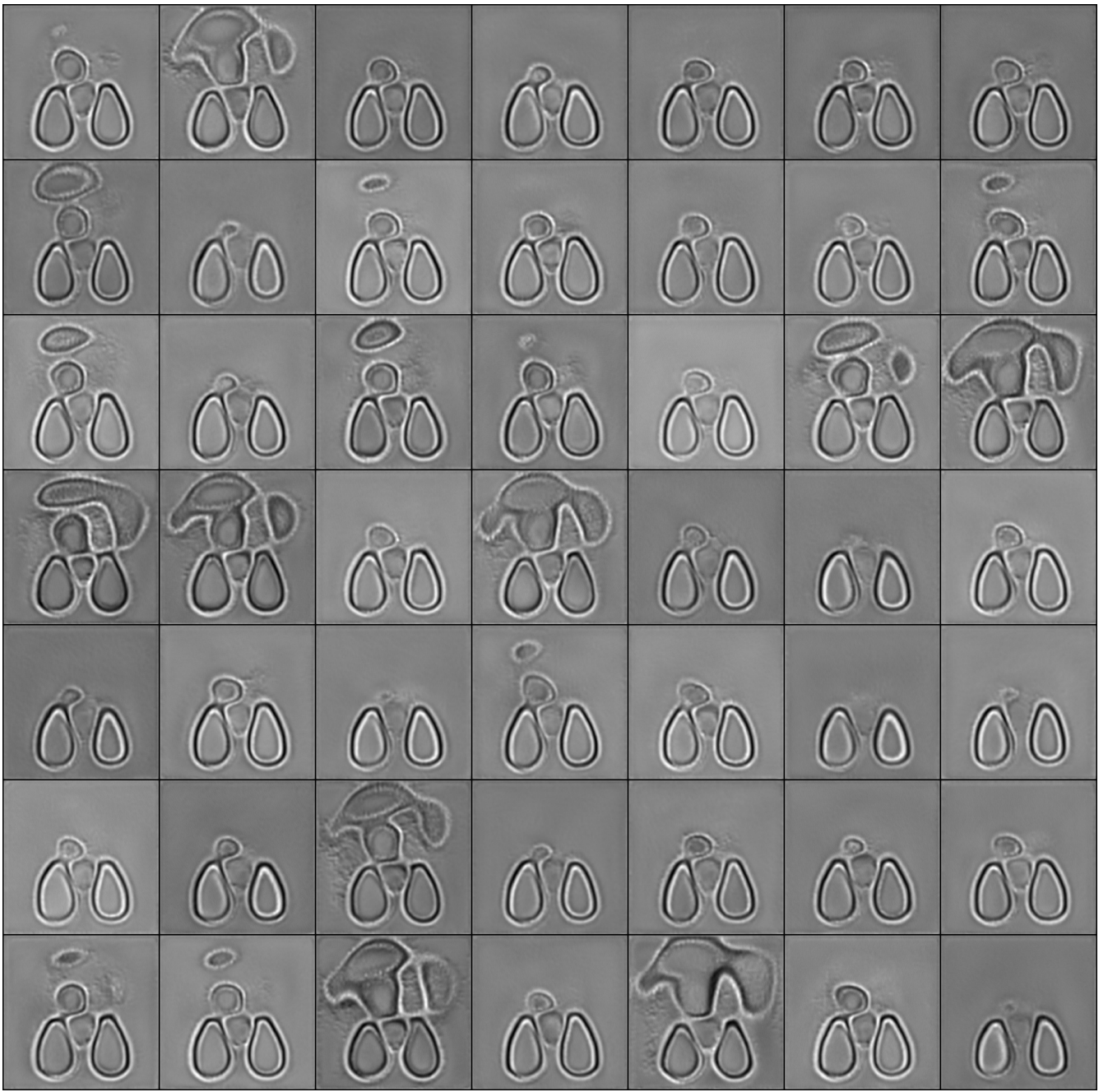

**Figure 9.16:** Unconditionalized generated samples from the Cell-GAN generator after 9 epochs.

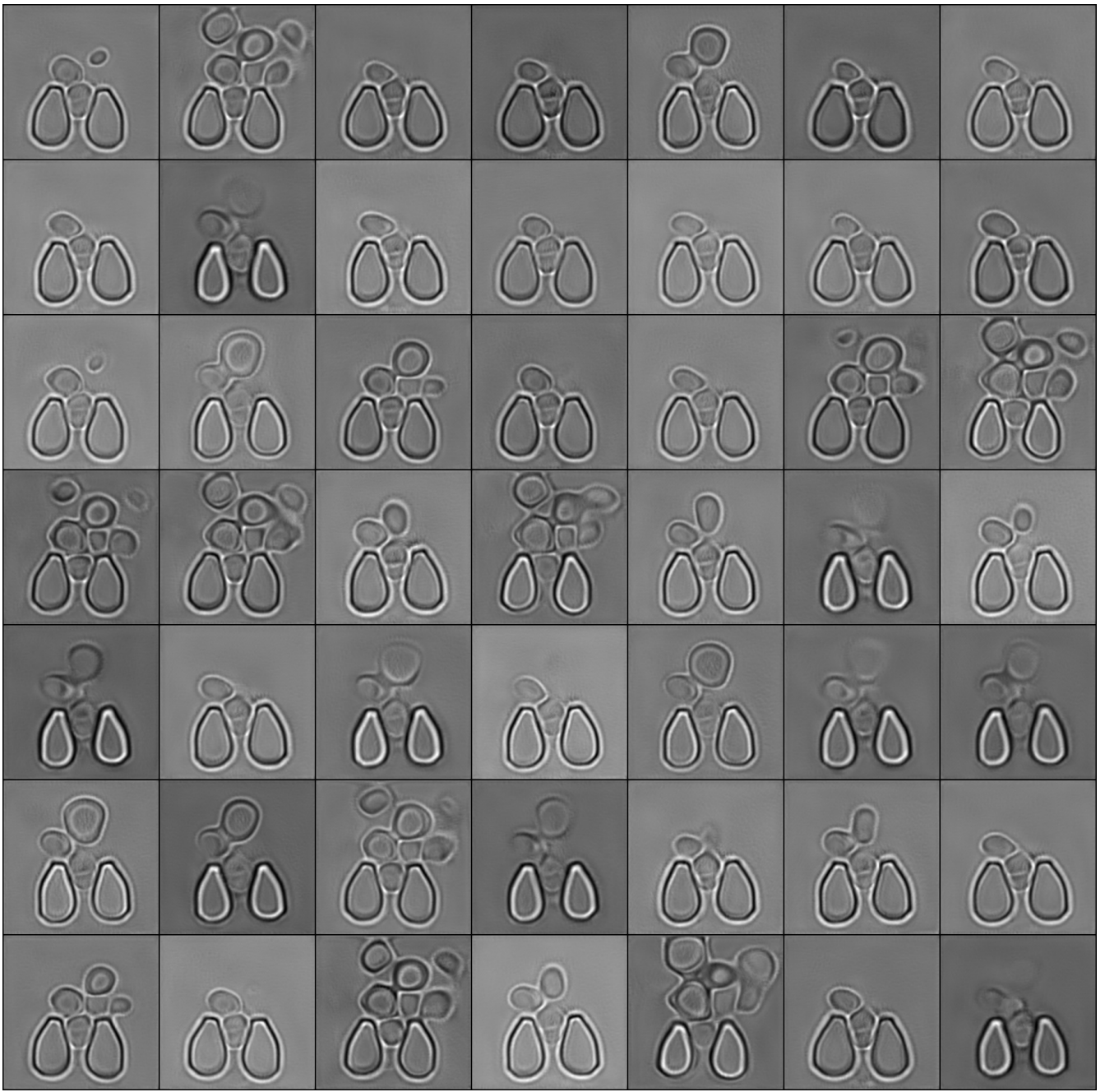

**Figure 9.17:** Unconditionalized generated samples from the Cell-GAN generator after 13 epochs.

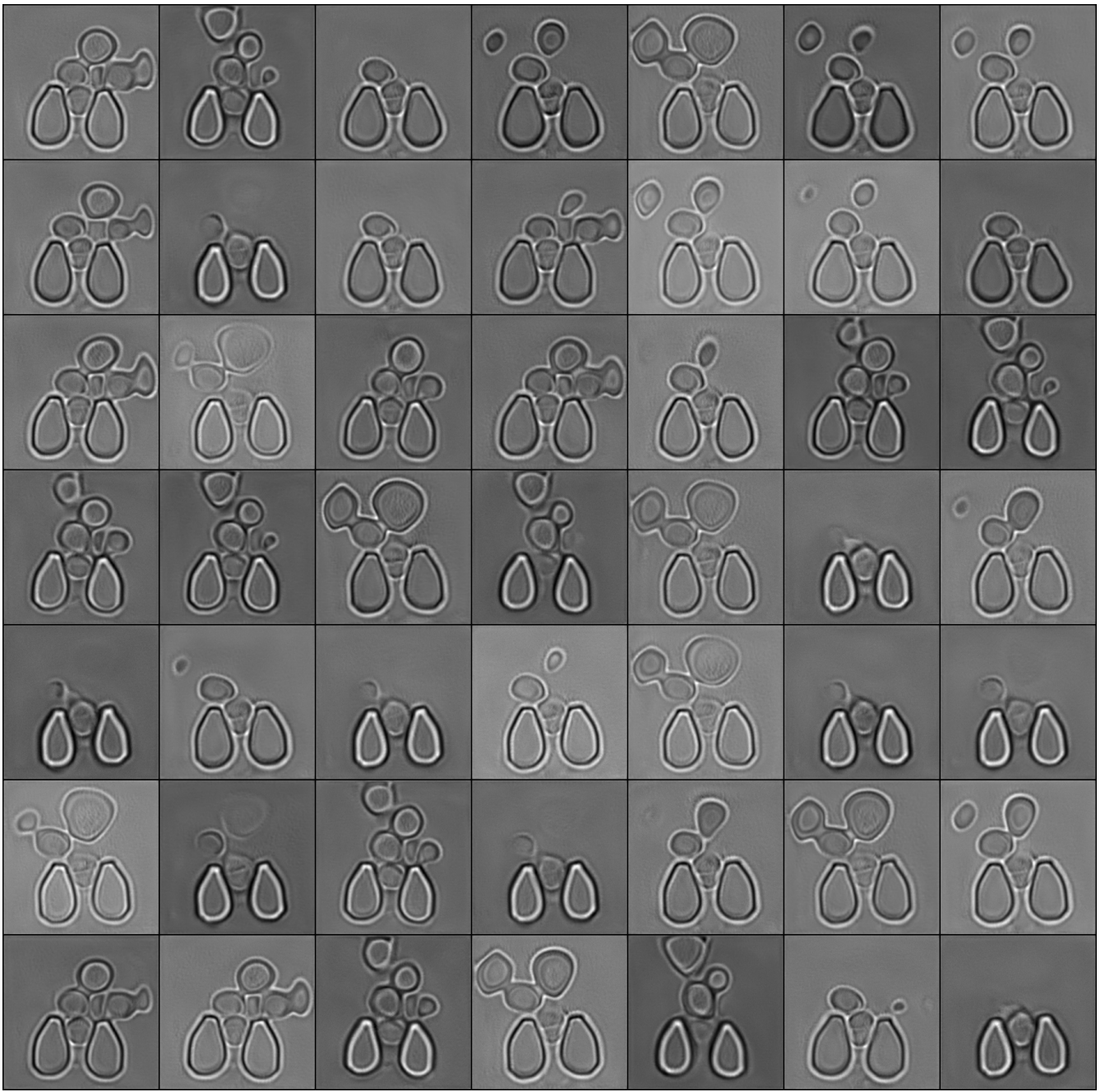

**Figure 9.18:** Unconditionalized generated samples from the Cell-GAN generator after 17 epochs.

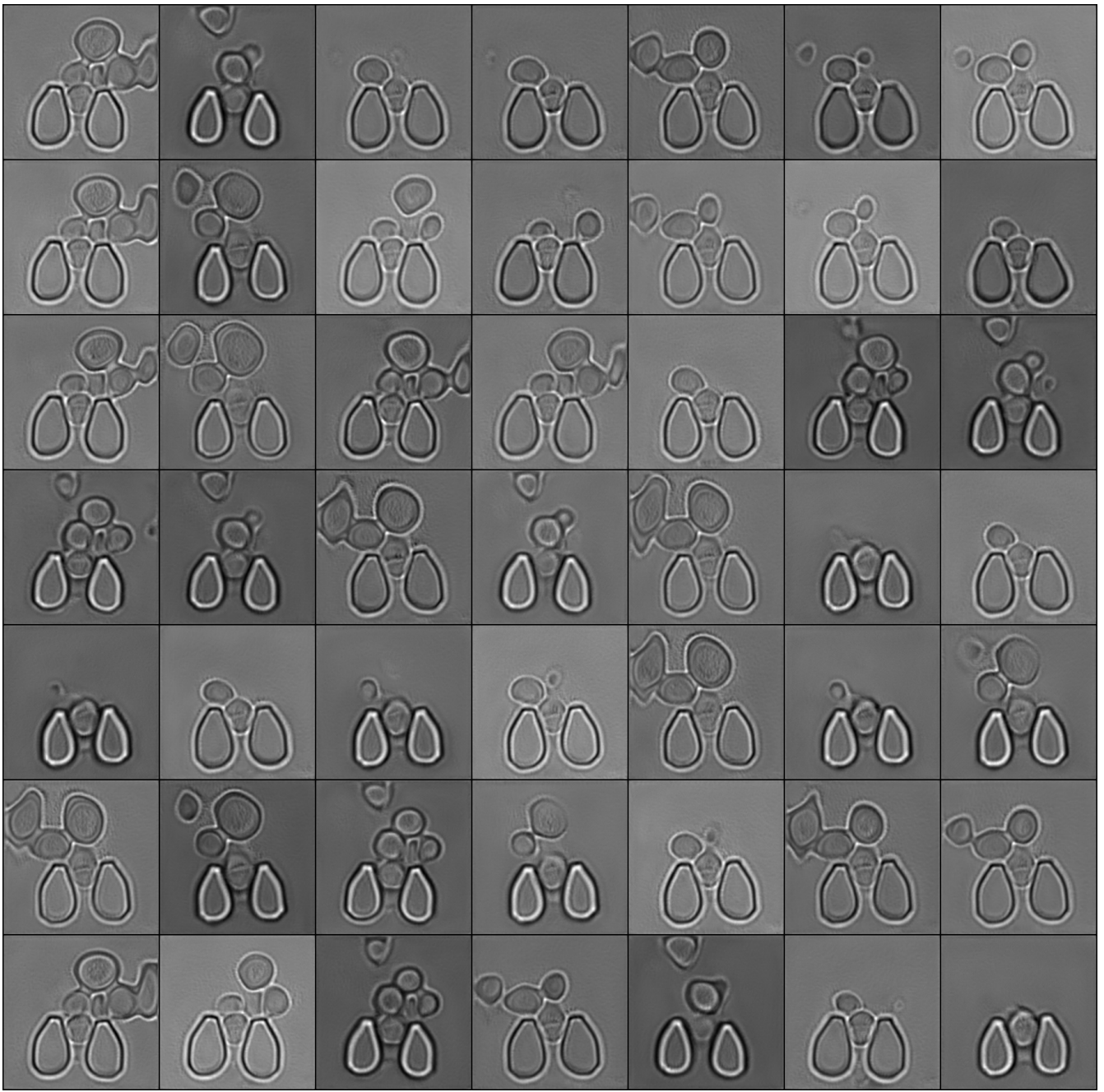

**Figure 9.19:** Unconditionalized generated samples from the Cell-GAN generator after 21 epochs.

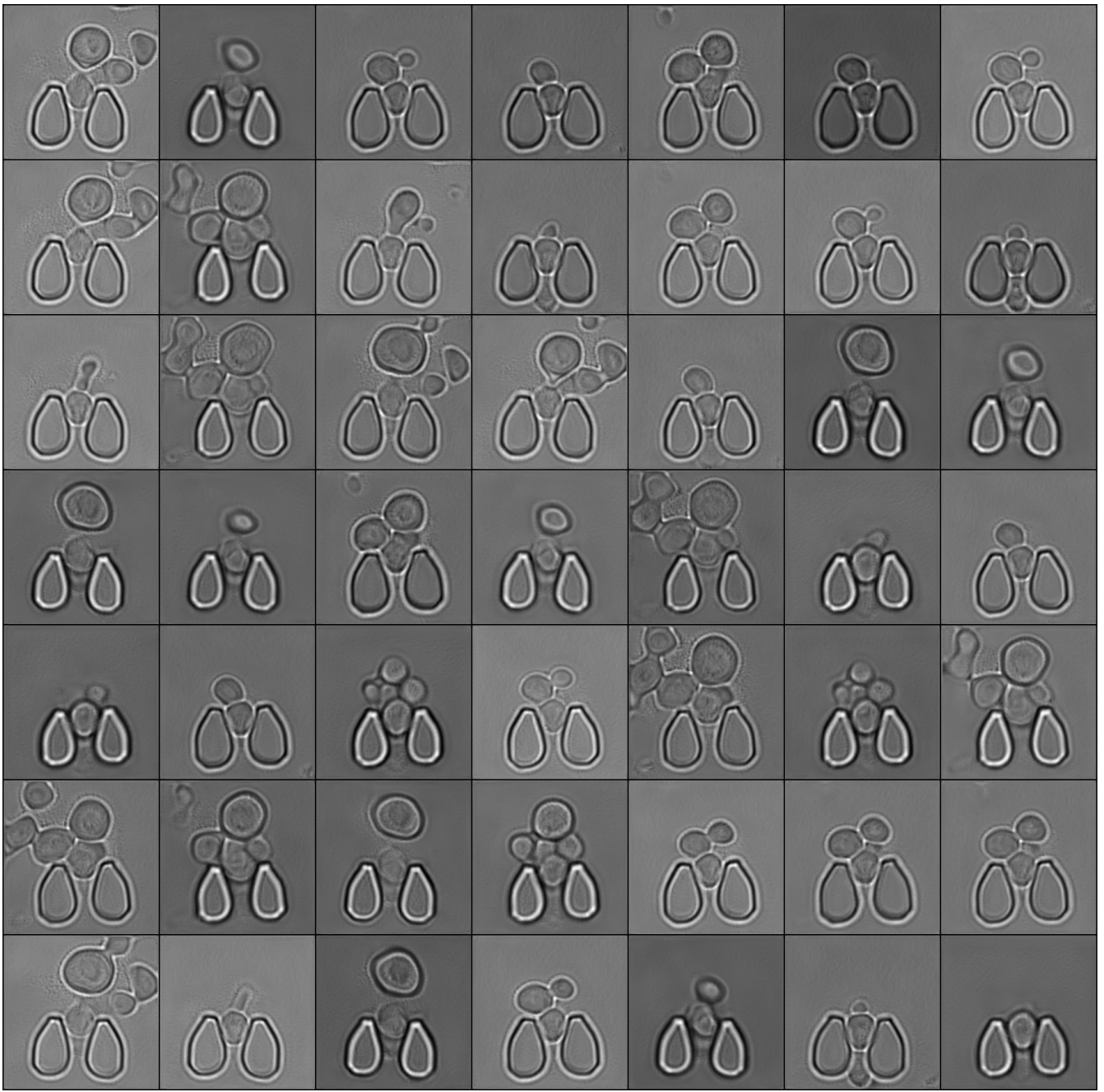

**Figure 9.20:** Unconditionalized generated samples from the Cell-GAN generator after 25 epochs.

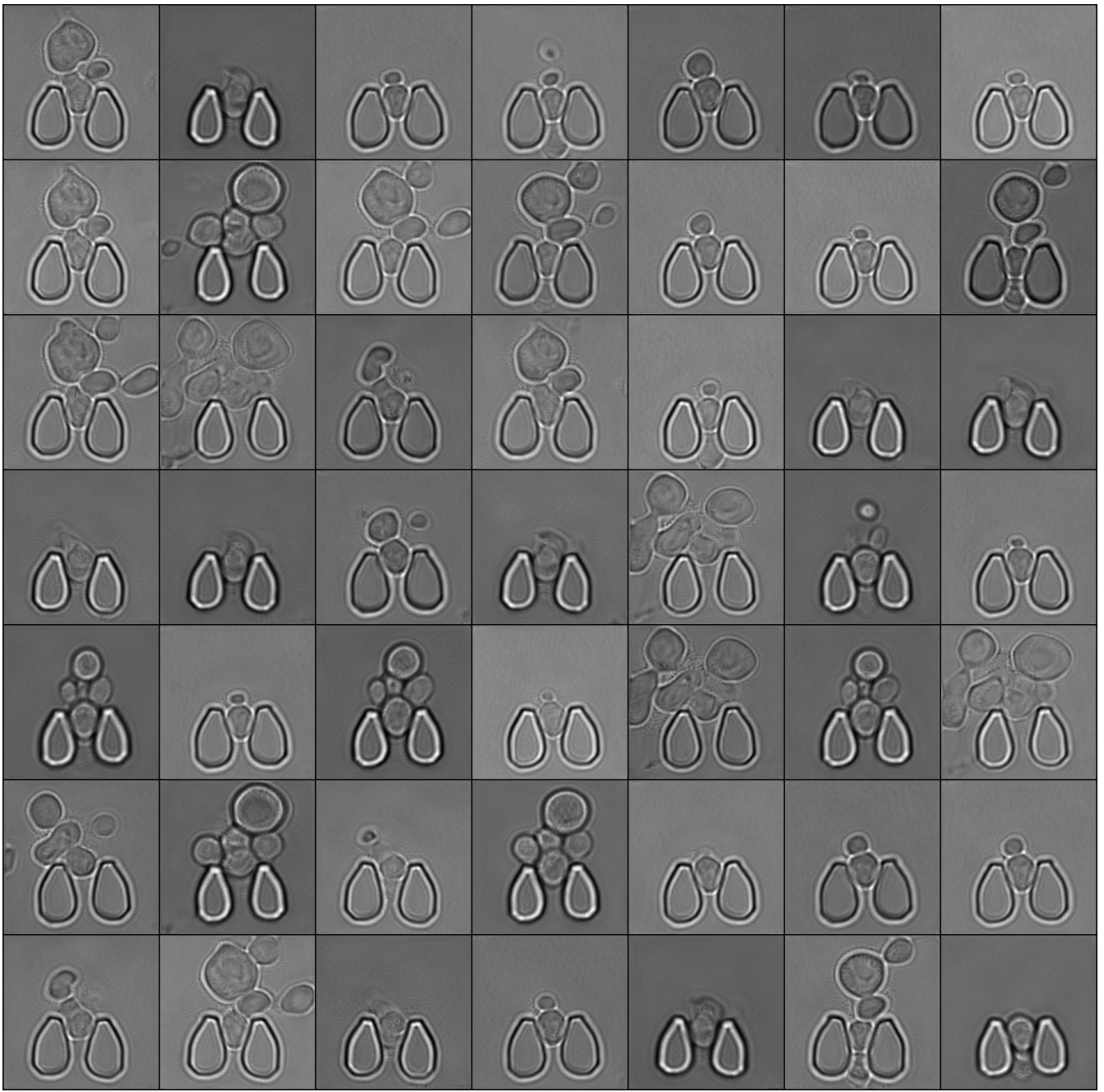

**Figure 9.21:** Unconditionalized generated samples from the Cell-GAN generator after 29 epochs.

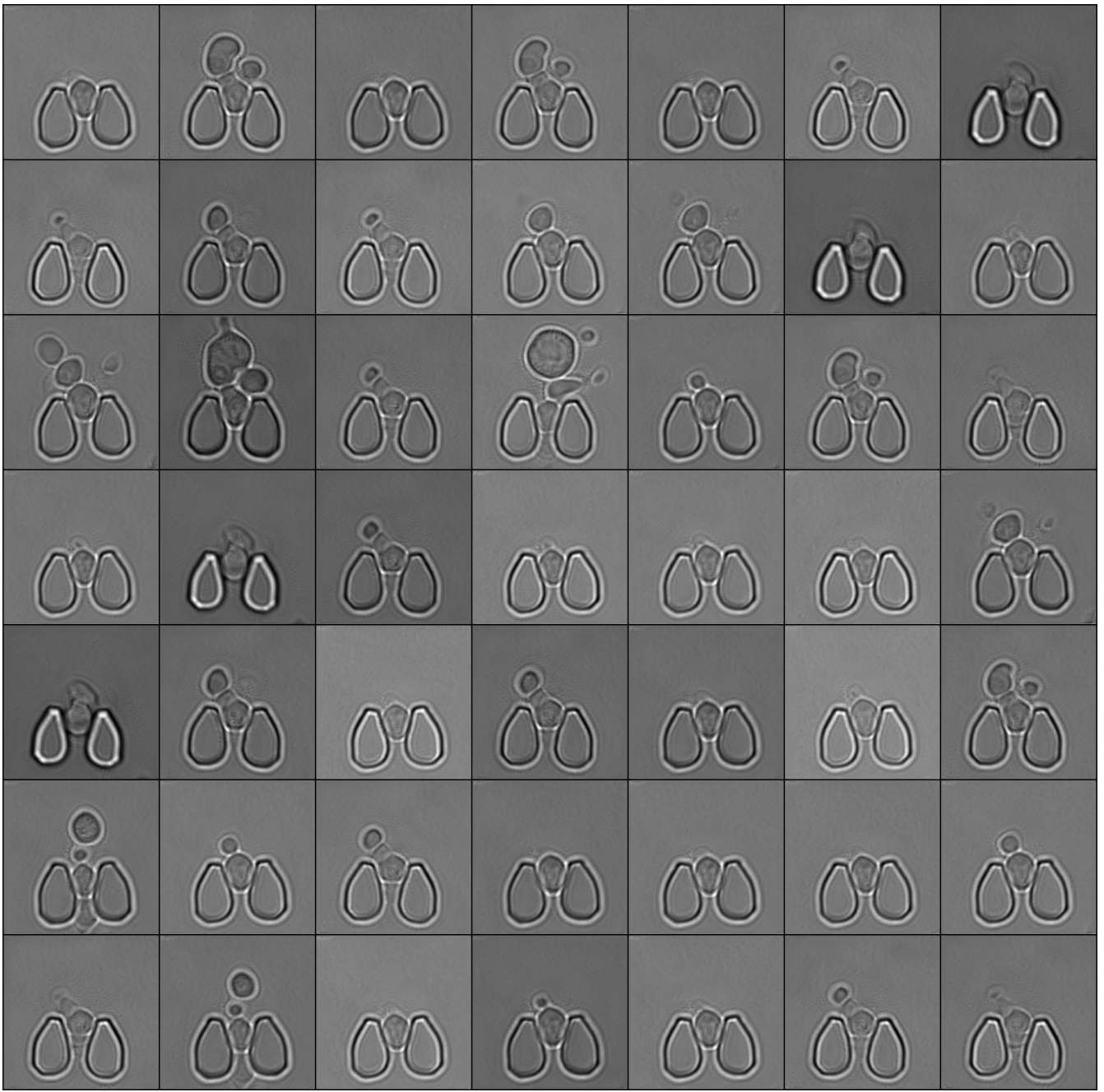

**Figure 9.22:** Conditionalized generated samples from the Cell-GAN guidance encoder and generator for real images.

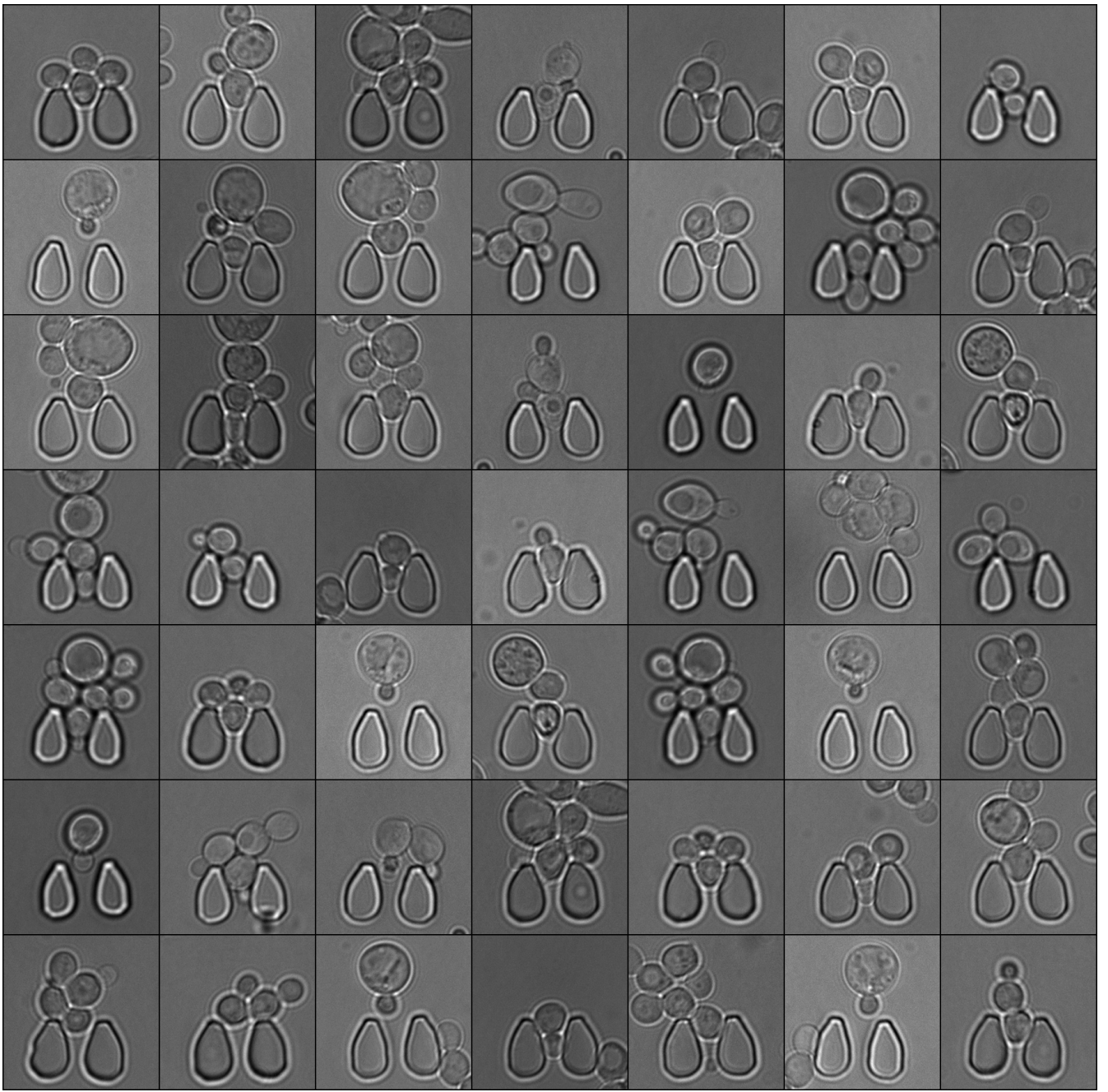

**Figure 9.23:** Real guidance images used to generate the samples in figure 9.22.

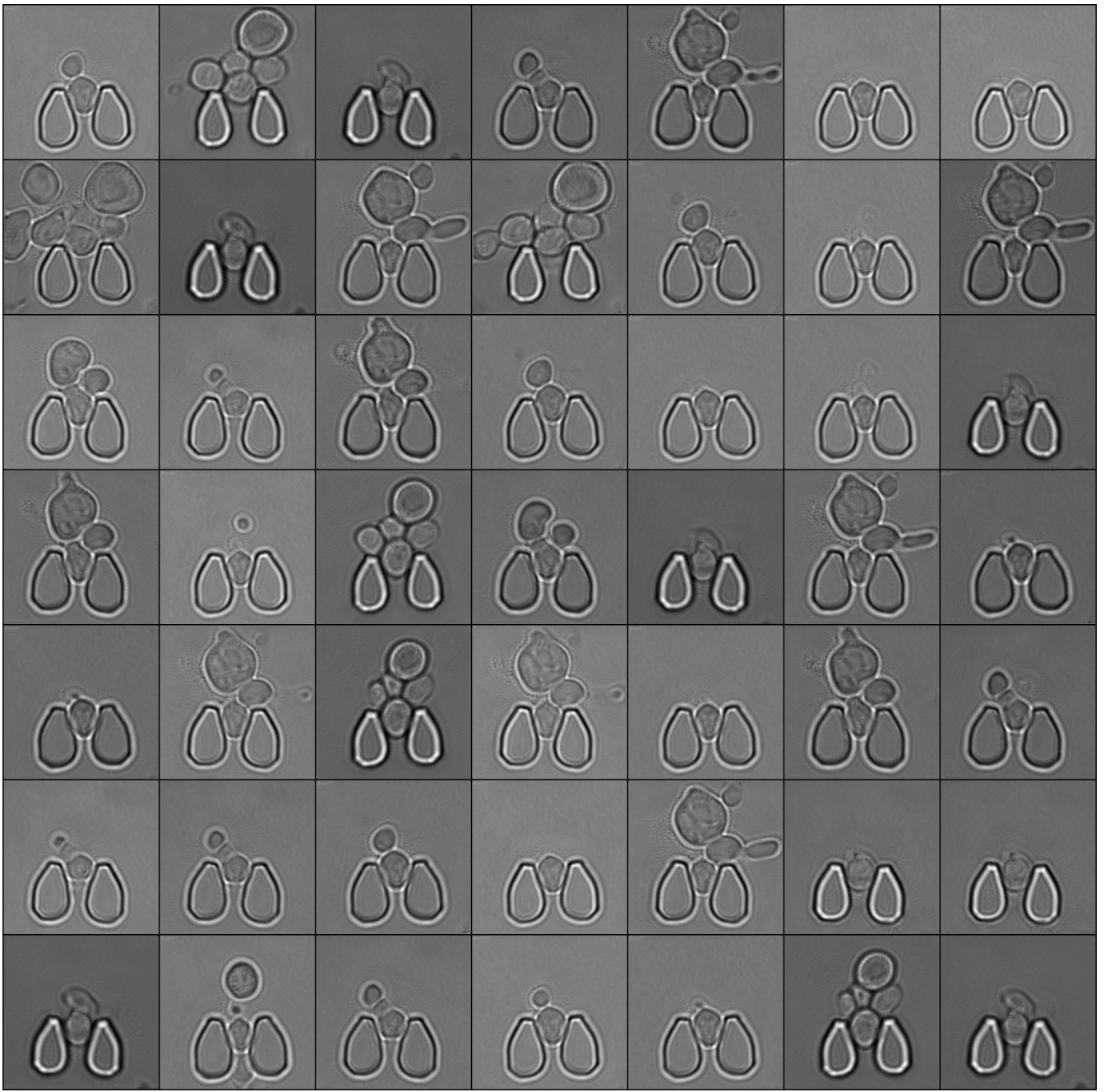

**Figure 9.24:** Conditionalized generated samples from the Cell-GAN guidance encoder and generator for real fake.

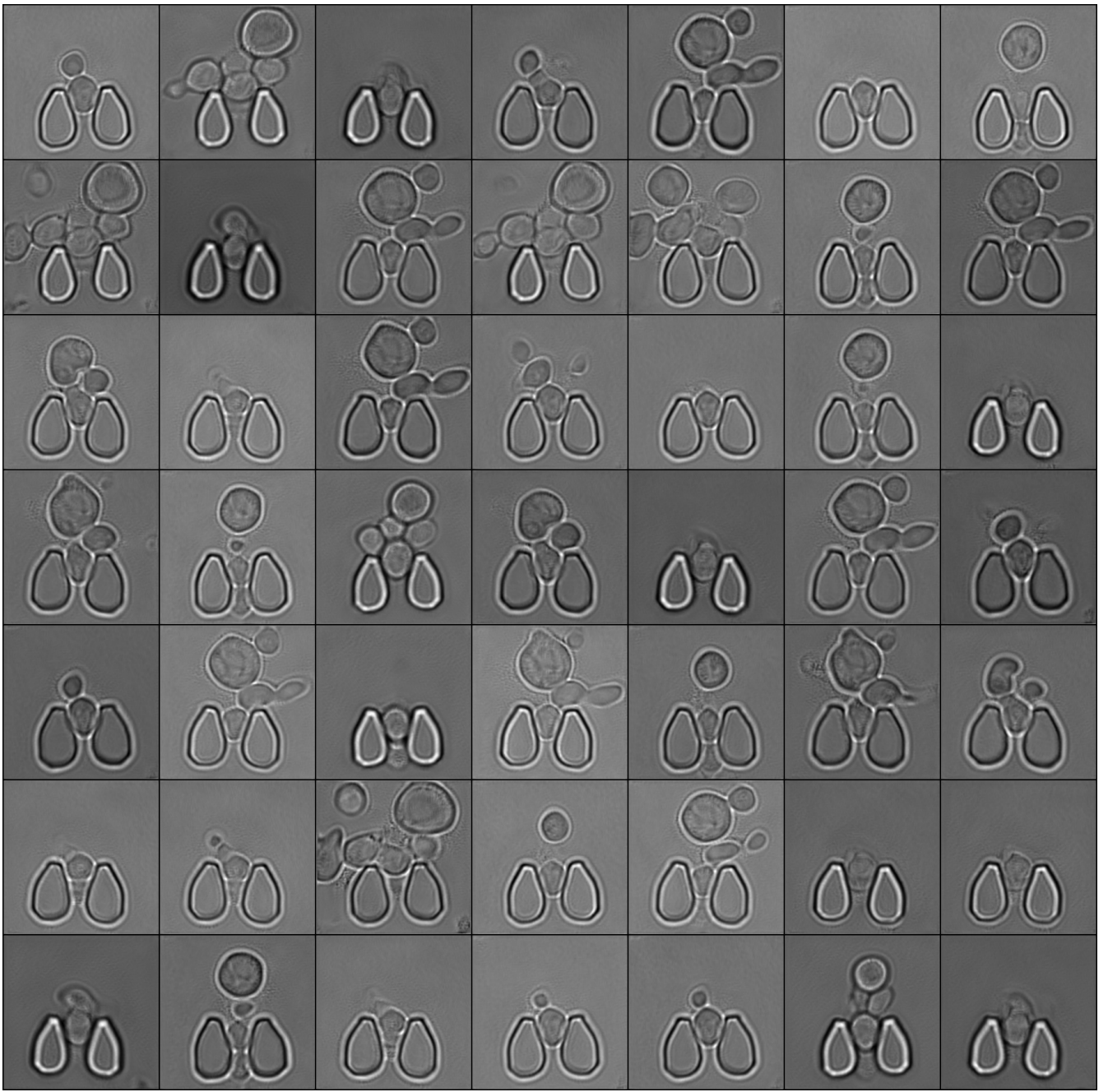

**Figure 9.25:** Fake guidance images used to generate the samples in figure 9.24.

## 9.5 SDC-Net++

This section shows additional predictions of the SDC-Net++ after different training steps.

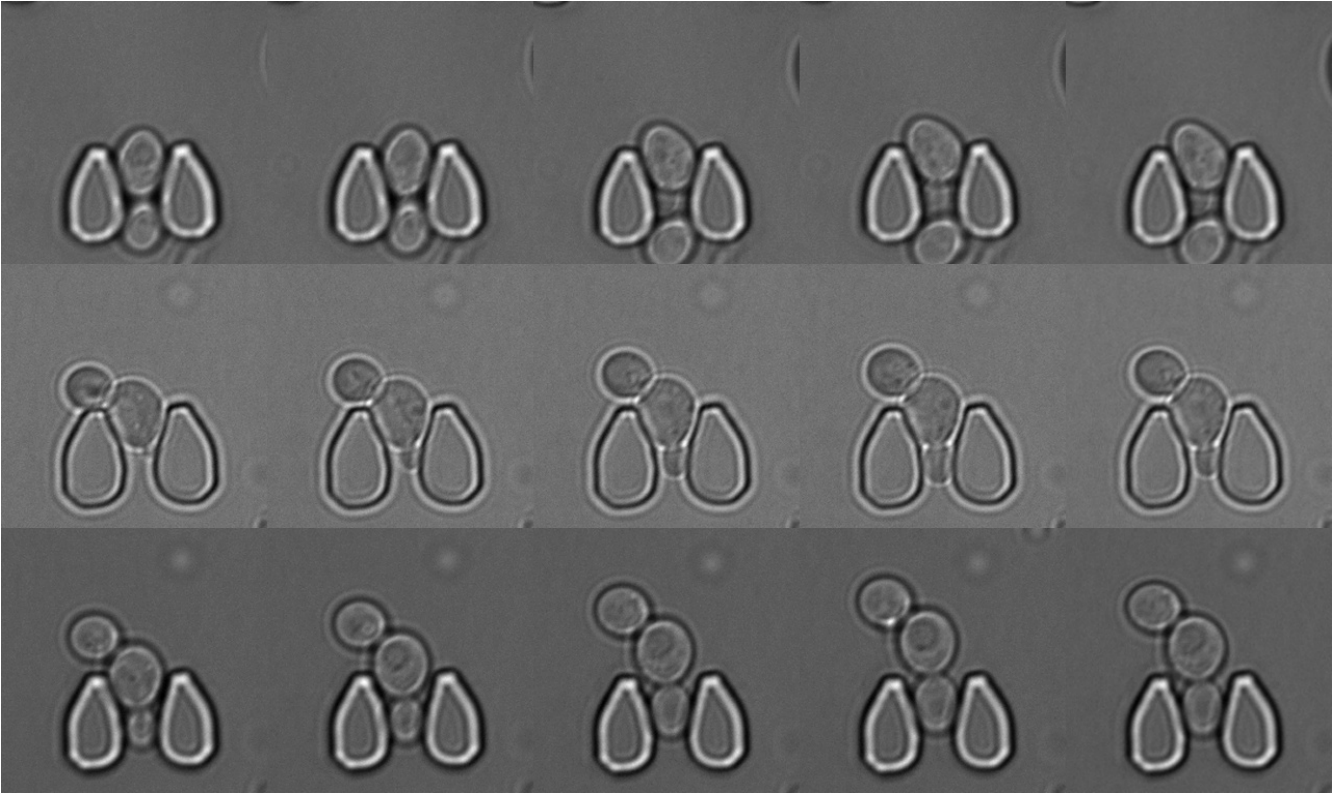

**Figure 9.26:** Future frame prediction result of the SDC-Net++ after flow training stage. Input frames in the three left columns, predicted future frames in the right column.

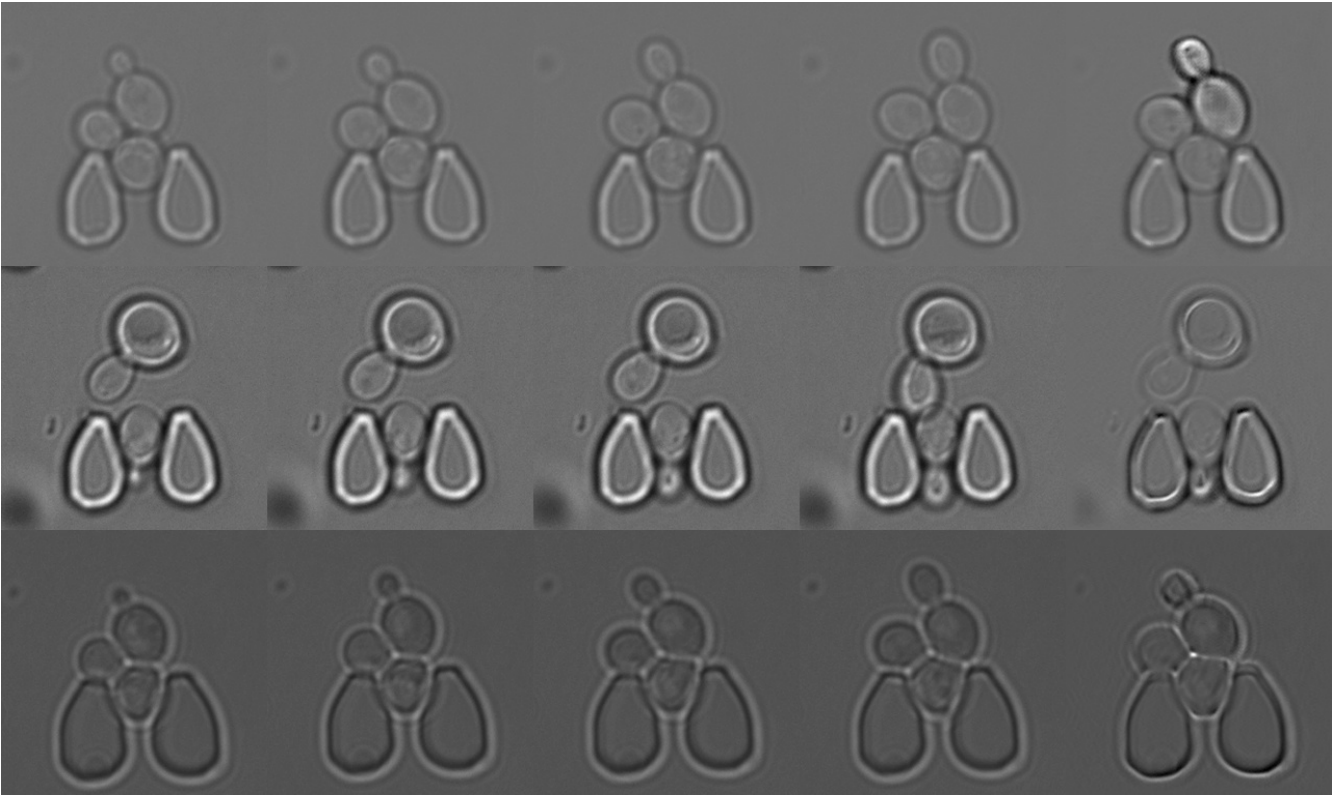

**Figure 9.27:** Future frame prediction results of the SDC-Net++ after the kernel training stage. Input frames in the three left columns, predicted future frames in the right column.

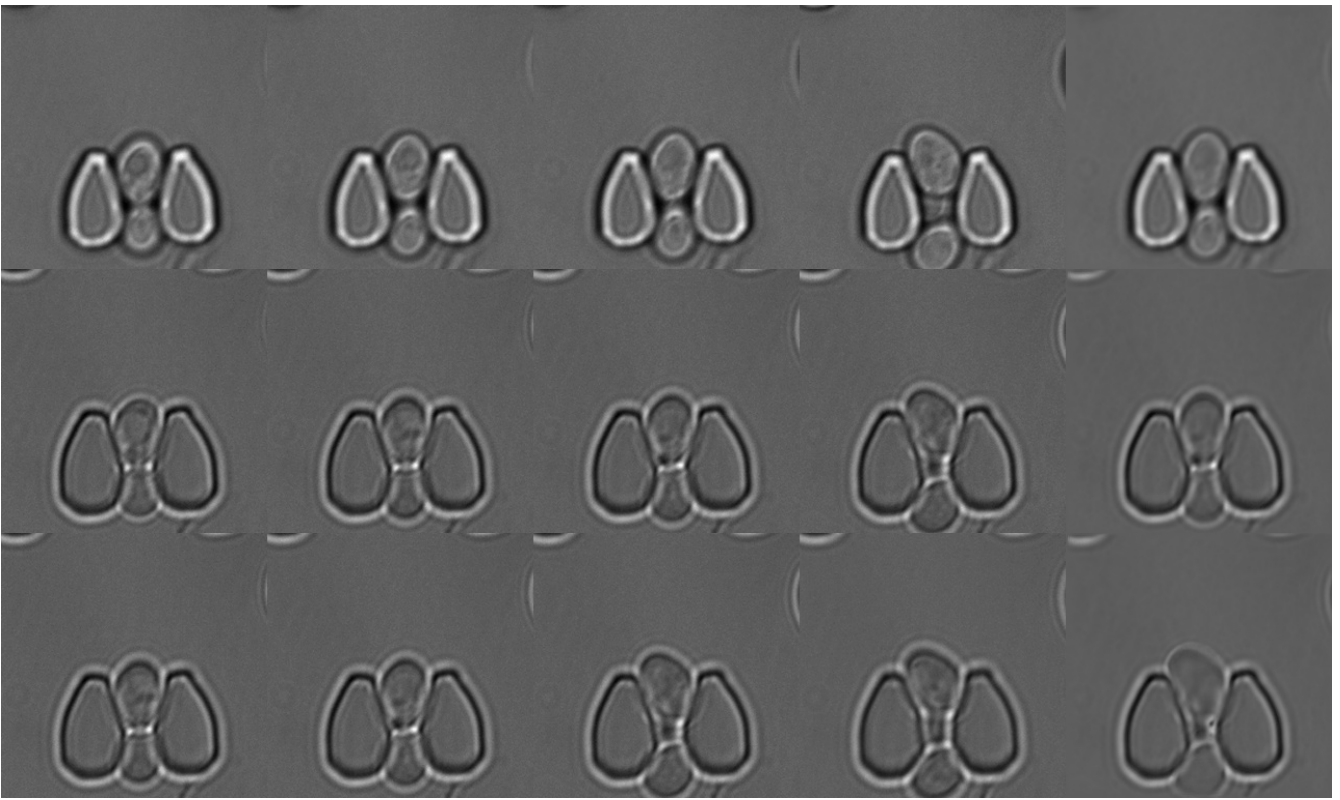

**Figure 9.28:** Future frame prediction results of the SDC-Net++ after the fine-tuning training stage. Input frames in the three left columns, predicted future frames in the right column.

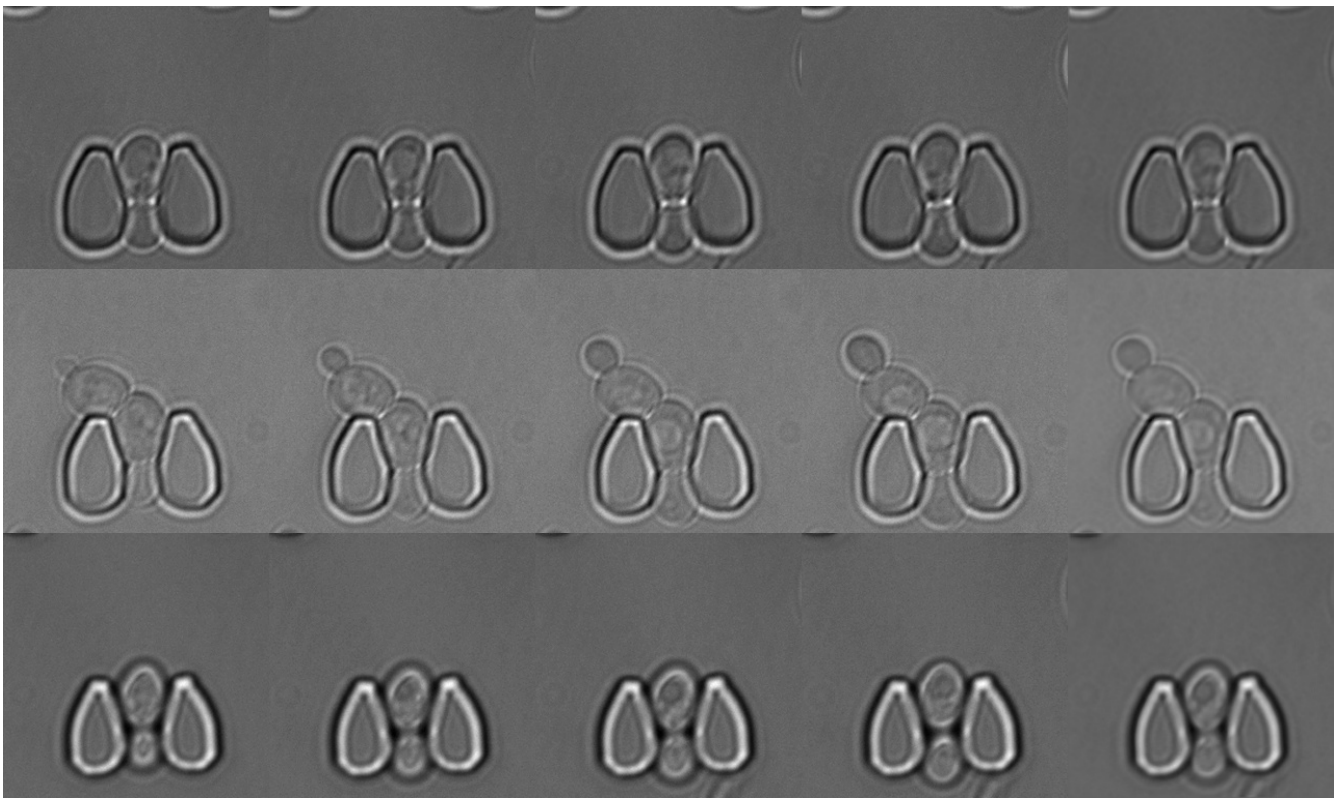

**Figure 9.29:** Future frame prediction results of the SDC-Net++ after the multi-prediction training stage. Input frames in the three left columns, predicted future frames in the right column.

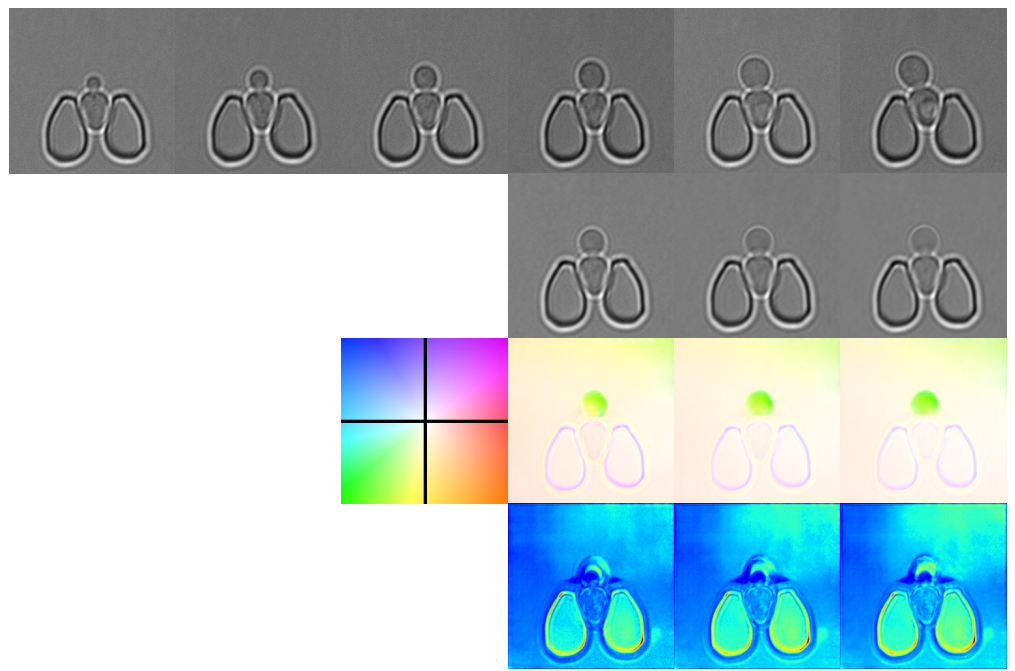

**Figure 9.30:** Multi future frame prediction of the SDC-Net++ with motion vectors and kernels. Motion and kernel visualization describe in 6.8.

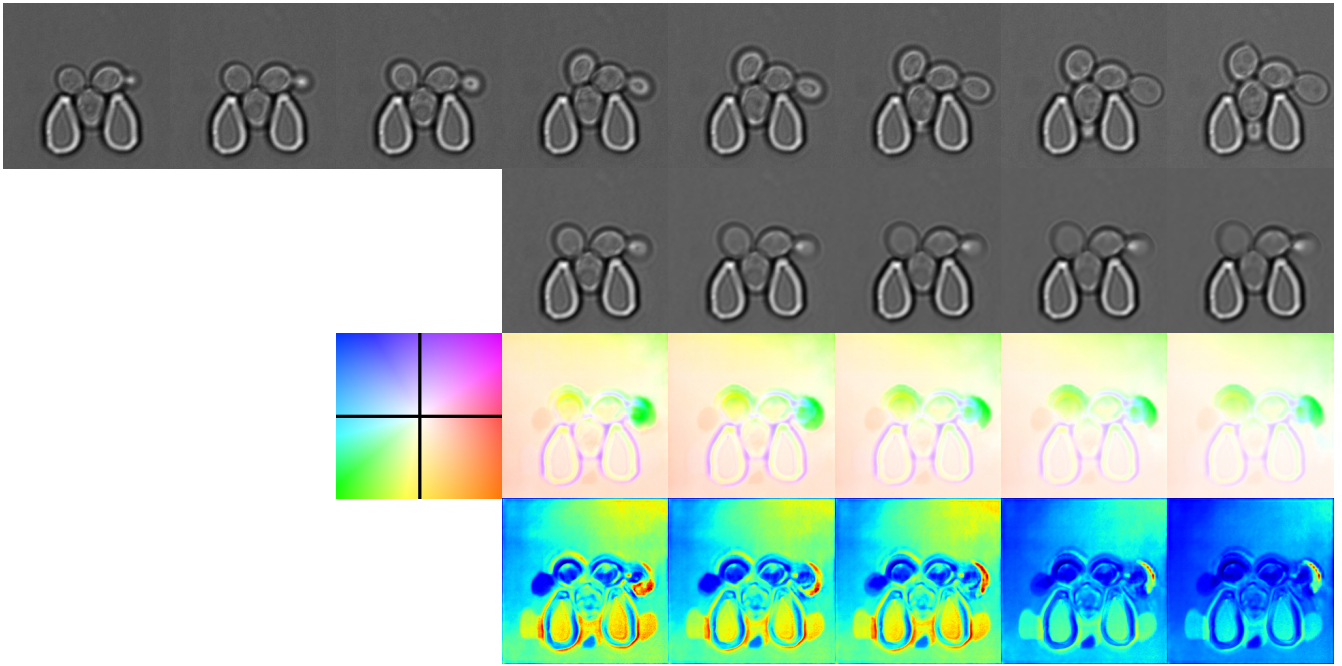

**Figure 9.31:** Multi future frame prediction of the SDC-Net++ with motion vectors and kernels.

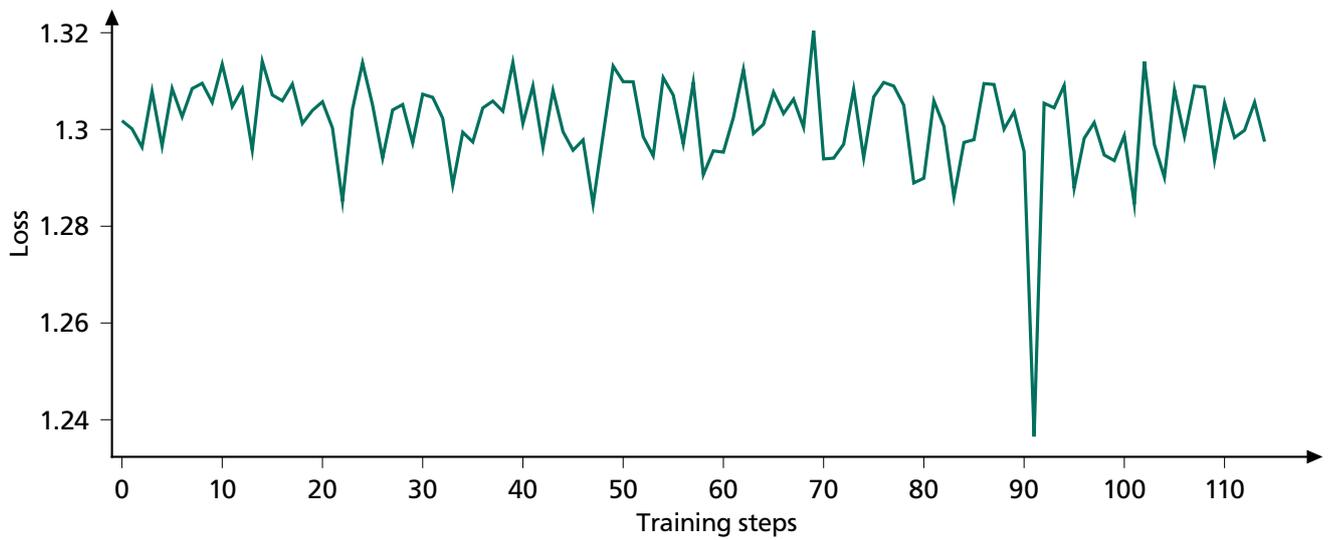

**Figure 9.32:** SDC-Net++ flow training loss curve.

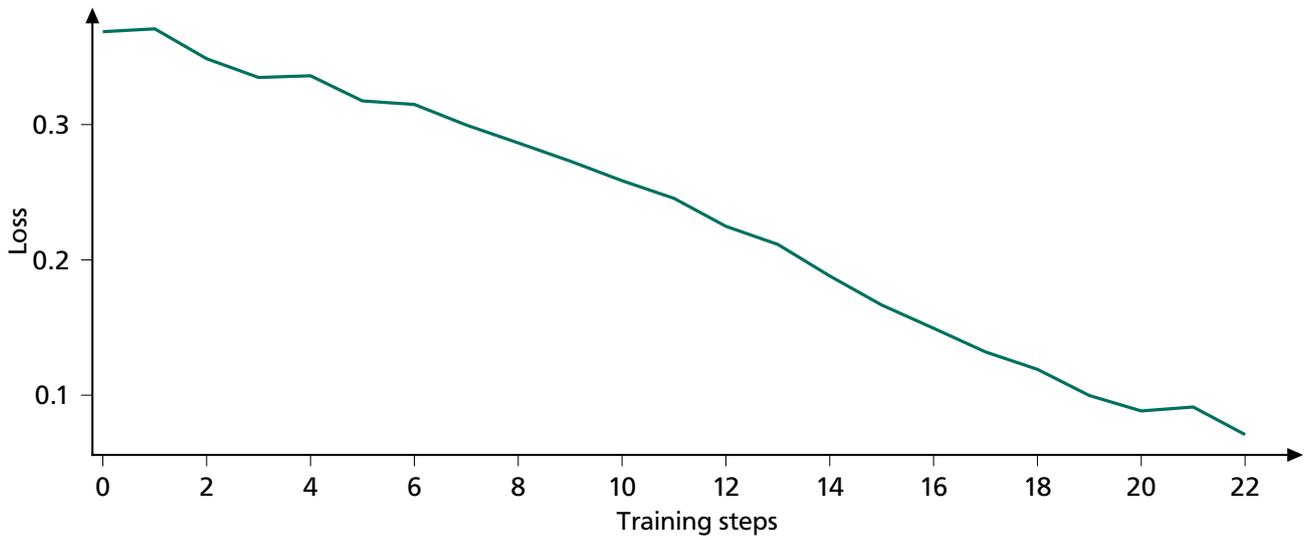

**Figure 9.33:** SDC-Net++ kernel training loss curve.

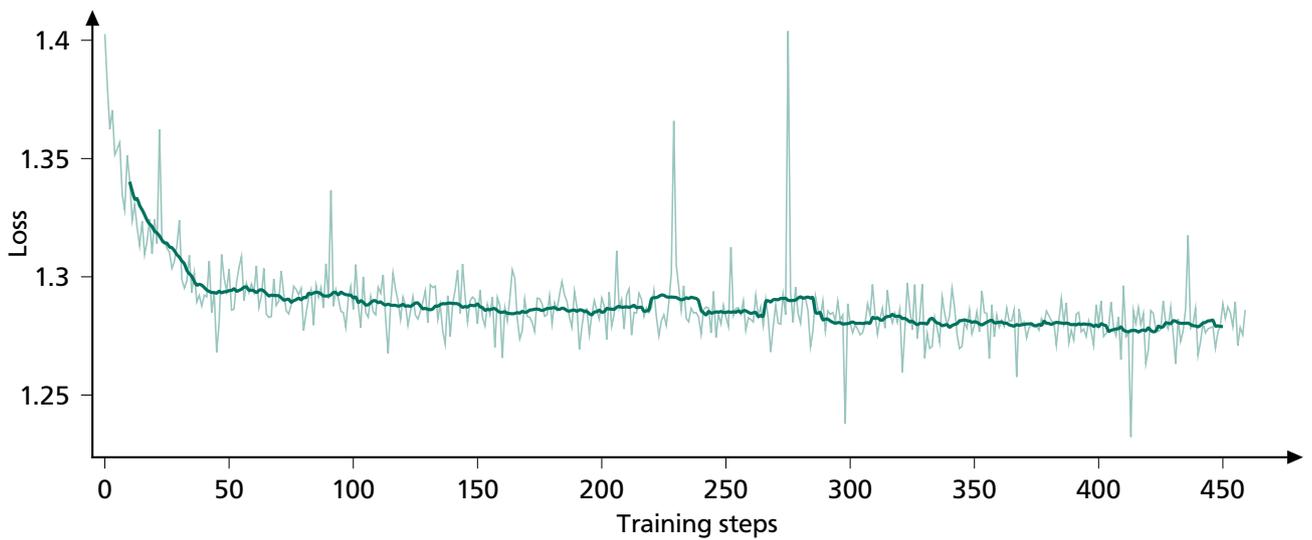

**Figure 9.34:** SDC-Net++ fine-tune training loss curve. Running average in green ■, computed with a window size of 20.

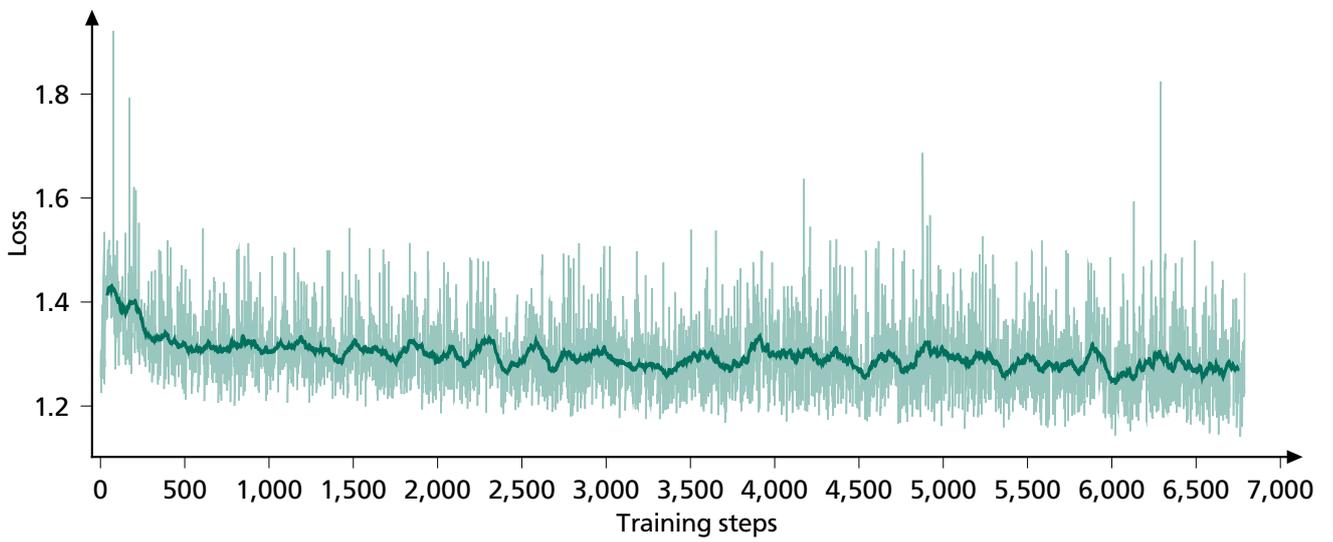

**Figure 9.35:** SDC-Net++ multi frame prediction training loss curve. Running average in green ■, computed with a window size of 80.